\documentclass[useAMS,usenatbib]{mn2e}
\usepackage{graphics}
 \usepackage{rotating}
 \usepackage{epsfig}
    \usepackage{times}
\usepackage{lscape}
\usepackage{longtable}   
\usepackage{tabularx}
    \usepackage[T1]{fontenc}
    \usepackage[latin1]{inputenc}
    \usepackage{fancyhdr}                    
    \usepackage{setspace}
 \usepackage{txfonts}
\usepackage[a4paper=true,ps2pdf=true,pagebackref=false]{hyperref}
\hypersetup{colorlinks = true,linkcolor = red, breaklinks= true,hyperindex=true,  anchorcolor = red, citecolor = blue, filecolor = red,   pagecolor = red,  urlcolor = red}
\title[X-ray identification of FIRST]{X-ray Identifications of FIRST Radio Sources  in the xBo\"{o}tes Field}
\author[K. EL Bouchefry 2009]{K. EL Bouchefry$^{1}$\thanks{E-mail:
kelbouchefry@gmail.com}\\
$^{1}$Astrophysics and Cosmology Reseach Unit, University of Kwazulu-Natal, Westville, 4000, South Africa}

\begin{document}

\date{Accepted 2009 April 14. Received 2009 April 09; in original form
  2008 June 03}

\pagerange{\pageref{firstpage}--\pageref{lastpage}} \pubyear{2009}

\maketitle

\label{firstpage}

\begin{abstract}
With the goal of investigating the nature and the environment of the faint radio sources (at ${\rm mJy}$ level), here  are presented results of  X-ray identifications of Faint Imaging Radio Survey at Twenty centimetres (FIRST) in the 9 square degrees Bo\"{o}tes field of the NOAO Deep Wide Field Survey (NDWFS), using data from the Chandra XBo\"{o}tes survey. A total of 92 ($10\%$) FIRST radio sources are identified above the X-ray flux limit $f_X\,(0.5-7)~{\rm keV} = 8\times10^{-15}~{\rm erg~s^{-1}~cm^{-2}}$, and 79 optical counterparts are  common to both the radio and X-ray sources.  Spectroscopic identifications   were available for 22 sources ($27\%$). Multi-wavelength optical/infrared photometric data  ($Bw\sim 25.5$ mag, $R\sim25.8$ mag, $I\sim25.5$ mag and $K\sim19.4$ mag) were available for this field and were used to derive photometric redshift for the remaining 57 sources without spectroscopic information. Most of the radio-X-ray matches are optically extended objects in the $R$ band with  a photometric redshift distribution peaking at $z\sim0.7$. Based on the hardness ratio and X-ray luminosity, 37 sources (89\%) were classified as AGN-1, 19 as AGN-2, 12 as QSO-1, 2 as QSO-2 and 9 sources as normal galaxies. While the majority of these sources have a hard X-ray luminosity $L_X(2-7)$ keV $>10^{42}$ erg s$^{-1}$, about one third of the sources have $L_X(2-7)$ keV $>10^{44}$ erg s$^{-1}$ and therefore classified as QSO-1. The majority (68\%) of the radio-X-ray matched population are found to have $\log\, f_X/f_{opt}$ within $0.0\pm 1$ region indicative of AGNs, $23\%$ with high X-ray-to-optical flux ratio ($\log\,f_X/f_{opt} > 1$), suggesting high redshift and/or dust obscured AGN, and $11\%$  of the radio-X-ray matches  that are X-ray faint optically bright sources with $\log\, f_X/f_{opt} <-1$, and most of these sources are optically extended. These objects  are low-$z$, normal galaxies or low luminosity AGNs (LINERS). 
\end{abstract}

\begin{keywords}
galaxies: active--galaxies: high-redshift--X-rays: diffuse back ground--X-rays:
galaxies
\end{keywords}

\section{Introduction}
X-ray surveys have played a crucial role in our understanding of the
nature of the sources that populate the X-ray universe. Early
surveys like the Einstein sensitivity survey \citep{gioia90},
ROSAT(Roentgen Satellite) International X-ray/optical  survey
\citep{ciliegi97} and the ASCA ( Advanced Satellite for Cosmology
and Astrophysics) Large Sky Survey \citep{akiyama2000} showed that
the vast majority of X-ray sources were AGN. In particular, in
shallow wide area surveys in the soft (0.5-2 keV) X-ray band, most
of the sources detected are unobscured, broad line AGN, which are
characterised by a soft X-ray spectrum with a photon index $\Gamma
= 1.9$ \citep{nandra94}. On the other hand, the number density of
AGN identified in the hard X-ray and mid-IR bands is far greater
than that  found in any optical surveys of comparable depth
\citep[e.g.][]{bauer04, stern02a}.  \\


A combination of radio data with optical and X-ray photometry provides a
considerable wealth of information on the nature of the faint radio/X-ray
population \citep{stocke91}.  Deep radio survey at $1.4\,{\rm GHz}$ observations consist of two main populations: AGN and star-forming galaxies \citep{condon84, windhorst85}. At this frequency the primary emission mechanism in virtually all radio sources is synchrotron  emission from relativistic electrons spiralling around
the magnetic field within the host galaxy. The radio emission can be related either to ongoing star-forming processes or to the accretion activity onto the central super-massive black hole within  the galaxy. At very faint flux densities ($\mu{\rm Jy}$), studies have shown that the radio population is provided by star
forming galaxies, which produce non-thermal radio continuum  at $1.4\,{\rm GHz}$ through synchrotron emission from supernovae remnants, while other studies suggested that the radio emission may be caused by `radio-quiet' AGN, rather than star-forming galaxies \citep[e.g. ][]{simpson06, jarvis04}. At the Jansky and mJy levels, the bulk of the radio source population is dominated by powerful AGNs. Observations of radio emission of AGNs provide clues to the accretion history of the universe and important information in our understanding of how
central black holes grow over cosmic time as well as the physical process behind these powerful objects, especially when combined with X-ray observations \citep[e.g.][]{merloni03}. They also provide clues to some properties of the interstellar and intergalactic medium.\\ 

Previous studies on cross correlations between radio and X-ray
surveys have determined a link between the star formation rate and
the  radio and X-ray luminosities of star forming galaxies
\citep[e.g.][]{condon92, ranalli03}, which can be combined to
determine  relationships between radio and X-ray luminosities for
star forming galaxies. \citet{bauer02} have conducted a study of
the faint X-ray and radio sources using a 1 Ms Chandra data set
and ultra deep VLA observations, finding a large overlap between
faint X-ray and radio sources. These authors have shown that the
X-ray and radio luminosities are correlated for nearby late type
galaxies, and therefore used it to estimate the star formation
rate in their sample. Other studies of the radio-X-ray correlations involve studies of the radio properties
of the hard X-ray sources. For example, \citet{barger01}, cross
correlated the VLA data ($S_{1.4\;GHz}=25\;\mu$Jy), with deep
Chandra survey ($f_X(2-10$ keV) $=3.8\times 10^{-15}$ erg
s$^{-1}$ cm$^{-2}$) finding about $50\%$ radio counterparts to
their X-ray sources. \citet{ciliegi03}, have performed a study of
the radio properties of the X-ray sources detected in the HELLAS
survey ($S_{5\;GHz}$= 0.3 mJy; $f_x(5-10 $ keV) $= 5-4\times
10^{-14}$ erg cm$^{-2}$ s$^{-1}$). They find an identification
rate of $30\%$ much higher than that of the soft band (0.5-2
keV). At somewhat fainter fluxes ($f_X(2-8$ keV)
$=7.7 \times 10^{-15}$erg cm$^{-2}$ s$^{-1}$, $
S_{1.4\;GHz}=60\;\mu$J), \citet{geor04} combined a single 50 ks
XMM-Newton pointing with an ultra deep radio
survey, finding an identification rate of $\sim 33\%$. \\

 In  previous papers \citep{khadija09, khadija08, elbouchefry07}, a detailed study on the optical/infrared identifications of FIRST radio sources in Bo\"{o}tes field of the NDWFS survey was performerd. A total of 688/900 FIRST radio sources have been identified in one or more bands (\textit{Bw, R, I, K}) in Bo\"{o}tes field. This field covers a large area of about 9.3 square degrees and  has been surveyed almost across the full electromagnetic spectrum: in UV (GALEX), optical (NDWFS), infrared \citep[FLAMEX;][]{elston06}, radio (VLA/FIRST; \citeauthor{becker95} \citeyear{becker95} and WSRT; \citeauthor{devries02} \citeyear{devries02}), and  X-ray \citep{murray05, kenter05}. The X-ray properties of FIRST radio sources have been studied in a few papers. For example, \citet{brinkmann00} compiled a large sample of X-ray selected AGN from the ROSAT All Sky Survey \citep{voges99} and cross correlate this sample with the FIRST radio survey, finding a relatively tight linear correlation between the logarithmic of radio luminosity and the X-ray luminosity in the soft X-ray 0.5-2 keV band. At bright flux limits, $f_X(2-10)$ keV $ = 10^{-13}$ erg s$^{-1}$ cm$^{-2}$, \citet{akiyama2000}  cross correlated the ASCA Large Sky Survey (LSS) with the FIRST catalogue, finding an identification rate of a $\sim$  $35\%$  and a fraction of radio loud hard X-ray selected sources of approximately $10\%$. At faint X-ray flux $f_X(0.5-2)$ keV=$1.5 \times 10^{-16}$ erg s$^{-1}$ cm$^{-2}$, four radio sources from FIRST survey have been identified as a radio counterparts to four X-ray sources in the Large Area Lyman Alpha survey (LALA) Bo\"{o}tes field \citep{wang04}. \citet{sutil06} cross correlated the FIRST radio sources with an ULX (Ultra Luminous X-ray) catalogue, and they found 70 positional coincidence and the majority of them  are associated with the galaxy nucleus.  The aim of this paper is to shed light on the nature and the environment of the faint radio population (at ${\rm mJy}$ level) detected in the FIRST radio survey, to characterise their X-ray counterparts and to distinguish between different groups of AGNs. Here, I combine data from the FIRST radio survey with a new, medium depth (5 ks/pointing) Wide field X-ray survey (known as the Chandra XBo\"{o}tes) of the Bo\"{o}tes field of the NDWFS survey. The X-ray data reach a limiting flux of $4\times 10^{-15}$ erg s$^{-1}$ cm$^{-2}$ in the soft band and $8\times 10^{-15}$ erg s$^{-1}$ cm$^{-2}$ in the full band. Compared to the previous studies this data set, the FIRST/XBo\"{o}tes, has the advantage of deep wide area optical/infrared  observations, radio and intermediate in depth X-ray observations.

The  paper is organised as follows: section 2 presents a  summary on the radio,  X-ray and optical  data. The matching procedure  is described in section 3. Optical morphology and optical magnitude of the radio-X-ray matches are investigated in section 4 and 5 respectively.  Section 6 presents the optical classification of the radio-X-ray matches, photometric redshift and also investigates the $K-z$ diagram. Section 7 investigates the radio-X-ray luminosities. Section 8 presents the X-ray-to-optical flux ratio and Extremely red objects. Section 9 is devoted to the optical and X-ray properties of the radio-X-ray matches. Section 10 presents the final data table,  and conclusions are summarised in section 11. Throughout this paper it is assumed that $\Omega_M=0.3$, $\Omega_\Lambda=0.7$ and H$_0$=70 km s$^{-1}$ Mpc$^{-1}$.

\section{The sample data}
\subsection{The FIRST catalogue}
The radio data are from the 2002 version of the  FIRST Very Large Array
catalogue\footnote{The FIRST catalogue is available online at
  http://sundog.stsci.edu} \citep[Faint Images of the Radio Sky at
Twenty-Centimetres; ][]{becker95}, and it is  derived from 1993
through 2002 observations. The FIRST  radio survey  has been
carried out in recent years   with the VLA in its B-configuration
to produce a map of 20 cm (1.4 GHz)  sky with  a beam size of 5.4
arcsec and an rms sensitivity of about 0.15 mJy/beam. The 2002
version of the catalogue covers a total of about 9033 square
degrees of the sky (8422 square degrees in the north Galactic cap
and  611 square degrees in the south Galactic cap); and contains
811,117 sources from the north and south Galactic caps. The
accuracy of the radio position depends on the brightness and size
of the source and the noise in the map. Point sources at the
detection limit of the catalogue have positions accurate to better
than 1 arcsec at $90\%$ confidence; 2 mJy  point sources typically
have positions good to 0.5 arcsec. The radio surface density is $\sim 90$
deg$^{-2}$; and about 900 sources fall within the NDWFS Bo\"{o}tes field.

The FIRST survey provides high spatial resolution (5 arcsec beam). As a result, many large radio sources are resolved out and split into multiple components. In order to identify groups of sources that are likely to be sub-components of a single source, an algorithm similar to that used by \cite{maglio98}  is adopted for identifying genuine double radio sources in the FIRST survey. All doubles with $\theta\leq 10\sqrt{{\rm F_{tot}}}$ are considered as a single object, where $\theta$ is their separation in arcsec and ${\rm F_{tot}}$ is their summed flux density in mJy.  The technique is  based on the $\theta \propto S$ relation found by \citet{oort87}. Among the 900 FIRST radio sources that cover the NDWFS survey (Bo\"{o}tes field,  145  (16\%) are double, 30  (3\%) have three components, 6 (0.6\%) sources have four components and 8 (0.8\%) sources have more than five components.

\subsection{The XBo\"{o}tes catalogue}
The X-ray data \citep{murray05, kenter05} used in this paper are from the Chandra
XBo\"{o}tes surveys. The XBo\"{o}tes catalogue contains $\sim 3213$ X-ray
point sources and is publicly available through the NOAO Deep Wide Field
Survey (NDWFS)
homepage\footnote{http://www.noao.edu/noao/noaodeep/XBootesPublic/index.html}. The
Chandra XBo\"{o}tes survey imaged a large and contiguous area of 9.3 square
degrees of the Bo\"{o}tes field of the NDWFS survey, and is centred on RA
(J2000) $\sim 14^h 32^m$ and Dec (J2000) $\sim 34^{\circ} 06'$. The field was
observed by the advanced CCD Imaging Spectrometer (AGIS-I) on the Chandra
X-ray Observatory, and the data have arcsecond resolution and broad energy
response up to 10 keV. The X-ray data was taken in 126 separate pointings, each
observed for 5 ks. The X-ray data are filtered in the three energy bands :
$0.5-7$ keV, $0.5-2$ keV, $2-7$ keV with a  limiting flux of $\sim 8\times
10^{-15}$ erg s$^{-1}$ cm$^{-2}$ in the full band ($0.2-7$ keV) and of $\sim 4
\times 10^{-15}$ erg s$^{-1}$ cm$^{-2}$ in the soft band ($0.5-2$ keV).

\subsection{The NDWFS survey (Bo\"{o}tes field)} 

The NOAO Deep Wide Field Survey (NDWFS) is a deep multi-band imaging (\textit{Bw, R, I, J, H, K}) designed to study the formation and evolution of large scale structures
(Jannuzi et al. 1999; Brown et al. 2003). This survey consists of two fields\footnote{http://www.noao.edu/noao/noaodeep/}; the first one  is
located in Bo\"{o}tes field centred on approximately $\alpha = 14^{h} \;
30^{'}\; 05.7120^"$, $\delta = +34^{\circ} 16^{'} 47.496^{"}$, covering a 3 by
 3 square degrees region, and the latter one is located in a 2.3 by 4
square degrees region in Cetus field.  The survey catalogue has been split by
declination range into four strips ($32^{\circ}\leq \delta <33^{\circ},
33^{\circ} \leq \delta <34^{\circ}, 34^{\circ} \leq \delta <35^{\circ},
35^{\circ}\leq \delta <36^{\circ}$); each strip observed in four bands (\textit{Bw, R, I, K}). The magnitude limits are: $Bw\sim 25.5$ mag, $R\sim25.8$ mag, $I\sim25.5$ mag and
$K\sim19.4$ mag.
\section {Cross identification method}
\begin{figure}
\begin{center}
\includegraphics{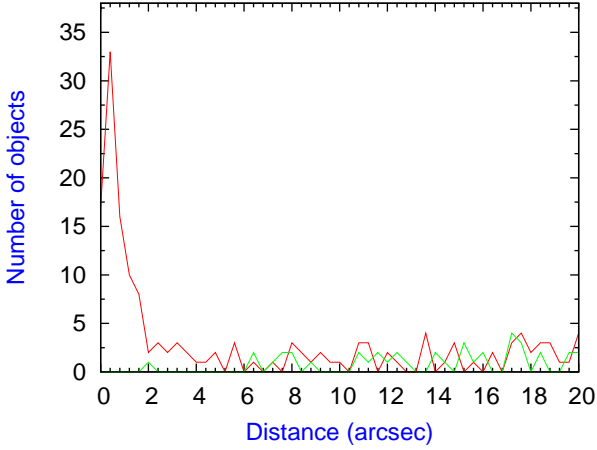}
\caption{Distribution of angular distances between radio and X-ray
positions (top line). The lower plot shows the expectation of
chance associations as a function of angular separation. Sources are considered to be associated if the separations are $\leqslant 2$ arcsec.} \label{xrayoffset}
\end{center}
\end{figure}

When searching for optical/infrared or X-ray counterparts of radio sources (or vice versa), one must decide on the matching criteria to be adopted.  A compromise is required in order to maximise both the completeness (i.e.  all potential radio sources are included) and the reliability (i.e all included identifications are genuine) of the data base. The best way to ensure completeness  and reliability is to determine a maximum position offset which includes all possible
identifications and to have some realistic estimate of the uncertainties in both the radio and X-ray positions. The FIRST radio survey provides  a high positional accuracy that is around $\sim 0.5''$.  The rms error of the X-ray positions  ranges from $0.0''$ to $6.1''$ depending on the off-axis angle. The mean uncertainty in the X-ray positions is estimated to be $\sim 1.68''$. Combining these in quadrature with the radio astrometry errors defines a circle about the predicted coordinates of the source; with a radius of $1.75''$ ($1\,\sigma$), within which the X-ray counterpart of the radio source is expected to be found.  \\
I first cross correlated both FIRST catalogue and XBo\"{o}tes based on simple
positional coincidence, and then took all the pairs whose radio and X-ray
positions differed by less than $20''$. The cumulative distribution of the
separations in angular distances between the X-ray and radio positions is shown in Figure
\ref{xrayoffset}. The upper curve  shows the distribution offset for all matches
between radio and X-ray positions. The lower curve shows the differential
distribution which would be expected if all identifications were chance
coincidence. To estimate the false match rate, all radio sources were shifted by $3'$, and the source matching algorithm was run again. As clearly seen from the plot, the radio-X-ray offset distribution exceed vastly the random distribution at offset $\leq2''$, meaning that real identifications of FIRST radio sources with X-ray objects dominate at smaller offsets. Based on this Figure, a $2''$ cut-off was chosen as a
good compromise that radio and X-ray sources can be considered as counterparts, which results in 92 radio-X-ray matches. Furthermore, only one random identification was made at offsets smaller than the chosen matching radius of 2 arcsec, corresponding to $1\%$ of the true matches.

\begin{figure}
\begin{center}
\begin{tabular}{c}
\includegraphics{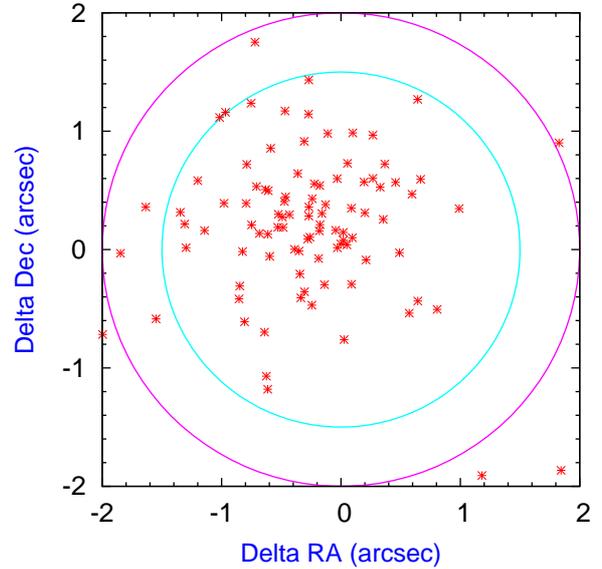} \\
\includegraphics{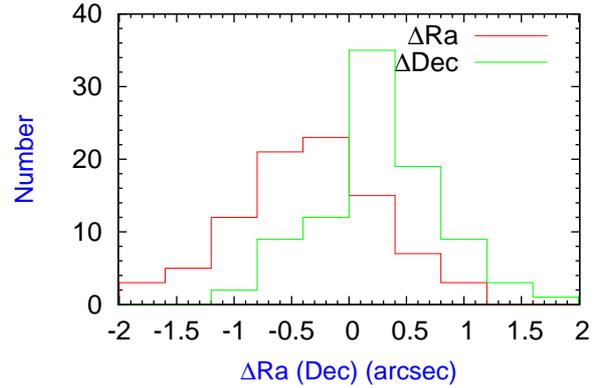} \\
\end{tabular}
\caption{Diagram of the offset in Ra and Dec (arcsec) of the FIRST radio sources and XBo\"{o}tes sources with a $2''$ and $1.5''$ circles overlaid to show how many of the matches are separated by more than $2''$. \textit{Lower panel}: offset histograms for $\Delta\,{\rm Ra}$ (solid histogram) and $\Delta\, {\rm Dec}$ (dashed histogram).}
\label{deltaxy}
\end{center}
\end{figure}

Figure \ref{deltaxy} shows the positional offset between the radio and their X-ray counterparts. There are 86/92 ($93\%$) X-ray-radio matches within $1.5''$ and only 6 sources lie beyond $1.5''$. It is clearly seen from the figure that the sources are not centered on (0,0) and there is a small systematic offset between the radio and X-ray reference frames. The mean offsets are $\langle \Delta\,{\rm Ra}\rangle=-0.32''$ and  $\langle \Delta\,{\rm Dec}\rangle=-0.27''$. I shifted the X-ray reference frame to match up to the radio using the median offset of matched objects and  again ran the source matching algorithm. Correcting this small offset made no difference to the total matched sample. \citet{bauer02}, used a $1''$  search radius for the VLA/Chandra cross correlation on the HDF-N; \citet{barger07} adopted a $1.5''$  search radius as a good compromise for the VLA/HDF-N. \\

The comparison of the positions of FIRST radio sources with the XBo\"{o}tes X-ray sources reveals 92 coincidence, corresponding to an identification rate of $\sim 10\%$ (92/900). This  clearly shows  that there is  little overlap between radio and X-ray  sources. A higher overlap between radio sources and X-ray sources has been found by \citet{tozzi09}. In their cross correlation between radio sources and X-ray sources in the  E-CDFS area, they found that 40\% of the radio sources are associated with X-ray sources. \citet{rov07} found that the radio detection rates for the E-CDFS and CDFS X-ray sources increases from 9\% to 14\% in the central region of this field (which has deeper X-ray data), while the X-ray detection rates for radio sources are 21\% for the E-CDFS and 33\% for the CDFS field. The radio flux density limit in \citet{tozzi09} ranges from $42\,\mu{\rm Jy}$ at the field center, to $125\,\mu{\rm Jy}$ near the field edge and in \citet{rov07} is $60\,\mu{\rm Jy}$, while the detection threshold in the FIRST survey is $1\,{\rm mJy}$. \citet{ciliegi03} reported a high fraction of $36\%$ in the HELLAS field. The identification rate measured in   \citet{akiyama2000} is 35\% for the FIRST radio sources in the LSS field ($f_X{\rm (2.0-10 keV)=1\times10^{-13}~erg~ s^{-1}~cm^{-2}}$). It is interesting to note that both radio and X-ray detection rates increases with deeper observations.  \citet{ciliegi03} have shown that the radio-X-ray association is a function of  the radio/X-ray limit ratio, $f_r/f_X$ (see their table 2), the deeper radio data compared to the X-ray flux limit, the higher the number of radio-X-ray associations. To summarise, the small fraction of radio detected X-ray sources in comparison to the previous results is due to a mismatch between the flux limit of the VLA FIRST 1.4 GHz (deep radio survey) and that of the Chandra XBo\"{o}tes survey (shallow X-ray survey).

\subsection{Optical counterparts to the radio/X-ray matches}

\citet{brand06} presented a catalogue of the optical/infrared counterparts to the 3213
X-ray point sources detected in the XBo\"{o}tes survey\footnote{The catalogue
  can be obtained through the NDWFS homepage at
  http://www.noao.edu/noao/noaodeep/XBootesPublic}. These authors found optical
counterparts for $98\%$ of the X-ray sources using a Bayesian identification
scheme. For each candidate the full catalogue provides  a number of parameters
containing information about the X-ray sources and their optical/infrared counterparts. A detailed analysis of  the matching criteria and a full description of the
catalogue is presented in \citet{brand06}. The following  provides a brief description of the parameters that have been taken into
account in this analysis:

\begin{itemize}
\item nopt: labels the number of optical sources with $>1 \%$ probability
of being the correct optical/infrared counterparts to the  X-ray source in the
case of multiple matches.

\item optrank: labels the optical rank when the X-ray
  sources are multiply matched. This parameter  ranges from 1 (most
  probable) in running order to the least probable.

\item bayprob: gives the Bayesian probability of an optical object being the true
  counterpart to an X-ray source.

\item Class: describes the optical morphology of the optical
  counterparts  to the X-ray sources (1: point-like object; 0: galaxy
  (extended or resolved)).
\item flag: In cases of no optical ID being the most probable identification, the flag parameter takes three values which have been applied manually \citep{brand06}: 
\begin{enumerate}
 
 \item ${\rm flag=-3}$: no optical ID but source obscured by nearby star / missing data.
\item ${\rm flag=-2}$: no optical ID, but X-ray position is close to optically bright galaxy (source is either obscured by or associated with the galaxy).
\item ${\rm flag=-1}$:`true' no optical ID (optical image is truly blank).

\end{enumerate}

\end{itemize}

A total of 90/92 of the X-ray counterparts to FIRST radio sources
have optical counterparts and the remaining two X-ray sources are
associated with a blank field (${\rm flag=-1}$). The majority ($75\%$) of
the X-ray candidates are associated with only one optical object
(nopt=1). In 22 cases, the X-ray candidates have more than one
optical match. Ten X-ray sources have 2 multiple  matches, three X-ray
candidates have three multiple matches, five X-ray sources are
associated with 4 optical objects, one source have 5 optical
counterparts, two X-ray candidates have 6 counterparts and one have
8 possible matches. In those multiple matches, only sources with
optrank = 1  (high probability) have been
considered as real matches.   \\

In \citet{khadija08}, I have presented  a cross correlation of FIRST radio sources and the NDWFS survey (Bo\"{o}tes field), using both the likelihood ratio and a simple positional coincidence. The identification procedure yielded  688/900 optical counterparts to FIRST radio sources in either one or more bands (\textit{Bw, R, I, K}). From these,  79 optical/infrared sources are common between the X-ray detected and radio detected samples. One should note that the Bo\"{o}tes field  is split by declination into four strips ($32^{\circ}< \delta \leq 33^{\circ}, 33^{\circ}< \delta \leq 34^{\circ}, 34^{\circ}< \delta \leq 35^{\circ}, 35^{\circ}< \delta \leq 36^{\circ}$), each is imaged in four bands (\textit{Bw, R, I, K}). The Bo\"{o}tes field is partially covered in \textit{K} band (especially the second strip and there is  no \textit{K} data for the first strip). In \citet{khadija08}, I cross correlated the FIRST radio sources and the FLAMEX survey \citep{elston06} in order to get infrared data (\textit{K} and \textit{J}) for the  second strip. FLAMEX survey covers about 4.7 deg${^2}$ of the Bo\"{o}tes field.\\

\begin{figure}
\begin{center}
 \begin{tabular}{c@{}c@{}c@{}c@{}c}
  \resizebox{16mm}{!}{\includegraphics{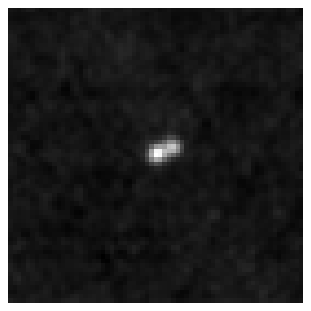}}  
& \resizebox{16mm}{!}{\includegraphics{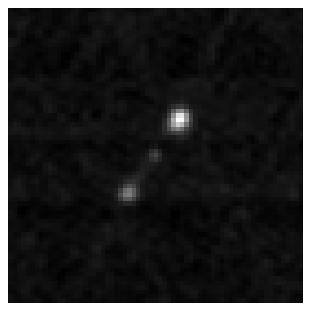}} 
& \resizebox{16mm}{!}{\includegraphics{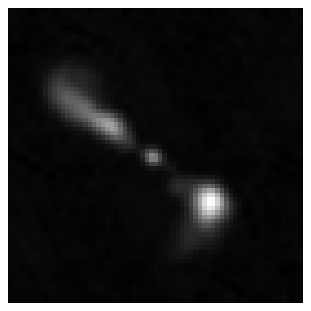}}
& \resizebox{16mm}{!}{\includegraphics{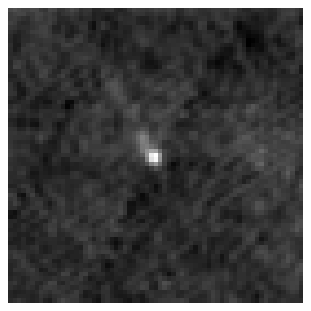}}
& \resizebox{16mm}{!}{\includegraphics{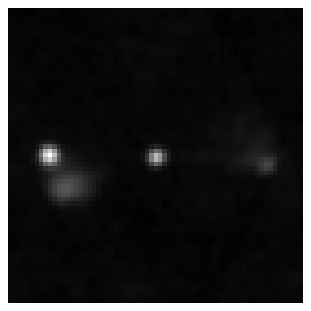}}\\
  \resizebox{16mm}{!}{\includegraphics{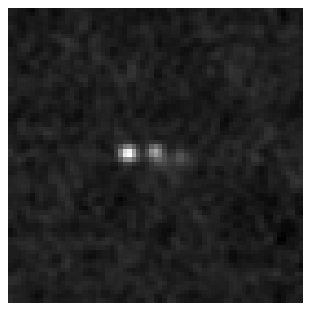}}
& \resizebox{16mm}{!}{\includegraphics{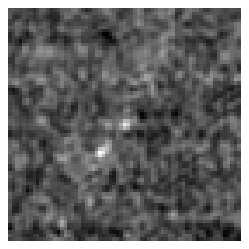}}
& \resizebox{16mm}{!}{\includegraphics{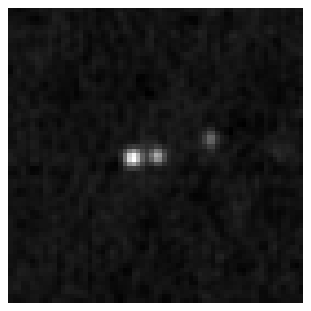}}
& \resizebox{16mm}{!}{\includegraphics{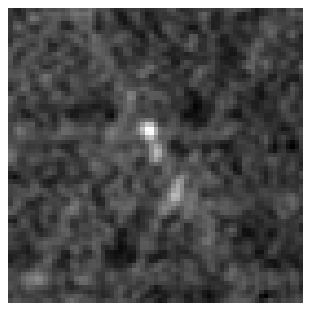}}
& \resizebox{16mm}{!}{\includegraphics{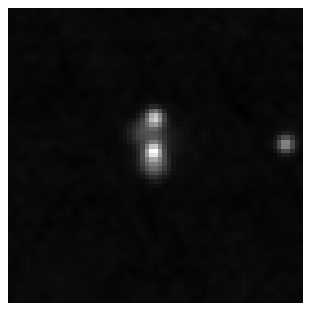}}\\
 \end{tabular}
\end{center}
\caption{The FIRST radio sources with multiple components identified by the selection algorithm \citep{maglio98}. Each FIRST stamp is $2'\times2'$, normalised to the maximum intensity.}
\label{multicomp}
\end{figure}

Among the 79 radio-X-ray matches,  4 double-radio sources are identified in XBo\"{o}tes field,  four radio sources with three components and one source with four components. But  only  one  component of each group  that is identified in the XBo\"{o}tes field,  Figure \ref{multicomp} shows images of the identified double radio sources recognised by the selection algorithm. These images are extracted from the FIRST website\footnote{http://third.ucllnl.org/cgi-bin/firstcutout}.

In summary, the cross correlation of FIRST radio sources and the XBo\"{o}tes survey yielded a total of 92 radio-X-ray associations. A total number of 79 optical/infrared sources is  common to both radio and X-ray samples.


\section{Stellarity}
\begin{figure}
\begin{tabular}{c}
\includegraphics{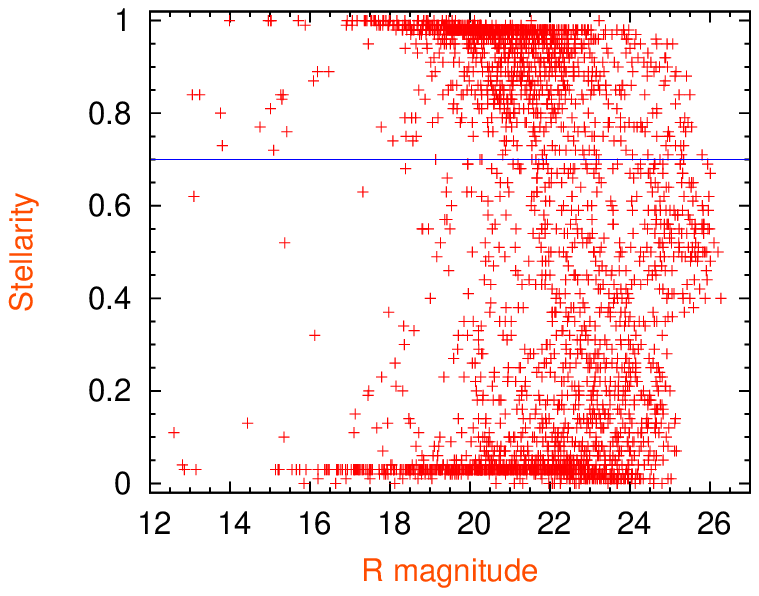}\\
\includegraphics{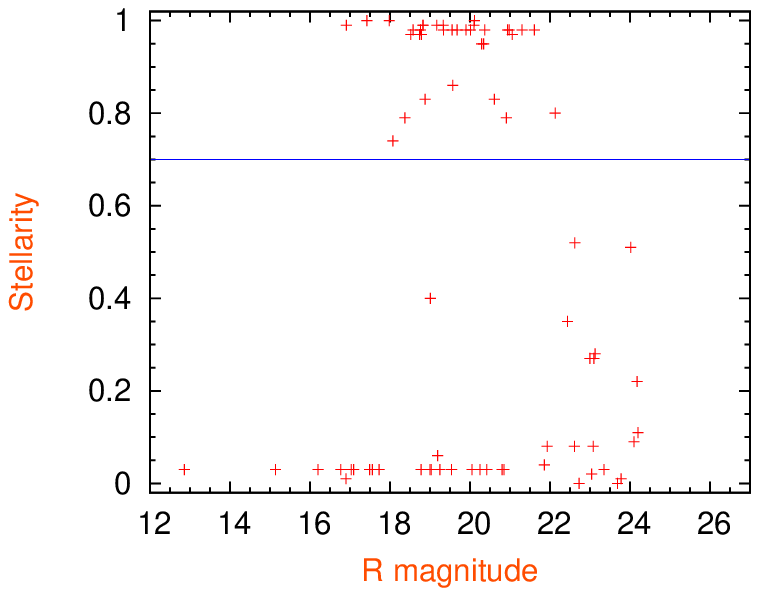}\\
\end{tabular}
\caption{The SExtractor stellarity parameter \citep{bertin96} versus $R$-band magnitude for the X-ray sources sources identified in Bo\"{o}tes field is shown in the top panel and for optical counterparts to the radio X-ray matches in the bottom panel. The horizontal line shows the boundary that has been assumed to classify the radio-X-ray matches into point sources (stellarity $\geqslant 0.7$) and extended sources (stellarity
$<0.7$). Stellarity varies from 0 for a galaxy to 1 for a star.}
\label{xstellarity}
\end{figure}
 Optical morphologies of these objects provide a further clue as to their nature. Here,
 I performed a simple morphological analysis of the identifications using the
 SExtractor stellarity parameter \citep{bertin96}; which has values between 0 (galaxies
 or resolved) and 1 (stars or unresolved). The SEXtractor software has difficulty in
 correctly classifying sources fainter than $R=23$ in the NDWFS data \citep{brown03, 
 brand06}. Typically, all objects brighter than $R\sim 23$ should be classified
 correctly. Beyond that magnitude the SExtractor cannot give a reliable classification
 and just assign random values between 0 and 1. Figure \ref{xstellarity} illustrates
 the stellarity parameter against the apparent $R$-band magnitude for all the X-ray
 sources identified in Bo\"{o}tes field in the upper panel, and for all radio-X-ray 
 matches in Bo\"{o}tes field in the lower panel. If all sources with stellarity $\geq 0.7$ are classified as
 point-like and all sources with stellarity $<0.7$ (extended) as galaxies, there are 33
 (42\%) optical point-like sources and 46 (58\%) galaxies. Taking a crude split between
 stars and galaxies at stellarity level of 0.9 and 0.1 for stars and galaxies
 respectively; all sources with stellarity $\ge 0.9$ are classified as point-like
 objects (mostly QSO), sources with stellarity $\le 0.1$ as galaxies and intermediate
 objects as sources with $0.1<$ stellarity $<0.9$. This criterion  yielded 27 (34\%)
 point-like sources, 35 (44\%) galaxies and 17 (22\%) intermediate sources.

\begin{figure*}
  \begin{center}
    \begin{tabular}{c@{}c@{}c@{}c@{}c@{}c}

 \resizebox{25mm}{!}{\includegraphics{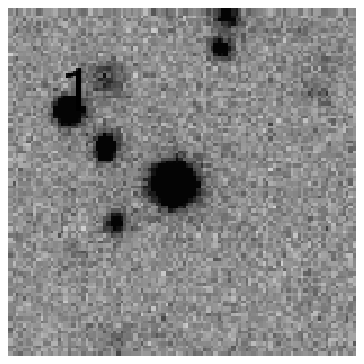} }
& \resizebox{25mm}{!}{\includegraphics{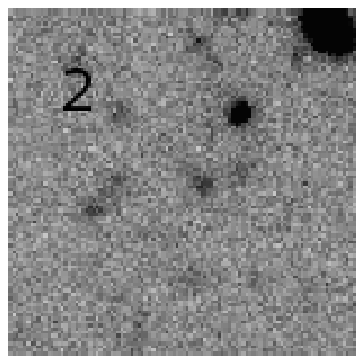} }
& \resizebox{25mm}{!}{\includegraphics{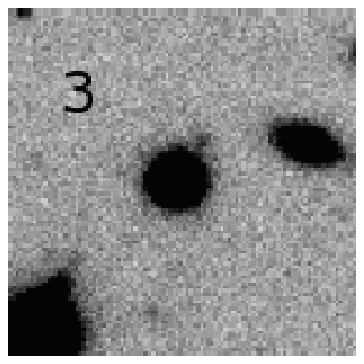} }
& \resizebox{25mm}{!}{\includegraphics{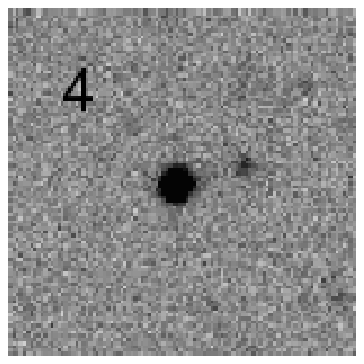} }
& \resizebox{25mm}{!}{\includegraphics{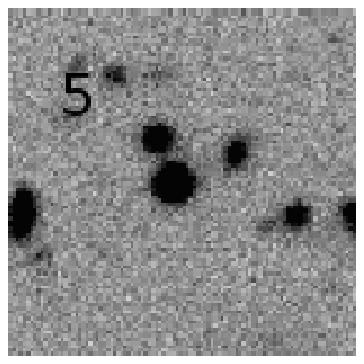} }
& \resizebox{25mm}{!}{\includegraphics{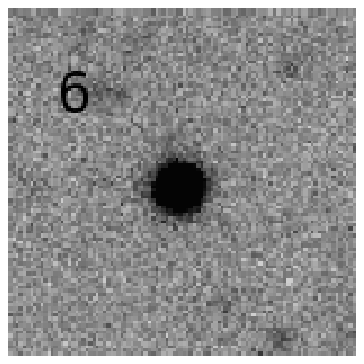} } \\
      \resizebox{25mm}{!}{\includegraphics{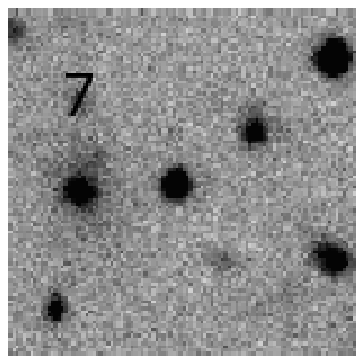} }
  &   \resizebox{25mm}{!}{\includegraphics{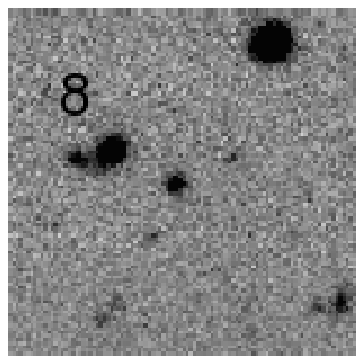} } 

  &   \resizebox{25mm}{!}{\includegraphics{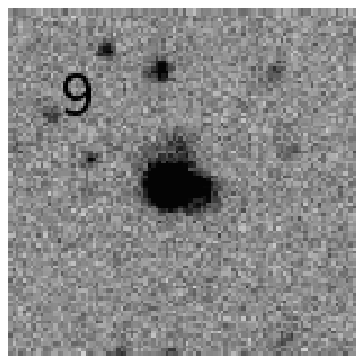} }
  &   \resizebox{25mm}{!}{\includegraphics{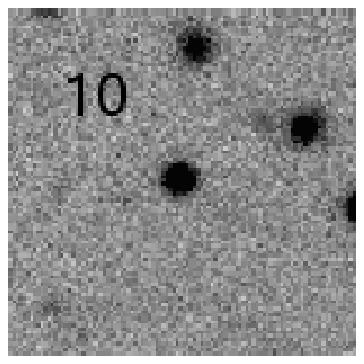} }

  &   \resizebox{25mm}{!}{\includegraphics{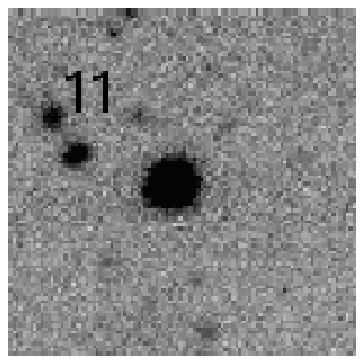} }
  &   \resizebox{25mm}{!}{\includegraphics{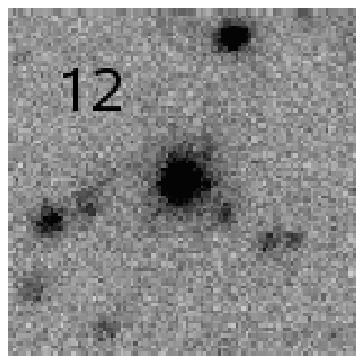} } \\
   \resizebox{25mm}{!}{\includegraphics{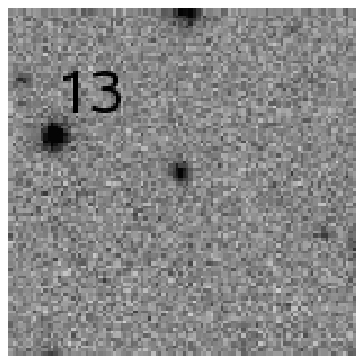} }
  &   \resizebox{25mm}{!}{\includegraphics{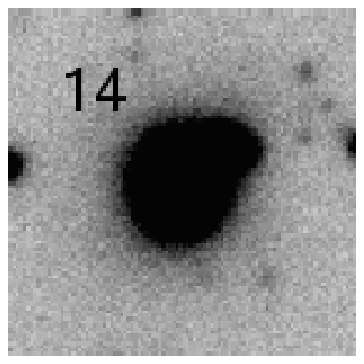} }
  &  \resizebox{25mm}{!}{\includegraphics{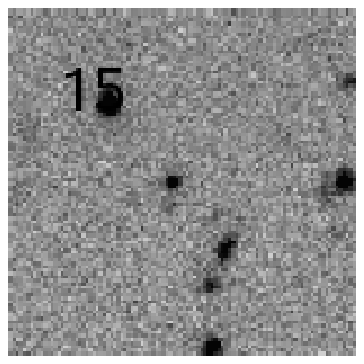} }
  &  \resizebox{25mm}{!}{\includegraphics{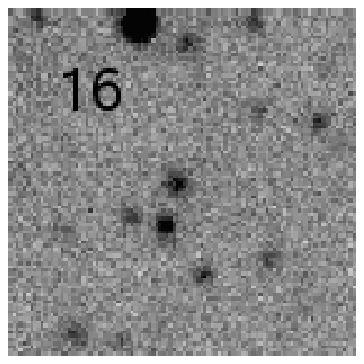} }
  &   \resizebox{25mm}{!}{\includegraphics{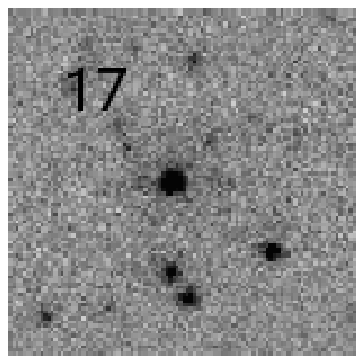}  }
  &   \resizebox{25mm}{!}{\includegraphics{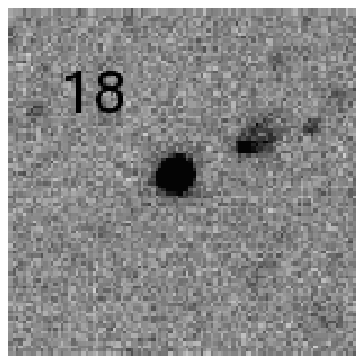} } \\
   \resizebox{25mm}{!}{\includegraphics{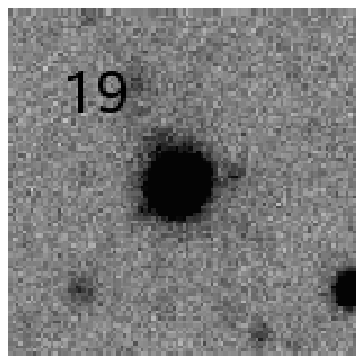} }
  &   \resizebox{25mm}{!}{\includegraphics{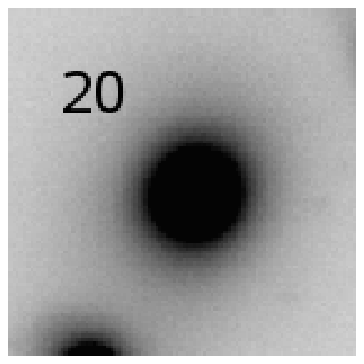} }
  &  \resizebox{25mm}{!}{\includegraphics{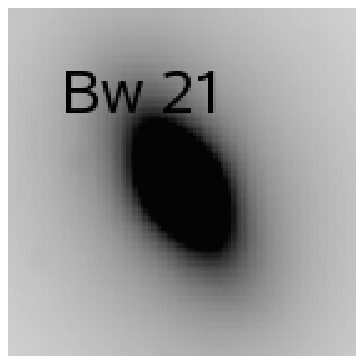} }
  &  \resizebox{25mm}{!}{\includegraphics{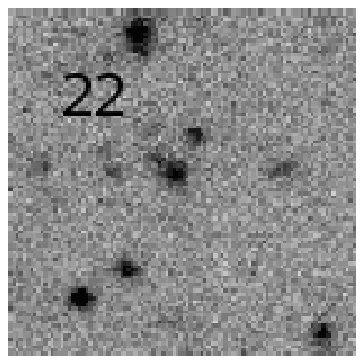} }
  &   \resizebox{25mm}{!}{\includegraphics{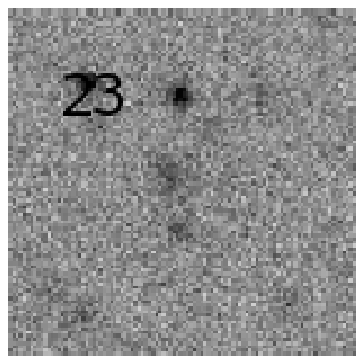} }
  &   \resizebox{25mm}{!}{\includegraphics{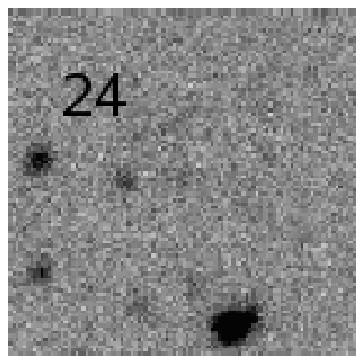} } \\
 \resizebox{25mm}{!}{\includegraphics{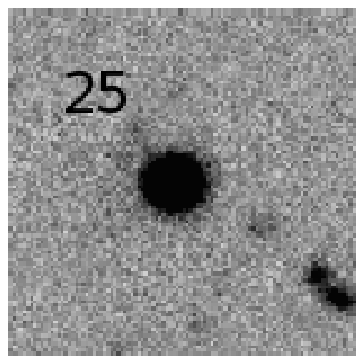} }
  &   \resizebox{25mm}{!}{\includegraphics{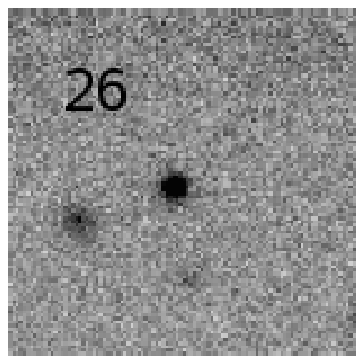} }
  &   \resizebox{25mm}{!}{\includegraphics{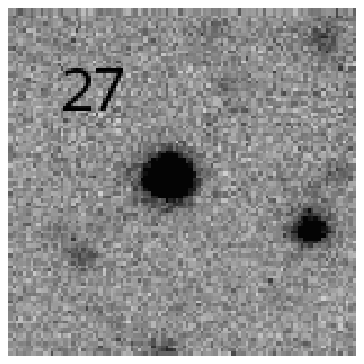} }
  & \resizebox{25mm}{!}{\includegraphics{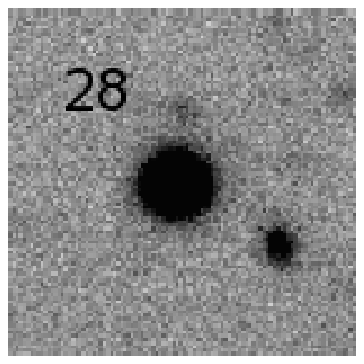} }
  &   \resizebox{25mm}{!}{\includegraphics{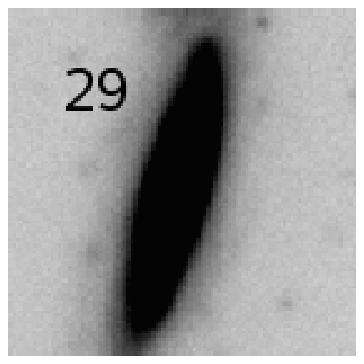} }
  &   \resizebox{25mm}{!}{\includegraphics{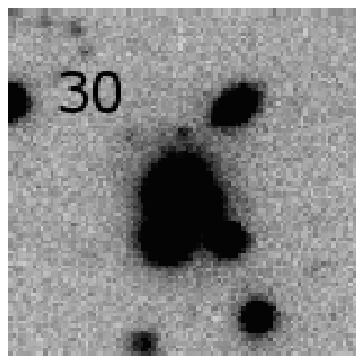} } \\
 \resizebox{25mm}{!}{\includegraphics{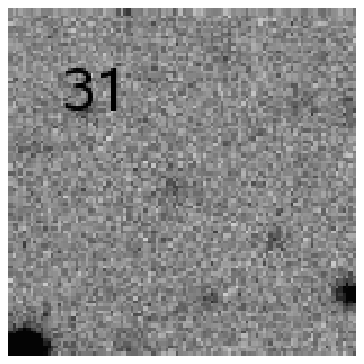} }
  &   \resizebox{25mm}{!}{\includegraphics{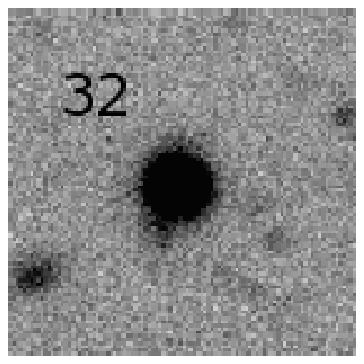} }
  &   \resizebox{25mm}{!}{\includegraphics{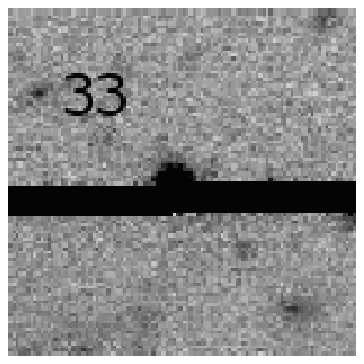} }
  &   \resizebox{25mm}{!}{\includegraphics{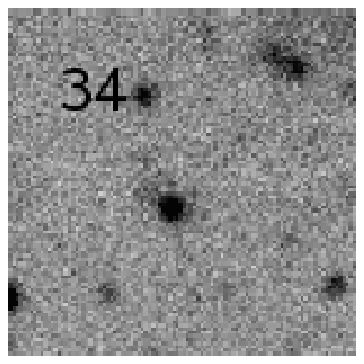} }
  &   \resizebox{25mm}{!}{\includegraphics{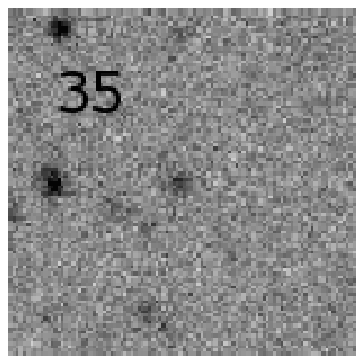} }
  &   \resizebox{25mm}{!}{\includegraphics{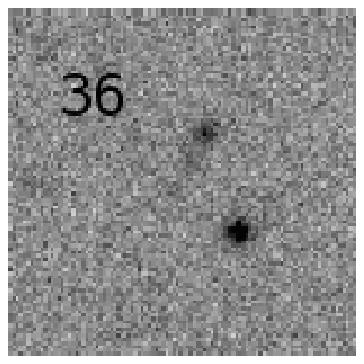} } \\
      \resizebox{25mm}{!}{\includegraphics{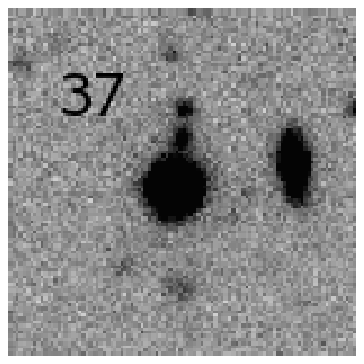} }
  &   \resizebox{25mm}{!}{\includegraphics{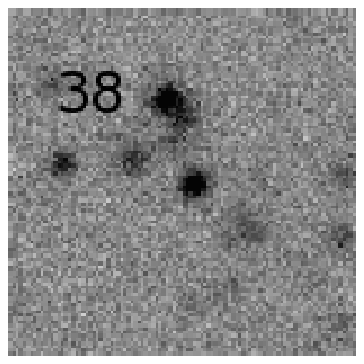} }
  &   \resizebox{25mm}{!}{\includegraphics{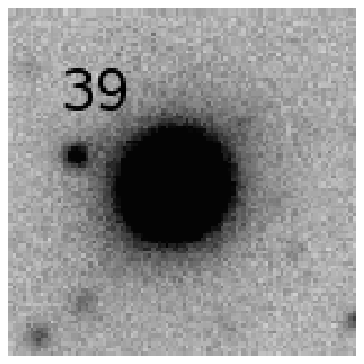} }
  &   \resizebox{25mm}{!}{\includegraphics{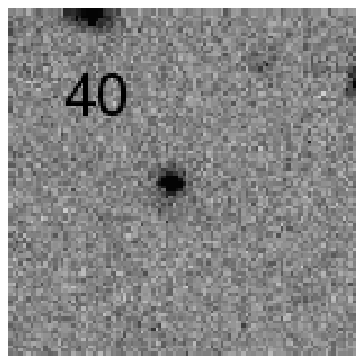} }
  &   \resizebox{25mm}{!}{\includegraphics{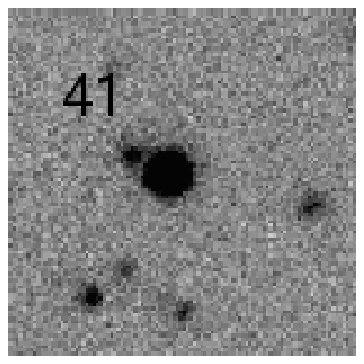} }
  &   \resizebox{25mm}{!}{\includegraphics{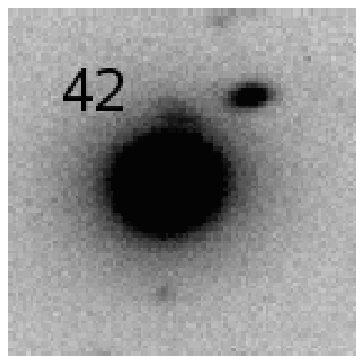} } \\

\end{tabular}
\end{center}
   \caption{The $R$ band images ($2'\times2'$) of all the 79 radio-X-ray matches in the 9.3
   deg$^{2}$ Bo\"{o}tes field.  All the images are centred on the FIRST radio sources.
 The FIRST radio sources reference number (shown in column 1 of Tables 1 and 2) is at
   the top left of each image. These cut-out images have been obtained from the NOAO
 cut-out service: http://archive.noao.edu/ndwfs/cutout-form.html.}
\label{cutout}
  
\end{figure*}

\begin{figure*}
  \begin{center}
    \begin{tabular}{c@{}c@{}c@{}c@{}c@{}c}
      \resizebox{25mm}{!}{\includegraphics{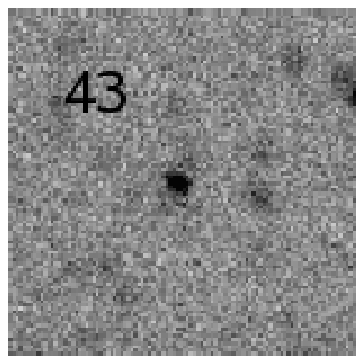} }
  &   \resizebox{25mm}{!}{\includegraphics{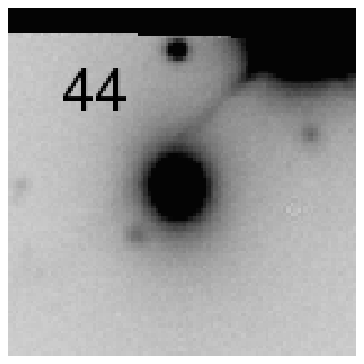} }
  &   \resizebox{25mm}{!}{\includegraphics{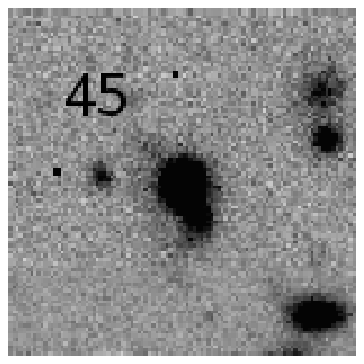} }
  &   \resizebox{25mm}{!}{\includegraphics{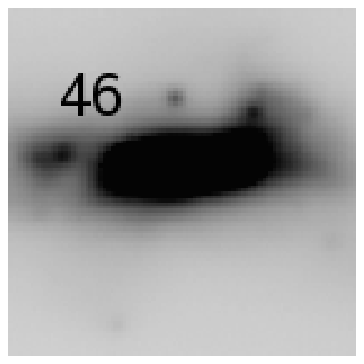} }
  &   \resizebox{25mm}{!}{\includegraphics{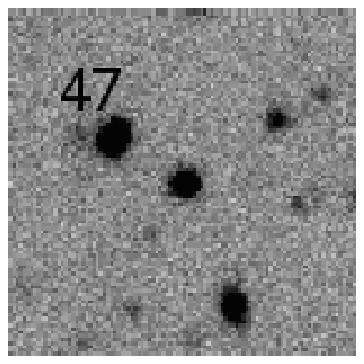} }
  &   \resizebox{25mm}{!}{\includegraphics{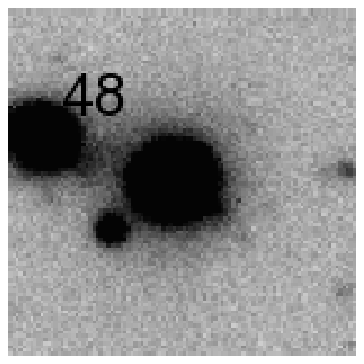} } \\

      \resizebox{25mm}{!}{\includegraphics{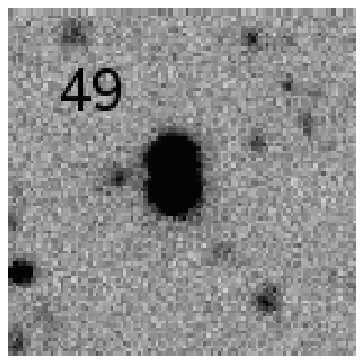} }
  &   \resizebox{25mm}{!}{\includegraphics{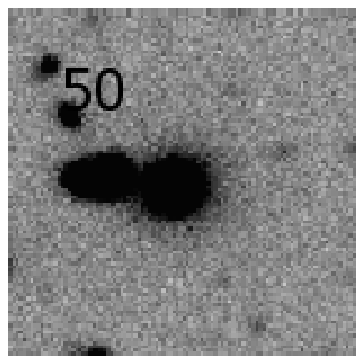} }
  &   \resizebox{25mm}{!}{\includegraphics{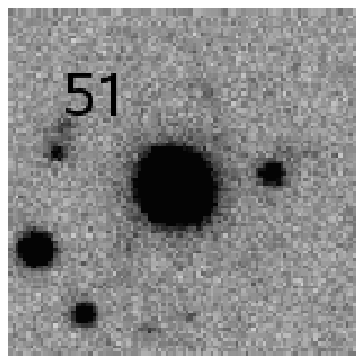} }
  &   \resizebox{25mm}{!}{\includegraphics{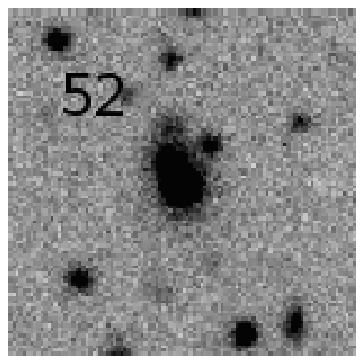} }
  &   \resizebox{25mm}{!}{\includegraphics{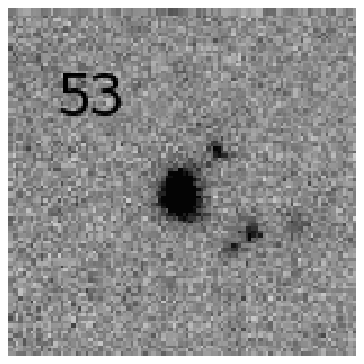} }
  &   \resizebox{25mm}{!}{\includegraphics{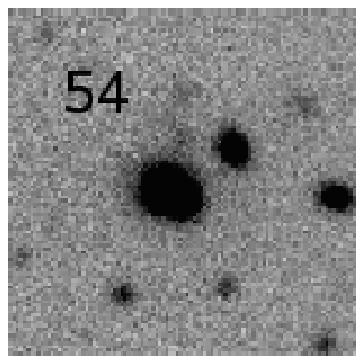} } \\
      \resizebox{25mm}{!}{\includegraphics{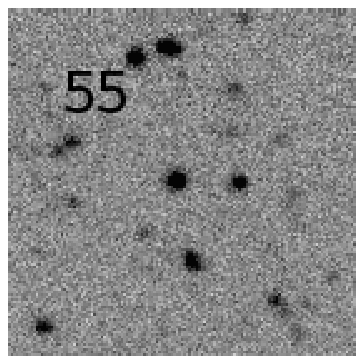} }
  &   \resizebox{25mm}{!}{\includegraphics{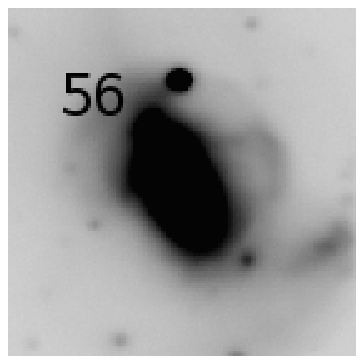} }
  &   \resizebox{25mm}{!}{\includegraphics{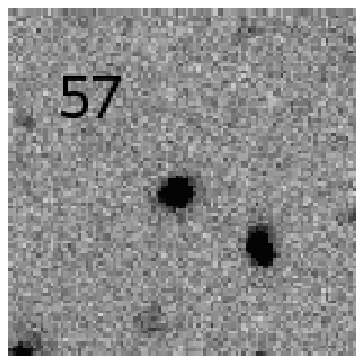} }
  &   \resizebox{25mm}{!}{\includegraphics{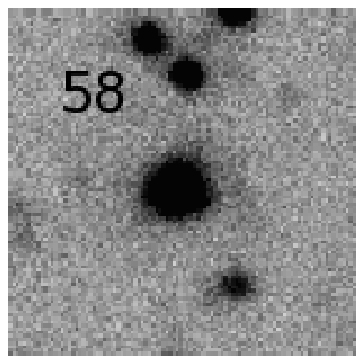} } 
  &   \resizebox{25mm}{!}{\includegraphics{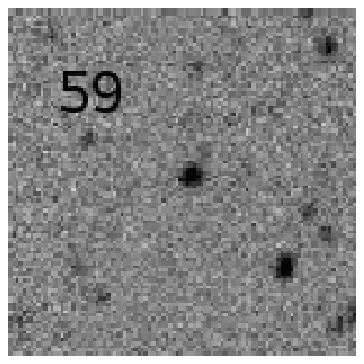} }
  &   \resizebox{25mm}{!}{\includegraphics{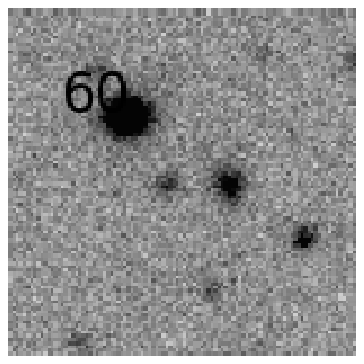} } \\

     \resizebox{25mm}{!}{\includegraphics{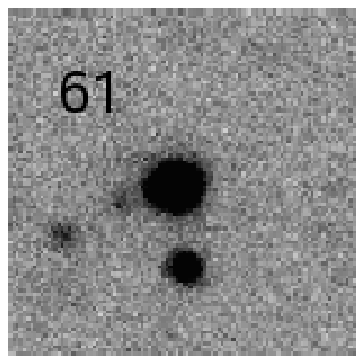} }
  &   \resizebox{25mm}{!}{\includegraphics{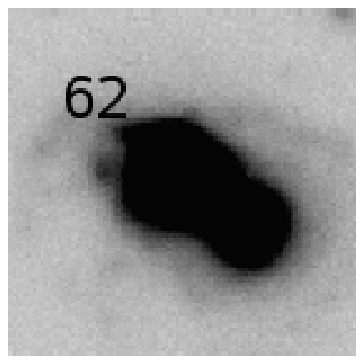} }
  &   \resizebox{25mm}{!}{\includegraphics{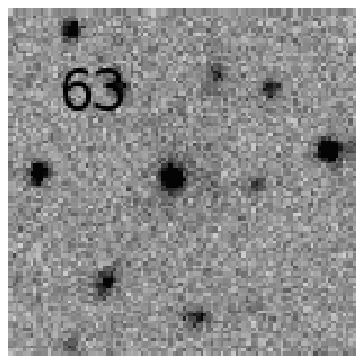} } 
  &   \resizebox{25mm}{!}{\includegraphics{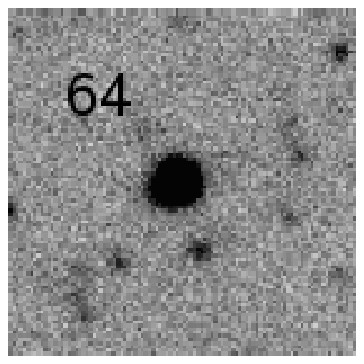} } 
  &   \resizebox{25mm}{!}{\includegraphics{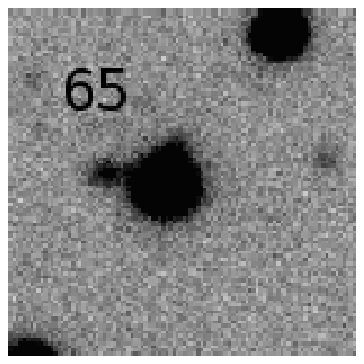} }
  &   \resizebox{25mm}{!}{\includegraphics{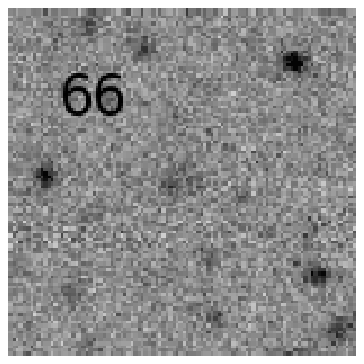} }\\

      \resizebox{25mm}{!}{\includegraphics{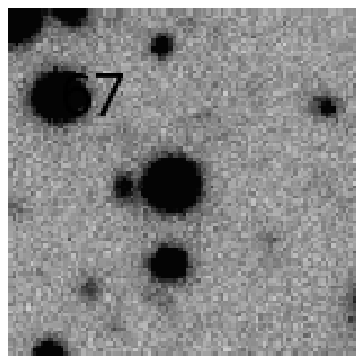} }
  &   \resizebox{25mm}{!}{\includegraphics{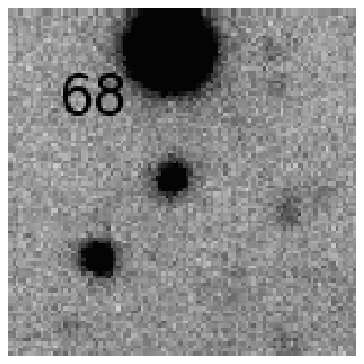} }
  &   \resizebox{25mm}{!}{\includegraphics{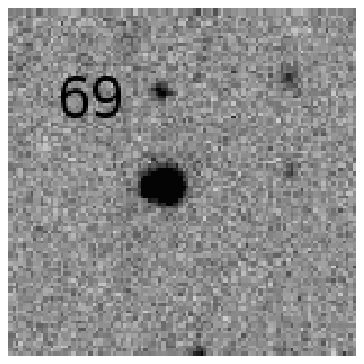} }
  &   \resizebox{25mm}{!}{\includegraphics{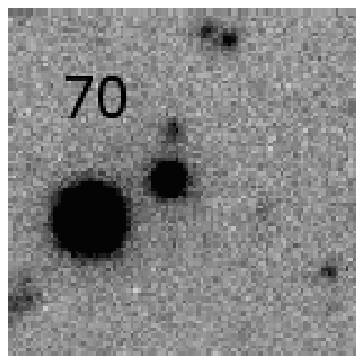} } 
  &   \resizebox{25mm}{!}{\includegraphics{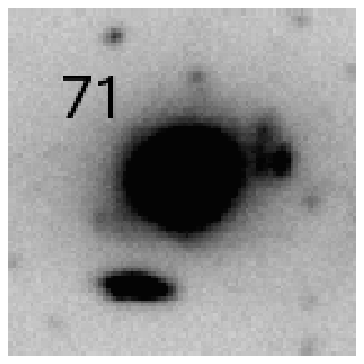} }
  &   \resizebox{25mm}{!}{\includegraphics{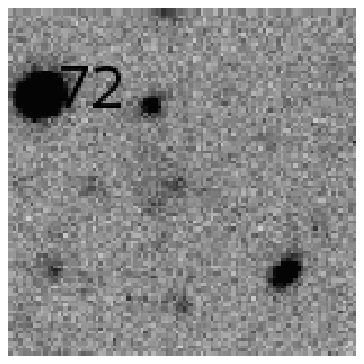} } \\

  \resizebox{25mm}{!}{\includegraphics{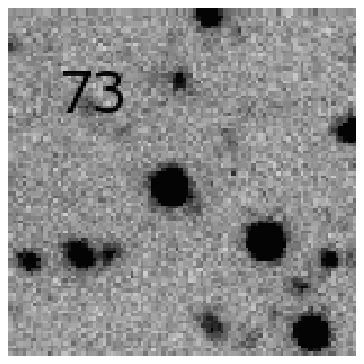} }
 & \resizebox{25mm}{!}{\includegraphics{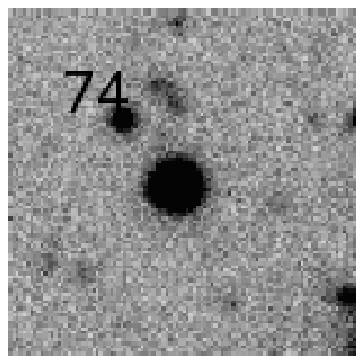} }
 & \resizebox{25mm}{!}{\includegraphics{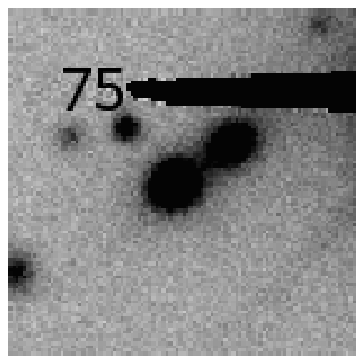} }
 & \resizebox{25mm}{!}{\includegraphics{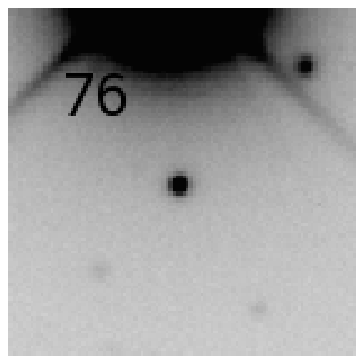} } 
 & \resizebox{25mm}{!}{\includegraphics{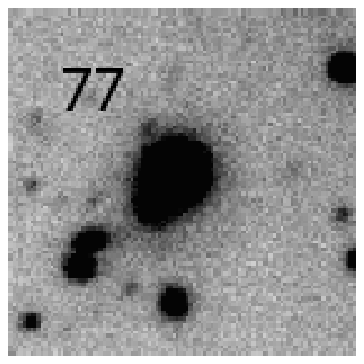} }
 & \resizebox{25mm}{!}{\includegraphics{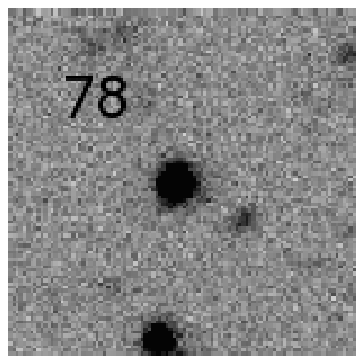} } \\

 \resizebox{25mm}{!}{\includegraphics{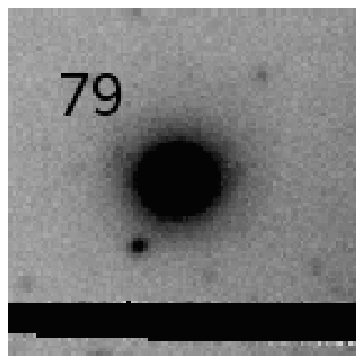} }
&
&
&
& \\
    \end{tabular}
 \end{center}
   \contcaption{ }
\end{figure*}
The $R$ band images for all the radio-X-ray matches are shown in Figure \ref{cutout}. These cut-out images have been obtained from the NOAO cut-out service: http://archive.noao.edu/ndwfs/cutout-form.html. The majority (63/75; $84\%$) of the radio-X-ray matches  have magnitudes brighter than $R=23$  from which $57 \%$ (36/63) have magnitudes brighter than $R=20$ mag. These sources appear  to be mainly elliptical,  and some lenticulars are also seen. A number of these sources also appear to be either   interacting or merging  galaxies. At fainter magnitudes, the sources tend to have low stellarity (see Figure \ref{xstellarity}), so they are likely  QSOs or compact galaxies.
\section{Optical magnitude distributions}
\begin{figure*}
\includegraphics{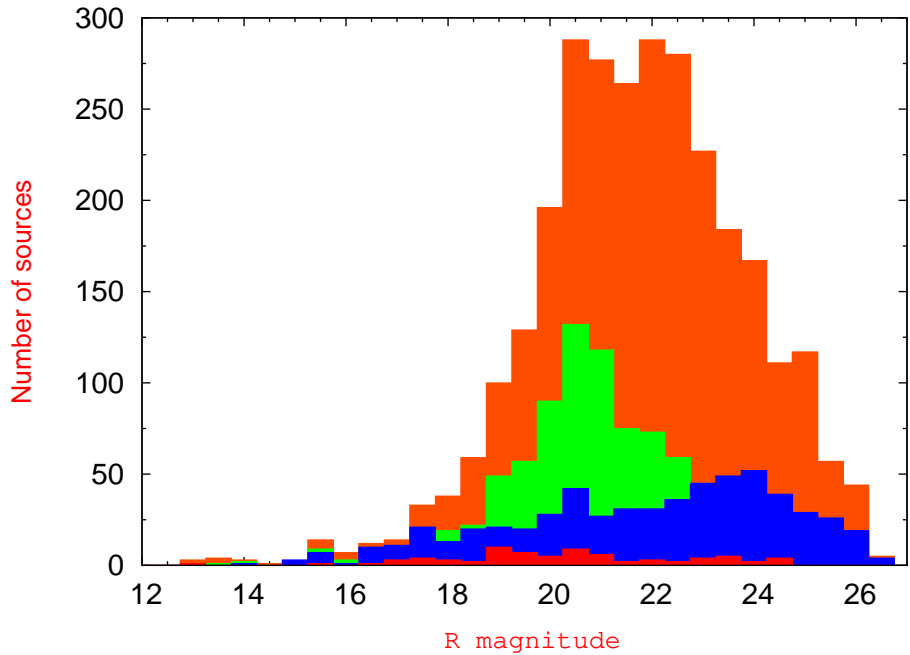}
\caption{The optical magnitude distributions in $R$ band for (from top to bottom): all the X-ray sources  identified in Bo\"{o}tes field, all  X-ray sources identified in  four bands (\textit{Bw, R, I, K}) in Bo\"{o}tes field, all FIRST radio sources identified in Bo\"{o}tes field and all the common optical counterparts to the radio-X-ray matches.} \label{magnitudeX}
\end{figure*}

In the absence of spectroscopic data, the magnitude and colour distributions of the optical counterparts can be used  to derive some information on the nature of the faint radio population. In Figure \ref{magnitudeX}, the optical magnitude distributions  in $R$ band of the  optical counterparts to  the radio-X-ray matches are plotted. The histograms shown are (from top to bottom) distributions for: 1) all X-ray sources optically identified in the Bo\"{o}tes field, 2) all X-ray sources identified in  four bands
(\textit{Bw, R, I, K}) in the Bo\"{o}tes field, 3) all FIRST radio sources identified in the Bo\"{o}tes field, and 4)  all the optical counterparts to the radio-X-ray matches. The magnitude distribution of all X-ray sources identified in Bo\"{o}tes field has two apparently weak peaks  with a small tail at brighter magnitudes and  few counterparts at fainter magnitudes. The radio sample has significantly flatter distributions compared to the global optical catalogue (orange histogram) and to the X-ray sources identified in four bands (green histogram), and falls off at $R \sim 24$. The mean magnitude of the optically extended radio-X-ray matches is $R=
20.68$ with a median of 19.91. For radio-X-ray matches that are associated with point-like objects the mean magnitude is $R = 19.62$ and the median is 18.58. The mean magnitude for all radio-X-ray matches is $R=20.16$ and the median is $R=22.13$.

\section{Photometric redshifts}
\subsection{Optical classification}

Optical spectroscopic information is available for 22 out of 79 sources (28\%). The spectroscopic data have been obtained from the SDSS (Sloan Digital Sky Survey). The spectroscopically identified sample comprises 14 point-like objects  (13 sources show  broad emission lines and one source shows a featureless spectrum) and 8 resolved radio galaxies (with  emission/absorption lines). Extragalactic radio sources are known to be a mixture of two main populations: AGN and star-forming galaxies. The AGN class can be divided into two sub-classes: QSOs (often unresolved in the optical image) and objects where the AGN does not dominate the entire SED (Spectral Energy Distribution), such as type 2 QSOs, low luminosity AGN (seyfert and LINER) and absorption line AGN. A total number of 13 point-like objects are spectroscopically classified  as QSOs by the SDSS and the point-like source which shows a featureless optical spectrum  could be classified as a BL Lac object. The presence of broad emission lines (width larger than 2000 km/s) like ${\rm Mg_{II}}$, ${\rm C_{III}}$ and at large redshifts ${\rm C_{IV}}$ and ${\rm Ly\alpha}$ classifies the source as a broad line AGN (BLAGN), type -1  AGN or QSO according to the simple unification model by \citet{antonucci93}. To classify emission line galaxies as star-forming or AGN, I used the line flux ratio diagram of [${\rm O_{III}]/H\beta}$ versus [${\rm N_{II}]/H\alpha}$ \citep[BPT diagram, ][]{baldwin81}. I used an empirical criterion to segregate AGN host galaxies from star-forming galaxies: $[{\rm O_{III}]/H \beta =0.61/([ N_{II}]/H \alpha-0.05) +1.3}$ given by \citet{kauffmann03}. With these diagnostics, the AGN population still remained mixed. To distinguish seyferts from LINERS I used an empirical relation given by  \citet{kewley06}: [${\rm O_{III}]/H\beta=1.18\,[ O_{I}]/H\alpha}$ + 1.3. Based on the 2 independent classification  of galaxies, 3 emission lines galaxies were classified as star-forming galaxies (sources \# 42, \#46, \#71) and four were classified as LINERS (sources \#29, \#39, \#56, \#62).
 
  In Summary, among the spectroscopically classified sources I find:  13 BLAGNs (59\% of the sample with  spectroscopic redshift), one BL Lac object (source \#11), 3 (14\%)  star-forming galaxies and  four (18\%)   LINERS galaxies. The optical spectra of these sources are shown in Figure  \ref{sdss-bootes}  with  spectroscopic redshift, type and source number printed on each figure. One also notes  the fair agreement  (98\%) between the  SDSS classification and the SExtractor stellarity parameter (provided  by the NDWFS  catalogue) used so far to classify the radio-X-ray matches into point-like objects and extended  objects. For  the remaining  sources (55) without spectroscopic information I turn to photometric redshift techniques to estimate redshifts  for these.

\subsection{Photometric redshift technique}
Photometric redshift techniques have been  applied by many authors to obtain photometric redshift for X-ray sources identified in optical/radio surveys \citep{barger01, barger02, ghandi04,  geor04, mobasher04,   zheng04}. X-ray sources may have complex spectral energy distributions  that arise from both the host galaxy and the AGN. Consequently estimating photometric redshifts for X-ray sources using templates for normal galaxies may be problematic. However, with the aim of studying the X-ray spatial correlation function in the NDWFS survey (Bo\"{o}tes field), and in the absence of spectroscopic redshifts, \citet{gonza02} have studied  the effectiveness of photometric redshifts based on galaxy spectral template fitting for 65 X-ray luminous objects detected by Chandra in the  Caltech Faint Galaxy redshift survey (CFGRs). The authors  used the two publicly available codes, \textit{Hyperz}\footnote{The code is publicly available at:   http://webast.ast.obs-mip.fr/hyperz/}\citep{bolzonella00} and BPZ \citep[Bayesian photometric  redshift,][]{benitez00}.  \citet{gonza02} have shown that the two codes produce similar results when compared to spectroscopic data, and confirm  that  photometric redshifts based on template fitting for X-ray sources are quite robust for $90\%$  of sources that have optical counterparts brighter than $R\sim 24.5$ \citep[see also][]{barger02, ghandi04, zheng04}. They have shown that photometric redshift estimates agree well with spectroscopic measurements for objects in which galactic light dominates the optical flux. Using multi-wavelength photometric data from the Great Observatories Origins Deep Survey, \citet{mobasher04} also attempted to estimate photometric redshifts for a sample of 434 galaxies with spectroscopic redshifts in the Chandra Deep Field-South with magnitudes in the range $18<R_{AB}<25.5$. The authors  have applied a Bayesian method to two subsamples of galaxies: Extremely Red Objects (EROs) and AGNs. \citet{mobasher04} found good agreement between photometric redshifts  based on template fitting and more accurate results for EROs when compared to the sample as  a whole ($\sigma=0.051$), while the results tended to be less accurate for X-ray sources (AGNs) ($\sigma=0.104$), but still acceptable for further studies. 

In this paper, I used the  publicly available code \textit{Hyperz} to obtain photometric redshifts for the radio-X-ray matches. The code is based on fitting template spectral energy distributions (SEDs) to broad band photometry fluxes in as many filters as possible. I used Gissel 98 galaxy templates which are  the 1998 update of the spectral synthesis models described by \citet{Bruzual93}, with different star formation histories and spanning a wide range of ages from 1 Gyr to 30 Gyrs. All the models include solar metalicity and \citet{miller79} initial mass function. The \textit{Hyperz} code also includes Lyman  forest absorption according to the  model of \citet{madau95}. Four quasars templates compiled from Le PHARE software\footnote{http://www.lam.oamp.fr/arnouts/LE\_PHARE.html} \citep{cristiani90, cristiani04} were included as  well. For each identified radio-X-ray source \textit{Hyperz} provides a possible $z_{phot}$ (and corresponding reduced $\chi^{2}$ probability) for each set of templates. The ``best'' highest probability ($\chi^{2} \leq 2.7$, $90\%$ confidence limit) was selected as the correct one, together with the corresponding spectral type.

\begin{figure}
\includegraphics{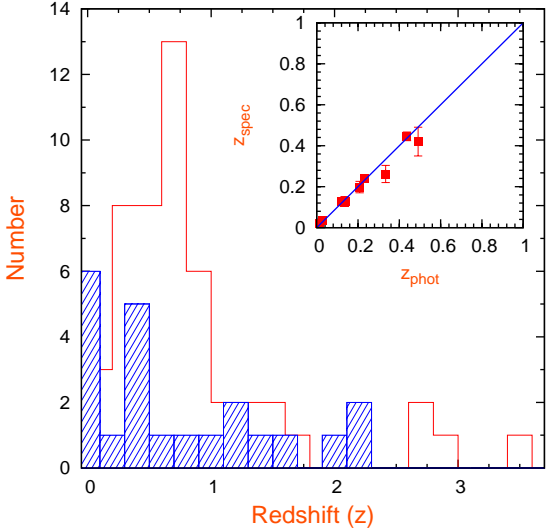} 
\caption{Photometric redshift distribution for the common optical counterparts to the radio-X-ray matches (open histogram) and spectroscopic redshift distribution for 22 radio-X-ray matches (hatched regions). The inset shows the ${\rm z_{phot}}$ vs. ${\rm z_{spec}}$. The solid line is the one-to-one relation.} \label{z}
\end{figure}

Four filters (\textit{Bw, R, I, K}) have been used to obtain photometric redshifts for 79 radio-X-ray matches. The majority of the radio-X-ray matches ($63\%$)  were detected in four bands. As mentioned previously, no infrared data were available for the first strip ($32^{\circ} \leq \delta < 33^{\circ}$),  8 radio-X-ray sources were identified in this strip. In this case, spectroscopic redshift was available for four radio-X-ray matches, and for the remaining four sources, photometric redshift was estimated based on three filters (\textit{Bw, R} and \textit{I}).  In 23 cases, it was not possible to assign an accurate photometric redshift ($\chi^{2}$ > 2.7, $90\%$ confidence limit), but 12/ 23 of these sources have spectroscopic redshifts (The majority of these sources were detected either in three bands or two bands).  Figure  \ref{z} displays the  photometric redshift distribution derived for the radio-X-ray matches. The hatched histogram shows the spectroscopic redshift distribution for 22 radio-X-ray matches and the open histogram displays the best estimate of photometric redshifts ($\chi^2$ < 2.7, $90\%$ confidence limit) as derived from \textit{Hyperz}. The inset shows the photometric redshift vs spectroscopic redshift for sources with secure redshift (i.e. $\chi^2$ <2.7). About $72\%$ of the sources are estimated to be at $z\leq 1$ with a small tail extending up to $z\sim 4$, and  the mean photometric/spectroscopic redshift is $z\sim 0.85$. Only a small fraction (22/79) of the optical counterparts to the radio-X-ray matches have a spectroscopically measured redshift. Comparing the $z_{phot}$ with $z_{spec}$, the accuracy of the photometric redshift technique applied so far is estimated to be $\Delta\,z /(1+z_{spec}) \approx 0.017$.

\subsection{The $K-z$ diagram}

\begin{figure}
\includegraphics{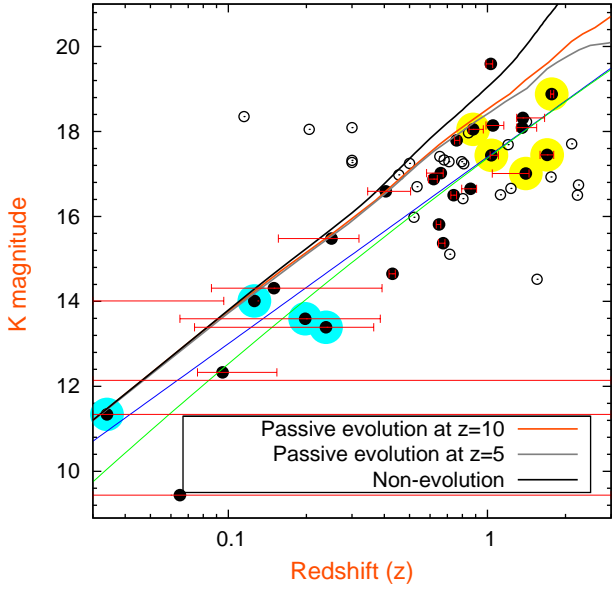}
\caption{The $K$ magnitude versus redshift $z$ for all the  radio-X-ray matches. Point-like objects (QSO, stellarity $\ge 0.7$) are represented by empty circles, small filled circles stand for resolved objects (galaxies; stellarity<0.7). Large blue circles denote spectroscopically identified sources, and large yellow cirles  stand for sources with $R-K>5$.  The green line shows the best fit   to $K-z$ relation of \citet{willott03} and the blue line shows the best
  fit $K-z$ relation for only AGNs with $\log\; f_X/f_{opt} > 1$
  for this study (point sources are not included in this fit). The three
  upper lines show the passive stellar evolutionary tracks of an $L_{\star}(K)$
  galaxy (where $L_{\star}(K)$ is the $K$ band $L_{\star}$) for an
   instantaneous starburst at $z=5$ and $z=10$ as well as a no-evolution curve,
    as derived by \citet{jarvis01}.}
\label{Xk_z}
\end{figure}

It is believed that powerful radio galaxies and quasars are associated with the most massive galaxies at a wide range of redshifts. A number of studies of these objects have shown that the near infrared $K$ band magnitudes follow a tight correlation with redshifts \citep[e.g.][]{willott03, jarvis01}. Studies of fainter radio sources at mJy level also  have shown a remarkably tight correlation between the $K$ magnitude and redshift \citep[e.g.][]{brookes06, elbouchefry07}. This diagram has been widely
used to study evolution of galaxies at high redshifts and is known to be  an excellent tool to measure stellar masses of galaxies up to high redshifts.  The $K$ magnitude versus  redshift of the optical  counterparts to the radio-X-ray matches is shown in Figure \ref{Xk_z}; point-like objects (QSO, stellarity $\ge 0.7$) are represented by empty circles, small filled circles stand for resolved objects (galaxies; stellarity<0.7). Large blue circles denote spectroscopically identified sources, and large yellow cirles  stand for sources with $R-K>5$.  The blue curve shows the best fit to the $K-z$ relation for the radio-X-ray matches (only extended objects were considered in this fit):
\begin{equation}
K=17.52 + 4.32 \log_{10}\,z,
\end{equation}
\noindent  and the green line illustrates the second order polynomial best fit of  \citet{willott03}: $K=17.37 + 4.53 \log_{10}\, z - 0.31 (\log_{10}\, z)^{2} $.  The three upper lines show the passive stellar evolutionary tracks of an $L_{\star}(K)$ galaxy (where $L_{\star}(K)$ is the $K$ band $L_{\star}$) for an instantaneous starburst at $z=5$ and $z=10$ as well as a no-evolution curve,  as derived by \citet{jarvis01}. The majority of the sources  lie brighter than $L_{\star}$. But it is important to note that the $K$ band catalogue is not as deep  compared to the study of \citet{willott03}. \citet{willott03} obtained complete $K$-band data of the complete radio samples. Here there are 28 sources that are not identified in the $K$ band, thus with a sample biased such as this the sources will all tend to be at the bright end of the magnitude range and this could be misleading. However, it is encouraging that the K-band magnitudes with the photometric redhsifts agree with the \citet{willott03} relation.

\section{X-ray and radio luminosities}
In the following subsection, the  photometric redshift estimates  have been used to
calculate the X-ray and radio luminosities. The spectroscopic redshift measurements have also been  used  whenever  available.

\subsection{X-ray luminosity}
\begin{figure}
\begin{center}
\begin{tabular}{c}
\includegraphics{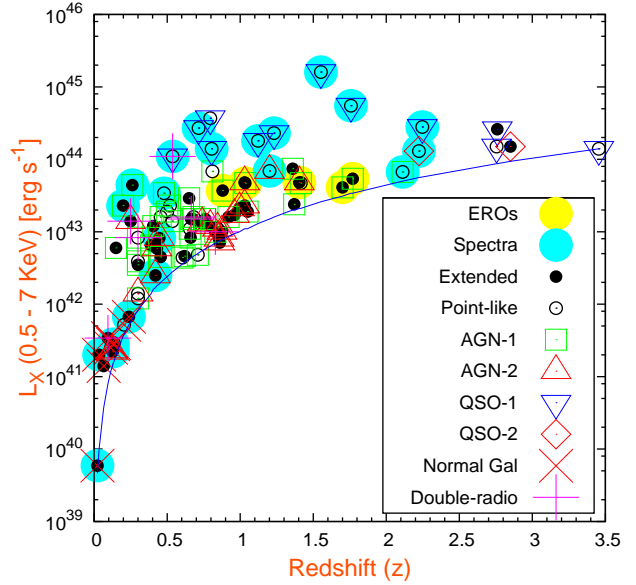} \\ 
 \includegraphics{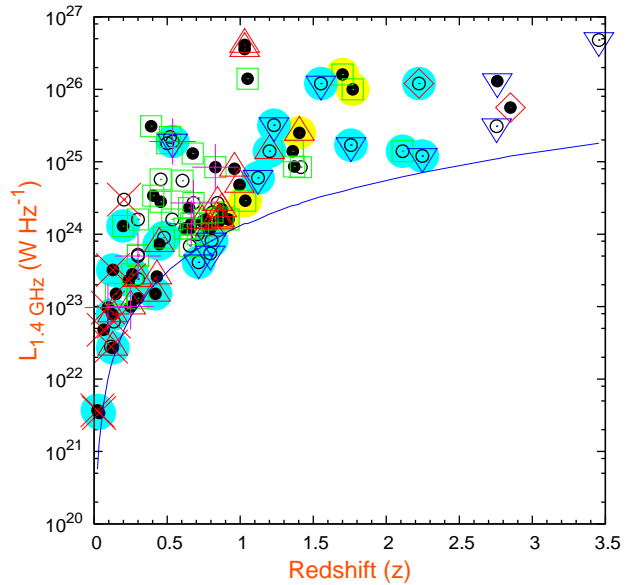}\\ 
\end{tabular}
\end{center}
\caption{Full $(0.5-7)$ keV X-ray luminosity versus redshift for the radio-X-ray matches  is shown in the top panel. Solid line  represents the faintest limiting X-ray luminosity for a source with a full flux of $\sim 8 \times 10^-{15}$ erg s$^{-1}$ cm$^{-2}$ as a function of redshift. The lower panel shows the 
radio luminosity as a function of redshift for all the radio-X-ray matches and the solid line corresponds to the radio flux limit of 1 mJy. Sources with spectroscopic redshift are shown with large filled circles (Cyan colour).  } \label{powersRX}
\end{figure}

 Figure \ref{powersRX} displays the X-ray ($0.5-7$) keV  luminosity (upper panel) as a
 function of redshift. The curve represents the detection limit. The rest-frame X-ray
 luminosity was calculated from:
\begin{eqnarray}
  \label{eq:xrayP}
  L_{X} = 4 \,\pi \, d_L^2\, f_{X}\, (1+z)^{-2 + \Gamma}~{\rm erg~s^{-1}},
\end{eqnarray}
\noindent where  $d_L$ is the luminosity distance (in cm), $f_X$ is the observed
 full X-ray flux (in erg s$^{-1}$ cm$^{-2}$ ). $\Gamma$  is the
photon index assumed to be $\Gamma= 1.8$ for all sources. \citet{barger02}, have shown that the use of individual indices result in only small differences in the rest-frame luminosities \citep[see also][]{barger07}.

In order to classify the X-ray counterparts to FIRST radio sources, I used the same procedure adopted by \citet{szokoly04}. Following \citet{szokoly04} the X-ray sources can be classified based only on their observed X-ray $L_X$ properties and the hardness ratio, ${\rm HR= h-s/h+s}$, where  $h$ is the number of counts detected in the ${\rm 2-7~keV}$ band and $s$ is the number of counts detected in the ${\rm 0.5-2~keV}$ band. The X-ray luminosity $L_X$ is derived from the ${\rm 0.5-7~keV}$ fluxes and spectroscopic/photometric redshifts. The criteria adopted by \citet{szokoly04} are as follows:

\begin{enumerate}
\item Galaxy: $L_X \leq 10^{42}$ erg. s$^{-1}$ with hardness ratio $HR <-0.2$.
\item AGN-2: $10^{41}\leq L_X < 10^{44}$ erg. s$^{-1}$ and the  hardness ratio $HR >-0.2$.
\item AGN-1: $10^{42} \leq L_X < 10^{44}$ erg. s$^{-1}$ and the  hardness ratio
  $HR\leq -0.2$.
\item QSO-2: $L_X > 10^{44}$ erg. s$^{-1}$ with hardness ratio $HR >-0.2$.
\item QSO-1: $L_X > 10^{44}$ erg. s$^{-1}$ with hardness ratio $HR <-0.2$.
\end{enumerate}

\begin{figure}
\begin{tabular}{c}
 \includegraphics{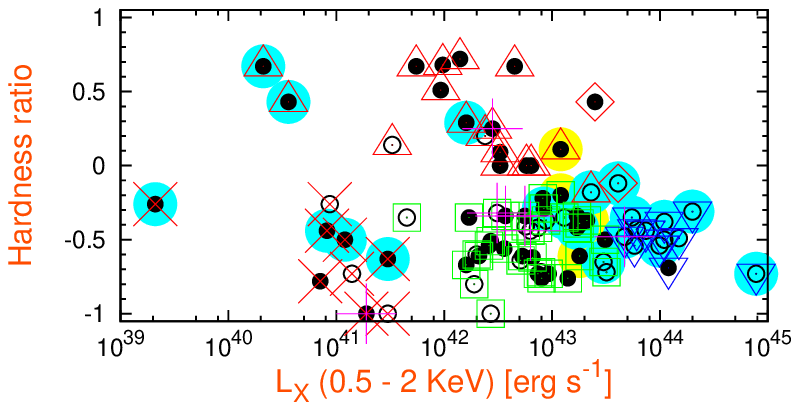}\\
 \includegraphics{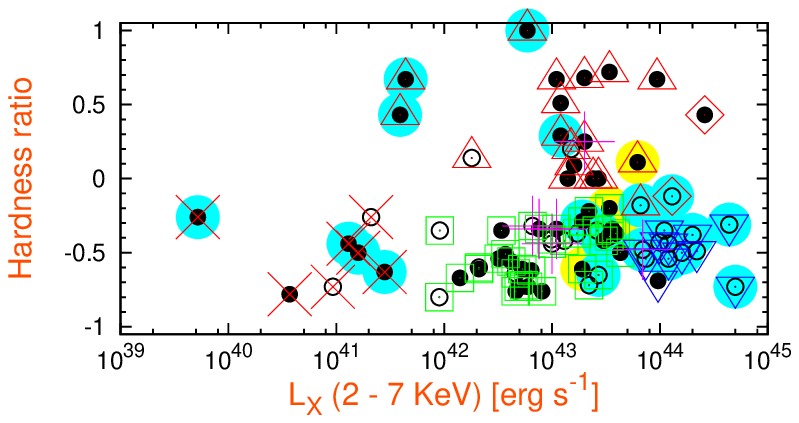}\\
\end{tabular}
\caption{The hardness ratio as a function of the soft X-ray
luminosity (top
  panel) and the hard X-ray luminosity (bottom panel) for the radio-X-ray
  matches. Same symbols as for the radio-X-ray matches in Fig \ref{powersRX}.}
\label{hr_LX}
\end{figure}

The classification criteria yielded  37  radio-X-ray  matches classified as AGN-1, 19 classified as AGN-2. Normal galaxies represent a small fraction of 9 sources while 12 sources were classified as QSO-1 and 10 sources of these were spectroscopically classified as QSOs and some of these sources were also identified in the ROSAT. The QSO-2 type constitutes the smallest fraction: two candidates were classified as QSO-2 sources.  Only  a few sources considered  as the prototype of type 2 quasars have been detected for example in \citet{norman02} and \citet{stern02b},  but many more have been discovered recently \citep[e.g.][]{fiore08}.

As discussed previously, different types of X-ray sources are separated in the
plot of X-ray hardness ratio versus  X-ray luminosity. This trend is shown
in Figure \ref{hr_LX}, where the hardness ratio as a function of the soft X-ray
luminosity is illustrated in the top panel and as a function of the hard X-ray
luminosity in the lower panel.

\subsection{Radio luminosity}
The rest-frame radio luminosity density was calculated from the following
equation:
\begin{eqnarray}
  \label{eq:radioP}
  L_{1.4 \,{\rm GHz}} = 4\, \pi\, d_L^2 \,S_{1.4 \,{\rm GHz}}\, (1+z)^{-1 + \alpha}\, 10^{-33} ~{\rm W~Hz^{-1}},
\end{eqnarray}

\noindent where $d_L$ is the luminosity distance (in cm), $S_{1.4}$ is the 1.4 GHz flux density (in mJy), and $\alpha$ is the radio spectral index ($f_{\nu} \propto \nu^{-\alpha}$), which is taken to be 0.8.  Figure \ref{powersRX} (bottom panel)  shows the 1.4 GHz luminosity as a function of redshift for all the
radio-X-ray matches. Here the curve represents the detection limits. Many of the objects are close to the lower radio flux limit and also many of the AGN-1 cluster around a redshift of $z\sim 0.7$. 
\subsection{X-ray-radio correlation?}

At the sub-mJ and $\mu$Jy levels, radio emission is known to be a highly accurate indicator of star
formation rate \citep[see the review by][]{condon92} because of the extremely tight FIR-radio correlation, which deviates by less than a factor of two over five orders of magnitude. A number of studies have investigated the radio-derived SFRs and found them to be in good agreement with the FIR derived values \citep[e.g. ][]{condon91, haarsma00,hopkins01, bell03}. The disadvantage of this method is that it is hard to account for the contamination of the radio flux by AGNs. Moreover, star-forming and spiral galaxies are also found to be powerful X-ray emitters with luminosities that sometimes exceed
$\approx 10^{42}$ erg s$^{-1}$ \citep[e.g. NGC 3265;][]{moran99}. This is believed to be due to a number of high mass X-ray binaries, young supernovae remnants and hot gas plasma associated with star forming regions \citep[see the review by][]{fabbiano89}. With the latest generation of X-ray satellites such as  Chandra and XMM, it is now possible to study the X-ray-SFR correlation \citep[][]{bauer02, franc03, grimm03, ranalli03}. Early results show good agreement with FIR and radio estimates of the SFR.

Correlations between X-ray and radio luminosities for radio-quiet AGN have been determined by \citet{brinkmann00}. These authors have cross correlated the FIRST radio survey ($S_{1.4 \,GHz}>1 $mJy) and the ROSAT ($f_X>10^{-13}$ erg m$^{-2}$ s$^{-1}$) and shown that the X-ray and radio luminosities for radio quite AGN follow a relatively tight linear correlation in the form:

\begin{eqnarray}
  \label{eq:brink}
  \log\,(L_X) = (-4.57\pm 2.55) + (1.012\pm 0.083)\, \log\,(L_{1.4\,{\rm GHz}})
\end{eqnarray}
\noindent \citet{simpson06} have  converted the previous
correlations (found for X-ray-radio AGN) to relationships between
fluxes in the form of :
\begin{eqnarray}
  \label{eq:simp1}
  S_{0.5 - 2\,{\rm keV}}\;{\rm (W~m^{-2})} = 10^{-15.5} \;S_{1.4\,{\rm GHz}}~{\rm (mJy)}
\end{eqnarray}

\begin{eqnarray}
  \label{eq:simp2}
  S_{2 - 10 \,{\rm keV}}\;{\rm (W~m^{-2})} = 10^{-15.3} \;S_{1.4 \,{\rm GHz}}~{\rm (mJy)},
\end{eqnarray}

\noindent and also converted correlations found for X-ray radio star-forming
galaxies into relations between fluxes:
\begin{eqnarray}
  \label{eq:simp3}
  S_{0.5 - 2 \,{\rm keV}}\;{\rm (W ~m^{-2}}) = 10^{-18} \;S_{1.4\,{\rm  GHz}}~{\rm  (mJy)}
\end{eqnarray}

\begin{eqnarray}
  \label{eq:simp4}
  S_{2 - 10 \,{\rm keV}}\;{\rm (W~m^{-2})} = 10^{-18}\; S_{1.4\, {\rm GHz}}~{\rm (mJy)},
\end{eqnarray}
and used these relations in their plot of  X-ray flux versus radio flux density in order to determine the nature of their radio-X-ray matches, depending on which of the two correlations they are closer to (radio quiet AGNs or starburst galaxies). The authors suggest that $20\%$ or more of the radio sources in the sample with ${\rm 100\,\mu Jy <S<300\, \mu Jy}$ are radio quiet AGNs.  They also claim that the radio-X-ray correlation found for radio quite AGN might
be biased to unusually high X-ray-luminous sources because it is derived from cross correlation of a deep radio catalogue (FIRST) and a shallower X-ray survey (ROSAT).

\begin{figure}
\begin{tabular}{c}
\includegraphics{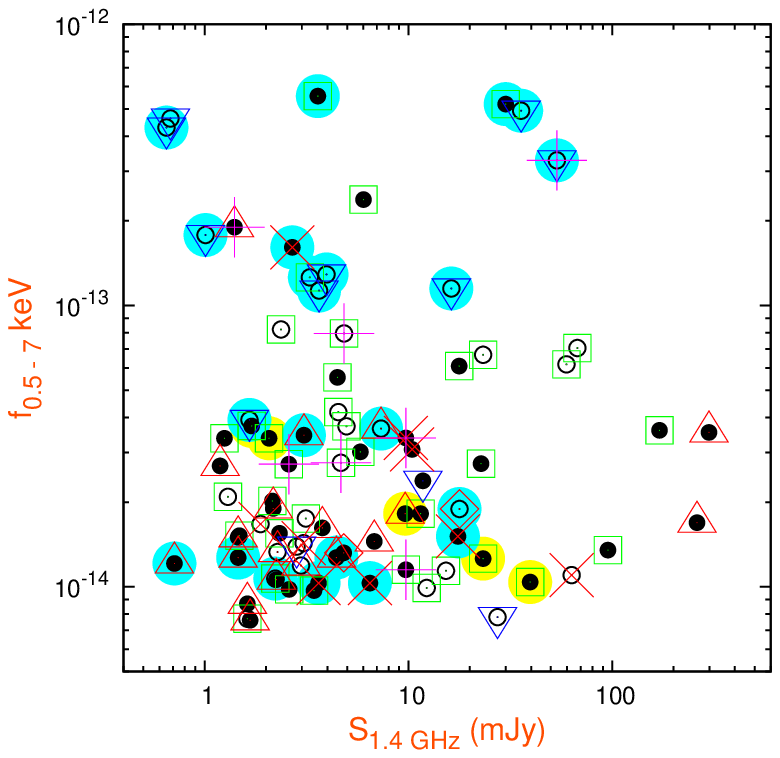} \\ 
 \includegraphics{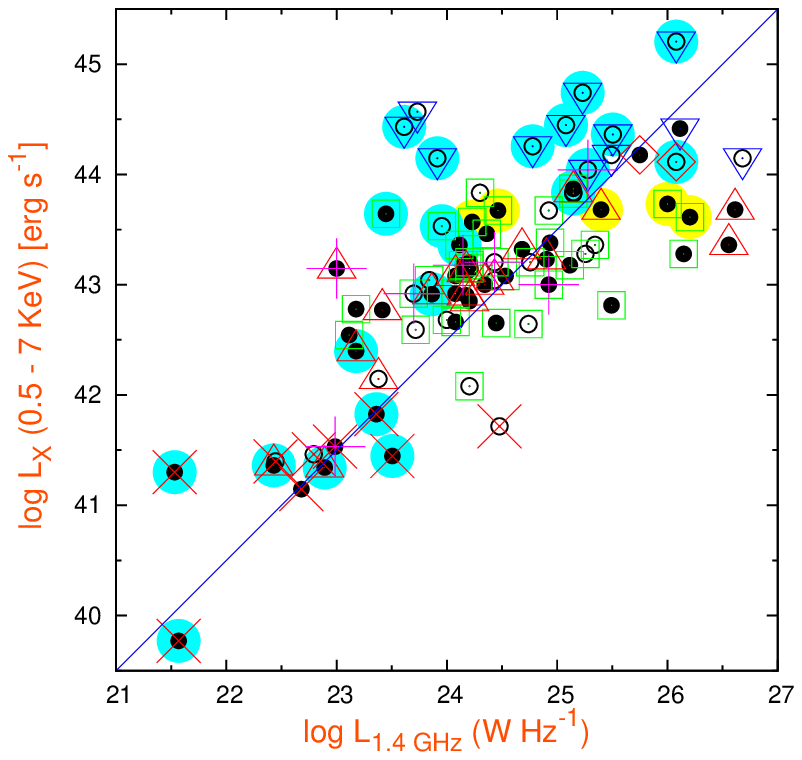}\\ 
\end{tabular}
\caption{The upper panel shows the full X-ray flux/radio flux density
  distribution with the class of source indicated by the symbol. The lower
  panel displays the full X-ray luminosity  versus the radio luminosity for
  the radio-X-ray matches. The symbols are the same as in Fig. \ref{powersRX}.}
\label{XR_V_R}
\end{figure}

The top panel in Figure \ref{XR_V_R} presents the distribution of
the full X-ray flux against radio flux density. It is immediately
clear from this plot that there is no obvious correlation between
radio and X-ray fluxes. The majority of the radio-X-ray matches have a
radio flux below 10 mJy. Three sources are detected below the
radio flux density limit (1 mJy); two of theses sources are
optically star-like and X-ray bright sources of which one is
spectroscopically identified as a QSO. The third source is an
optically extended source and is spectroscopically identified as
galaxy showing a narrow emission line. This source is classified
as an AGN-2 based on its X-ray luminosity and hardness ratio. The
lower panel in Figure  \ref{XR_V_R} shows the full X-ray
luminosity versus  the radio luminosity, galaxies extend to lower radio/X-ray luminosities from the quasar population with some overlap. The majority of the radio-X-ray matches tend to have a small radio to X-ray luminosities ratios, corresponding to radio-quiet AGN. The small sample size and incompleteness of the spectroscopy do not allow more quantitative conclusions.

\section{X-ray-to-optical flux ratio}
The X-ray to optical flux ratio is an important tool that can be used to
explore the nature of the X-ray sources. Previous studies \citep[e.g.][]{maccacaro88, elvis94} have shown that there is an empirical relationship between X-ray and optical emission in AGN, and  this can yield important information on the nature of  AGN activity. In the X-ray optical plane, powerful unobscured AGNs \citep[e.g.][]{stocke91, barger03} typically have flux ratios of $-1 <\log\,f_{(2-7)\,{\rm keV}}/f_{{\rm opt}} <+1$, while star forming galaxies and low luminosity AGNs tend to have $\log\,f_{(2-7)\,{\rm keV}} /f_{{\rm opt}}\leq -1$ \citep[e.g. ][]{horn01, gaic02, alexander01}. The X-ray-to-optical flux ratio is estimated from the relation \citep{horn01}:

\begin{eqnarray}
      \label{eq:ratioflux}
\log \frac{f_{X}}{f_{{\rm opt}}}&=&\log f_{X} + 0.4 \, R + 5.53 ,
  \end{eqnarray}
\noindent where the flux is measured in units of $10^{-15}$ erg s$^{-1}$ cm$^{-2}$   in the considered X-ray band  and $R$ is the apparent $R$-band  Vega magnitude.

\begin{figure*}
\begin{center}
\begin{tabular}{c}
\includegraphics{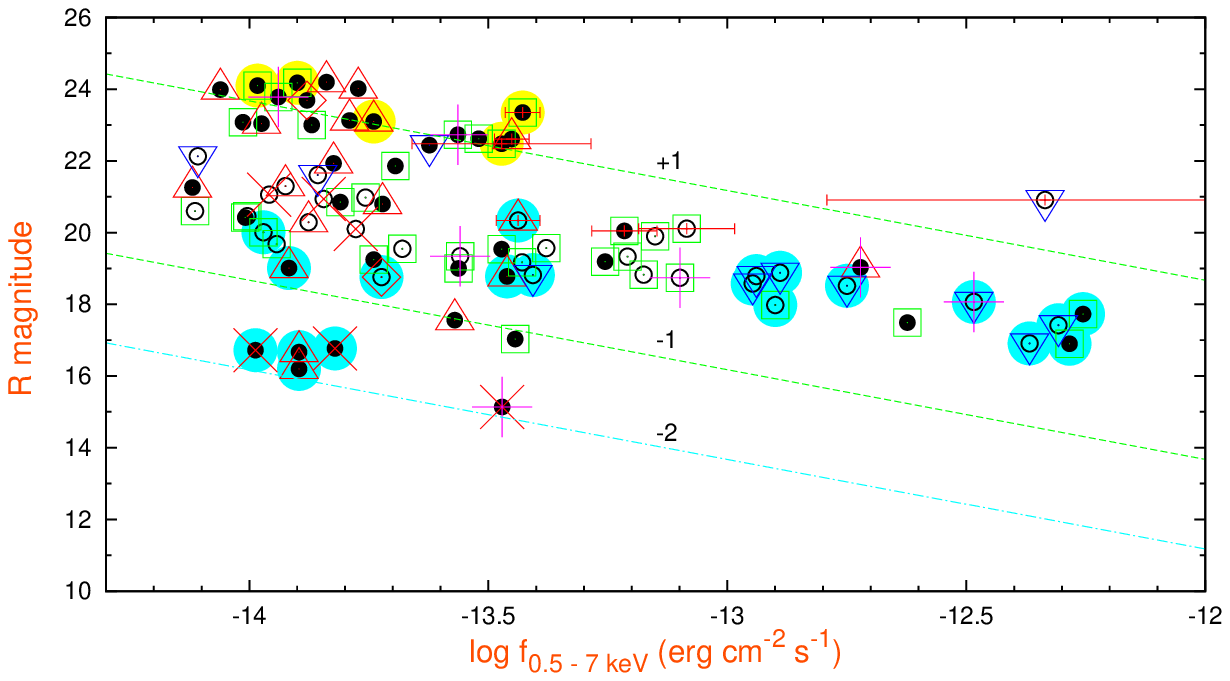} \\
\includegraphics{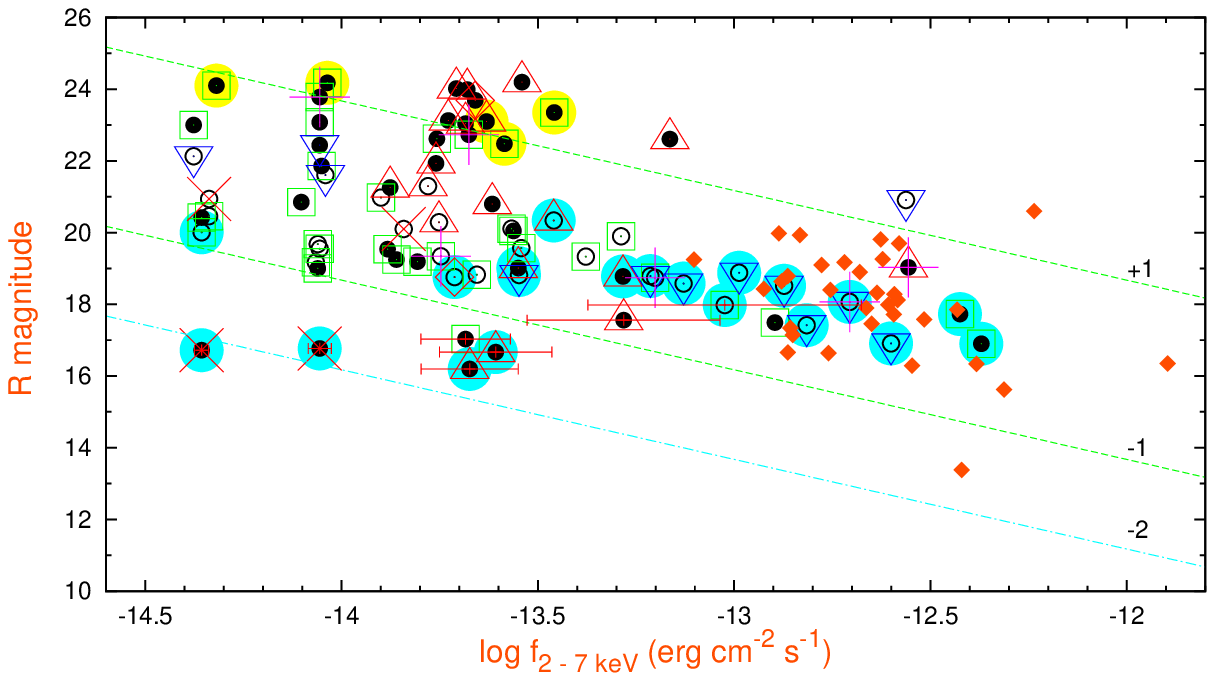} \\
\includegraphics{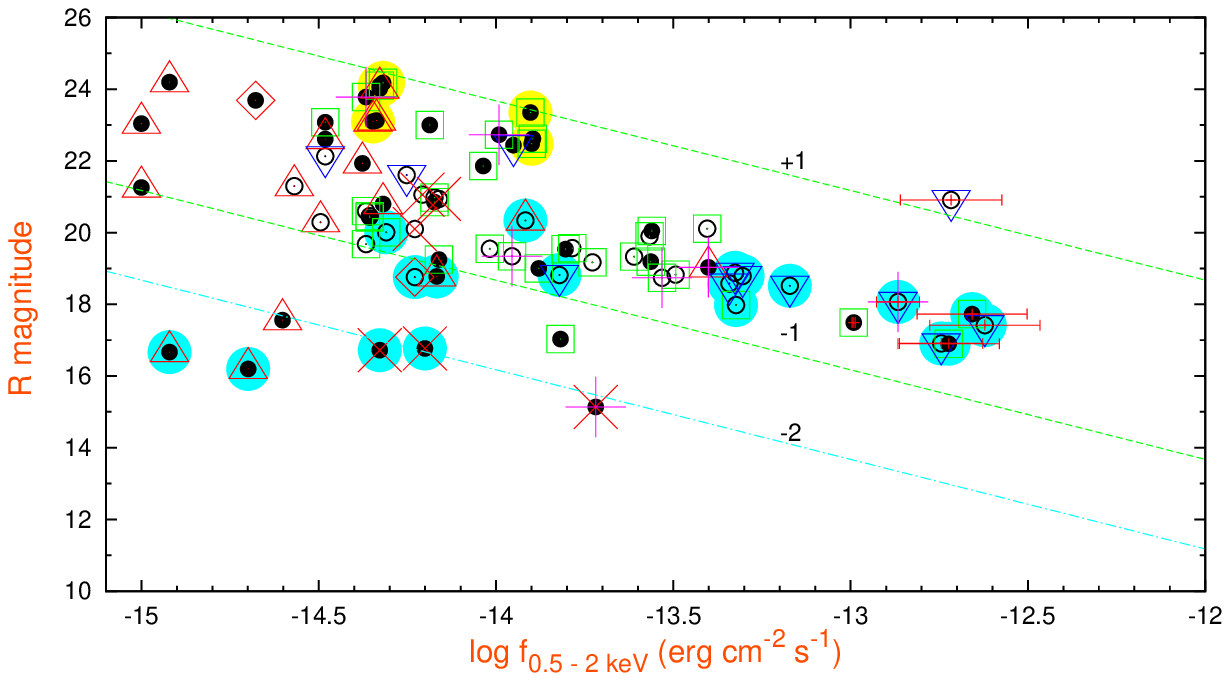}\\
\end{tabular}
\end{center}
\caption{Optical $R$-band magnitude of the radio-X-ray matches plotted against the
  full (0.5-7 KeV) X-ray fluxes in the top panel, hard (2-7 keV) X-ray fluxes
  in the middle panel and soft (0.5-2 keV) X-ray fluxes in the lower panel. The filled
  diamonds in the middle panel  correspond to the sources in \citet{akiyama2000}. The
 solid lines denote  the location of constant X-ray-to-optical flux ratios of $\log\,
  f_X/f_{opt}$=1, 0, -1  from top to bottom as given by equation \ref{eq:ratioflux}.
 Same symbols as for the radio-X-ray  matches in Fig \ref{powersRX}.}
\label{magR_flux}
\end{figure*}

 Figure \ref{magR_flux}  plots the $R$ magnitude as a function of the  ${\rm 0.5-7~keV}$ X-ray flux for the radio-X-ray matches (top panel). The middle panel displays the ${\rm 2-7~keV}$ X-ray flux for the radio-X-ray matches. The filled diamonds illustrate  the \citet{akiyama2000} ASCA Large Sky Survey data (two clusters and one source without  optical identifications have been excluded). The bottom panel shows the ${\rm 0.5-2~keV}$ soft X-ray flux for the radio-X-ray matches. The figures  show that
there is a large fraction of the sources that span the typical X-ray-to-optical
flux ratio of the AGN region. The majority of the radio-X-ray matches ($68\%$)
fall within $\log f_{\,(2-7)\,{\rm keV}}/f_{{\rm opt}}=0.0\pm1.0$. One notes that there is also a significant population ($\sim 23\%$) of sources that are X-ray over-luminous for their optical magnitudes [$\log f_{\,(2-7)\,{\rm keV}}/f_{{\rm opt}}>1$], suggesting  high redshift sources and/or dust obscuration \citep{alexander01,fiore03, brusa04, gergan04, mignoli04,   ghandi04, geor04}. Among the sources with a high X-ray-to-optical flux ratio, five are EROs shown with  large yellow filled circles. Two optically non-identified sources fit within this category but are not plotted here.

Recent studies have revealed a population of optically bright
X-ray faint sources. Such a population was unveiled by the Chandra
deep survey \citep{horn01, tozzi01} and appear to be at $z
\lesssim  1$. These sources appear mainly at very faint X-ray
fluxes ($\lesssim{\rm 10^{-15}~erg~s^{-1}~cm^{-2}}$), and were found to comprise star-forming galaxies,  normal galaxies and low
luminosity AGN (with $L_X<10^{42}$ erg s$^{-1}$ in the $0.5-10$
keV band). However, in this analysis 7 ($8\%$) radio-X-ray
matches are found to be optically bright-X-ray  faint [$\log
f_{\,(2-7)\,{\rm keV}}/f_{{\rm opt}}<-1$]. All are extended objects in the
$R$ band image (sources \# 20, \# 29, \#39, \# 42, \# 44, \# 48, \#
71) and their $R$ magnitude is in the range $15<R<18$ (and $I<17$).
Four of these sources are classified spectroscopically as
  galaxies: source \# 29, source \# 39, source \# 42 and source
\#71, with spectroscopic redshifts of $z=0.128$, $z=0.126$,
$z=0.238$ and  $z=0.129$ respectively (see their spectra in Figure
\ref{sdss-bootes}). Thus, these sources should be nearby, bright
normal galaxies \citep[see e.g.][]{barger01,   tozzi01, horn01}.
In addition, two sources (\# 29 and \# 39)  of the spectroscopically identified sources are classified  as AGN-2 (based on the X-ray luminosity) with $L_X[0.5-7\, {\rm keV}]= 2.3\times 10^{41}$ erg s$^{-1}$,  and $L_X[0.5-7\, {\rm keV}]=  2.2\times 10^{41}$ erg s$^{-1}$ respectively. The two sources are classified as LINERS using the BPT diagram. So it is interesting that the two classification methods, optical and X-ray, are in good agreement.  The other  two  sources (\# 42 and \# 71) are classified as normal galaxies based on their X-ray luminosity and classified as star-forming galaxies using the optical classification. Their  X-ray luminosity are: $L_X[0.5-7\, {\rm keV}]= 6.7\times  10^{41}$  erg s$^{-1}$,  and  $L_X[0.5-7 \,{\rm keV}]=  2.8\times 10^{41}$ erg s$^{-1}$ respectively.

\subsection{Extremely red objects}
\begin{figure}
\includegraphics{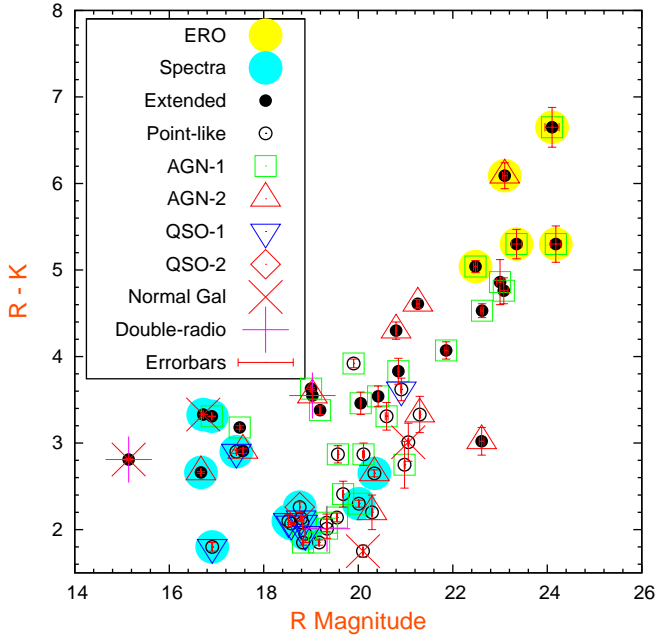}
\caption{Colour magnitude diagram  $R-K$ versus $R$-band magnitude for all radio-X-ray matches identified in both $R$ and $K$ bands in Bo\"{o}tes field.}\label{rk_r}
\end{figure}

Figure \ref{rk_r} shows the $R-K$ colour of the optical counterparts to the radio-X-ray matches as a function of the $R$ magnitude.
 One notes that the $R-K$ colour tend to increase to  fainter magnitudes. A Similar trend has been reported by several studies \citep[e.g.][]{lehman01, alexander01, mainieri02}. Galaxies (stellarity $<0.7$) rapidly become redder than point-like objects (stellarity $\ge0.7$). One also notes the clear separation between the two groups. This can be seen in a more pronounced way in figure \ref{bi_ik}, where the $Bw-I$ colour is plotted versus the $I-K$ colour. The optical point sources tend to occupy a different region from the optically extended ones in the colour-colour diagram.  The blue colour of point-like objects is consistent with those of quasars. Five interesting objects are associated with bright near infrared sources with $17<K<19$ and have colours $R-K>5$ that would classify them as Extremely Red objects (EROs). In this subsample of the EROs, four are classified as AGN1 and one source is classified as AGN2 based on their X-ray properties. The red colour of the radio-X-ray matches can be explained by  either obscured AGN or a high redshift cluster of galaxies (Lehmann et al. 2001). The extremely red colour of X-ray sources can be used as a good tracer of red galaxies at high redshifts. Near infrared spectroscopy observations have confirmed this \citep[e.g. ][]{cowie01}.

The ERO population is well known to be a mixture of old passively evolving elliptical and star-forming galaxies strongly reddened by dust extinction at high redshift  ($z>1$). Photometric and spectroscopic classifications \citep[see e.g.][]{cimatti03, smail02} have shown that the ERO population is almost equally divided between the  two components. 

The photometric redshift of the five EROs ranges from 0.88 to 1.77 and the spectral type classifies two EROs as ellipticals, two as  starburst galaxies and the last one as lenticular. Moreover, these sources have a high X-ray to optical flux ratio $\log\,(f_X/f_{opt})> 1$ (see Figure \ref{hr_ratio_color}, yellow filled circles) and this signature can be used to identify high redshift X-ray sources. Obviously spectroscopic confirmation is required to explore this further. The X-ray luminosities in the 0.5-7 keV rest-frame energy band of the EROs are in the range $10^{43}-10^{44}$ erg s$^{-1}$, suggesting that the EROs X-ray emission is most likely powered by AGN activity.

\section{General properties of the radio-X-ray sample}
\subsection{HR-X-ray-to-optical flux ratio-colour}
\begin{figure}
\begin{center}
\begin{tabular}{c}
\includegraphics{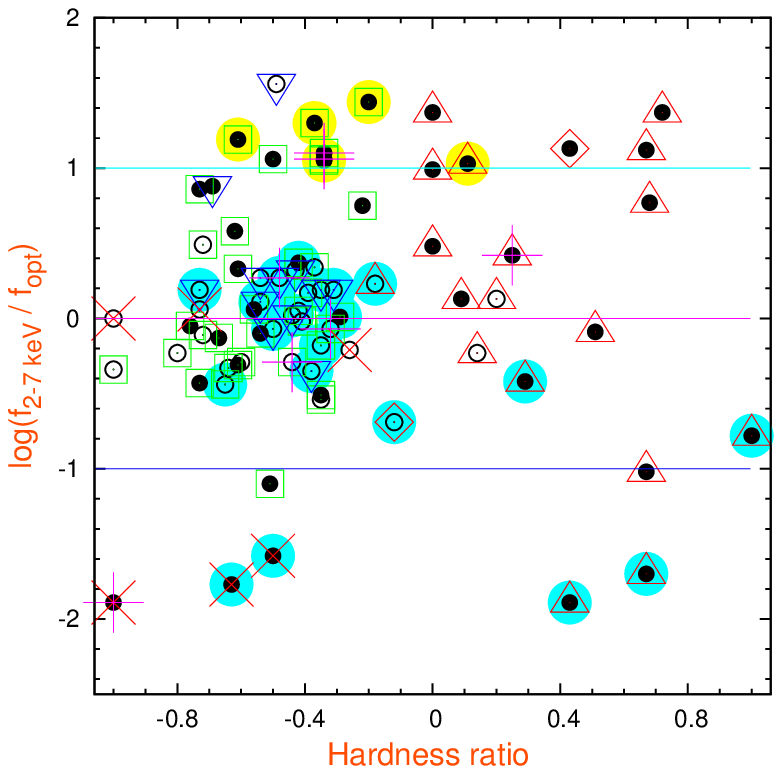}\\
\includegraphics{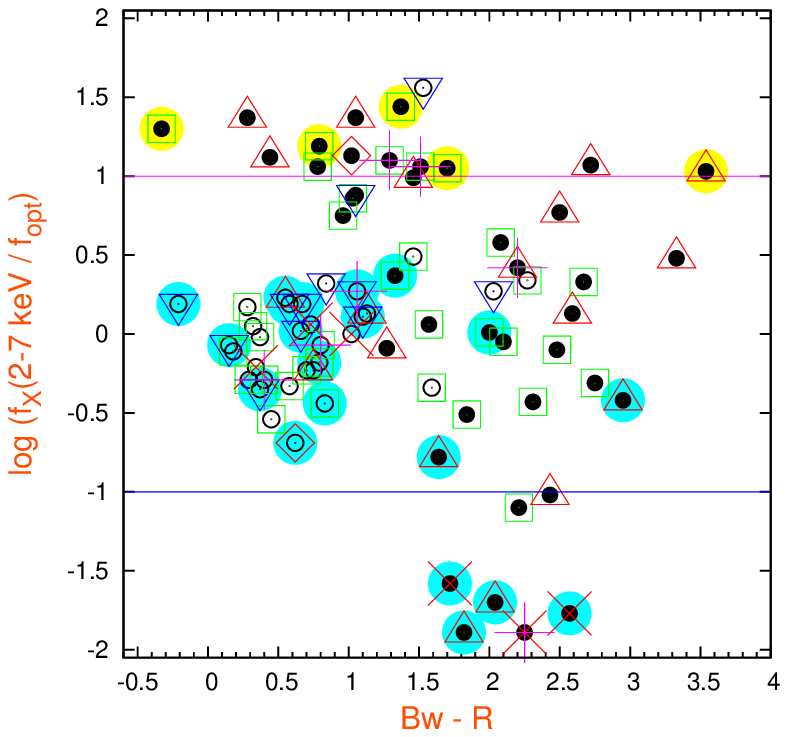}\\
\end{tabular}
\end{center}
\caption{The X-ray-to-optical flux ratio as a function of the
Hardness ratio HR (HR is defined as (h-s)/(h+s) where h and s are
the count rates in the $0.5$ keV and $2-7$ keV bands respectively)
(top panel) and as a function of the optical colour $Bw-R$ (bottom
panel). Horizontal lines show location of constant
  X-ray-to-optical flux ratio of +1, 0, and -1. Same symbols as for the radio-X-ray matches in Fig \ref{powersRX}.} \label{hr_ratio_color}
\end{figure}

\begin{figure}
\begin{center}
\includegraphics{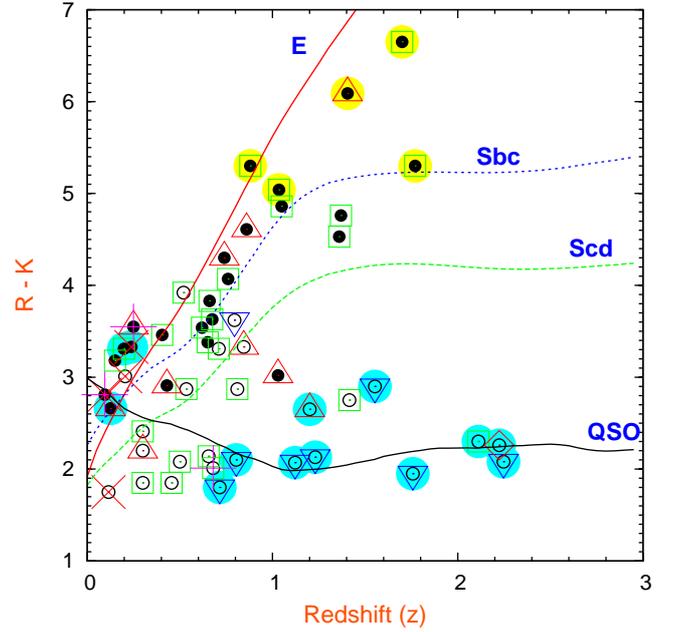}
\caption{Optical/near-infrared colours as a function of photometric/spectroscopic redshift for the radio-X-ray matches. Same symbols as for the radio-X-ray matches in Fig\ref{powersRX}. The curves are different template SEDs for E/S0 (continuous), Sbc(dotted), Scd(dashed) and QSOs (black continuous). The galaxy templates are observed SEDs from \citet{coleman80}. The QSO template  is obtained from the set of QSOs SEDs of the LE PHARE software.}\label{rk_z_xray}
\end{center}
 
\end{figure}

Figure \ref{hr_ratio_color} displays the X-ray-to-optical flux ratio
(calculated using equation \ref{eq:ratioflux}) as a function of the hardness
ratio (upper panel) and as a function of the  $Bw-R$ colour  (lower panel). Although the radio-X-ray matches span a large range of hardness ratios, most sources cluster around the soft values (${\rm HR\sim-0.4}$). The majority of these sources have point-like optical morphology and 12 of them are
spectroscopically identified as  QSOs with broad emission lines and high ${\rm f_x/f_{opt}}$, while two sources are spectroscopically identified as  galaxies showing narrow emission lines with high X-ray-to-optical flux ratio ($\log f_X/f_{opt}>-1$) with  high hardness ratio, possibly suggesting   high column densities. The lower panel in Figure \ref{hr_ratio_color}   shows that BLAGNs have bluer color compared to not BLAGNS. The same figure shows that the optically extended sources have, on average, a red colour ($Bw-R \geq   1.5$), suggesting that the optical light is dominated by the host galaxy   rather than the central AGN \citep[e.g.][]{barger03,     ghandi04}.  Additional evidence that the optical/NIR light in many sources   is dominated by the host galaxy can be seen in Figure \ref{rk_z_xray}  where   the   optical-NIR colour is plotted versus redshift and the magnitude respectively. Figure \ref{rk_z_xray} displays the $R-K$ colour as a function of redshift. Overlaid are the optical/NIR colours of a QSO spectrum \citep{cristiani90, cristiani04} obtained from the template SED of the LEPHARE software and the mean observed spectra of three different galaxy types (E/S0, Sbc, Scd). Broad line AGNs, most of which exhibit soft X-ray spectra, have colours consistent with the QSO template prediction, while the extended sources follow the galaxy tracks. Again this figure  illustrates an interesting segregation between optically extended and point-like objects, and clearly shows that optically extended objects  have a redder colour while point-like objects tend to have a blue colour consistent with QSOs.

\begin{figure}
\begin{center}
\begin{tabular}{c}
\includegraphics{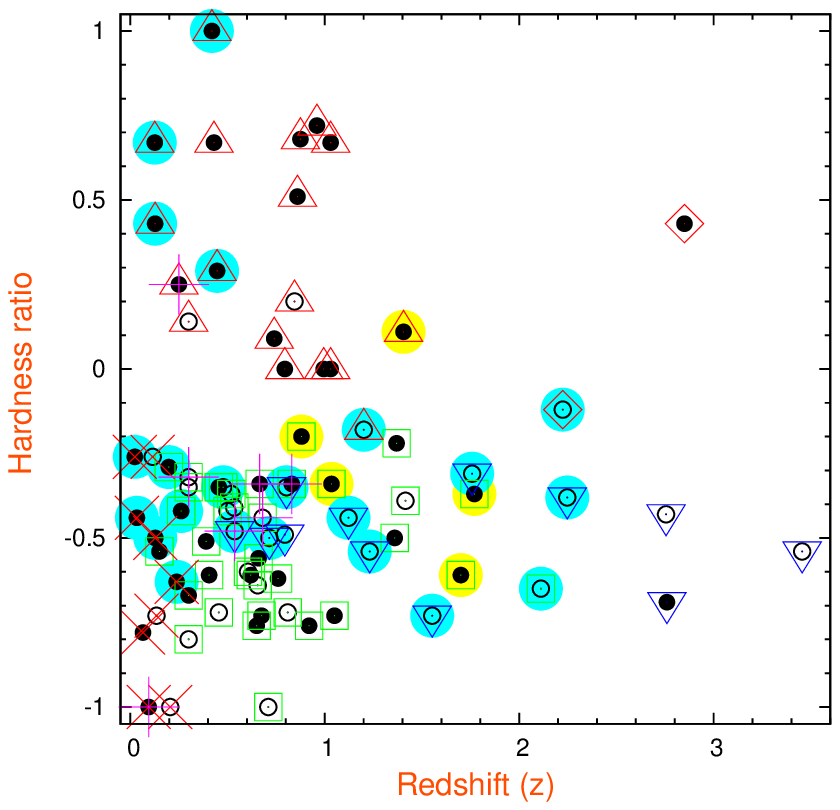} \\
\includegraphics{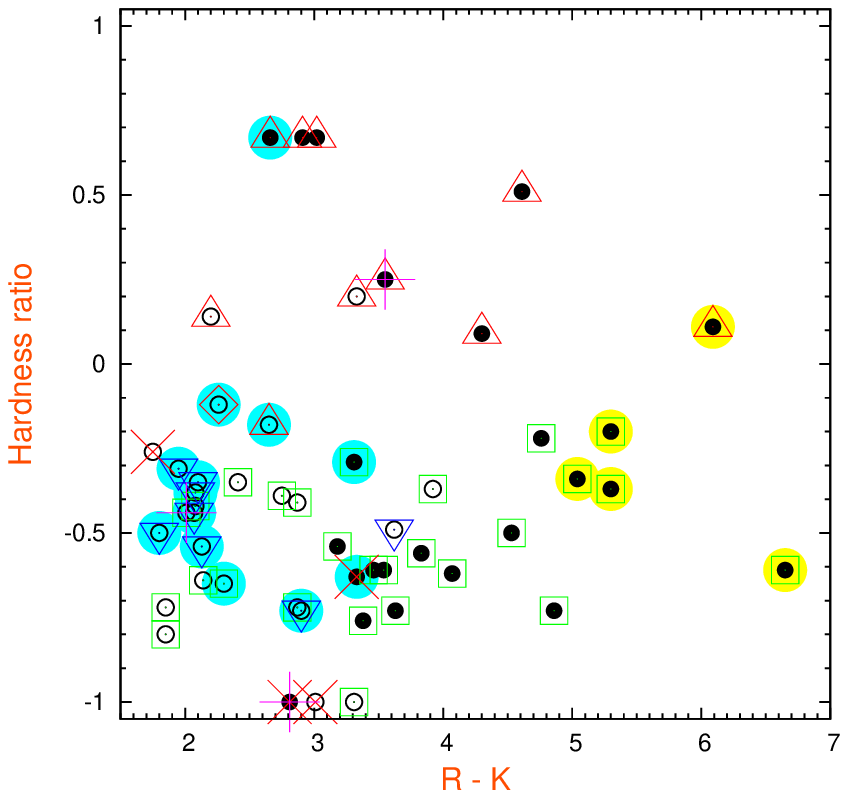} \\ 
\end{tabular}
\end{center}
\caption{The hardness ratio as function of redshift  is plotted in
the top panel, and as a function of the optical/near-infrared
colour is plotted in the bottom panel. Same symbols as for the
radio-X-ray matches in Fig \ref{powersRX}. A clear correlation
between HR and the optical/near-infrared colour is not observed.}
\label{hr_z-color}
\end{figure}

Figure \ref{hr_z-color} illustrates the hardness ratio as function of  redshift (top panel) and the optical/near-infrared colour (bottom panel). This figure  shows an interesting evolution of the hardness ratio with redshift of the radio-X-ray matches. The hardness ratio declines with redshift, and the majority of the hard X-ray counterparts to FIRST radio sources are at redshift lower than 1.5. This could be due to a combination of two effects. First, the fraction of AGN which are obscured is higher at lower luminosities (equals lower redshifts in a flux-limited sample). Sceond, the \textit{k-}correction for obscured AGN pushes the less obscuration sensistive hard X-rays into the observed band at higher redshift.
The hardness ratio versus $R-K$ optical colour is presented in the lower panel of Figure \ref{hr_z-color}. In this diagram, no clear correlation between the hardness ratio and optical colour is observed.  Ignoring the QSOs, it could be noted that  the very hard sources all have low $R-K$. This is a consequence of the correlation  of hardness ratio with redshift in Figure \ref{hr_z-color} (upper panel) and the fact that low $R-K$ galaxies are mostly at low redshift. One should note, however, that,  in general,  sources optically classified as obscured AGN are redder than unobscured AGN and also tend to have higher values of HR.

\subsection{Colour-colour diagram} 
\begin{figure}
\begin{center}
\includegraphics{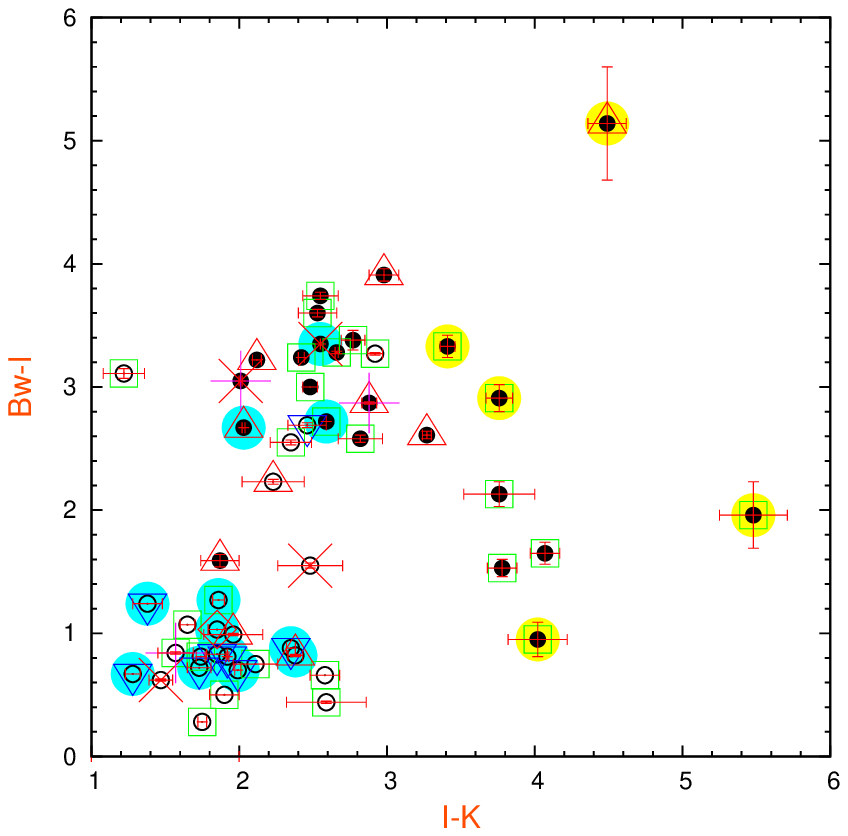}
\caption{Colour-colour diagram for all radio-X-ray matches identified in \textit{Bw, I} and \textit{K} bands in Bo\"{o}tes field. Same symbols as for the radio-X-ray matches in Fig \ref{powersRX}.}
\label{bi_ik}
\end{center}
\end{figure}

\begin{figure}
\begin{center}
\includegraphics{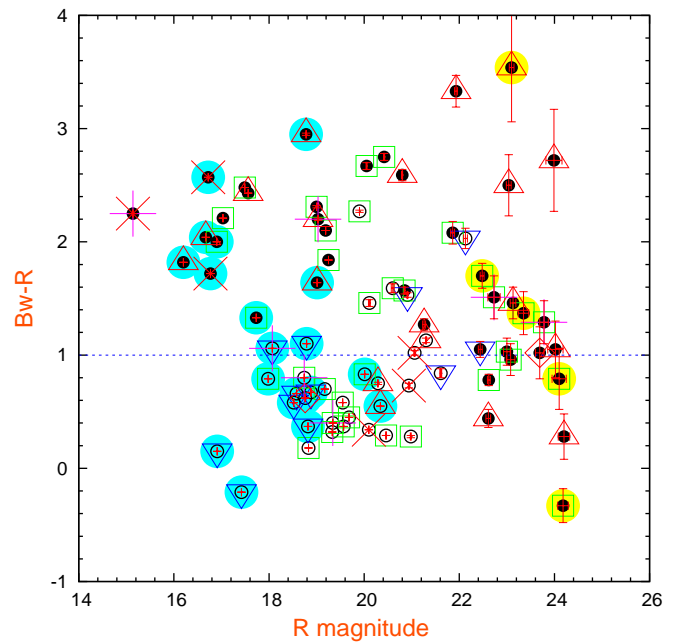}
\caption{ Colour magnitude diagram  $Bw-R$ versus $R$-band magnitude
for all radio-X-ray matches identified in both $Bw$ and $R$ bands in
Bo\"{o}tes field. The horizontal line corresponds to $Bw -R =1$.
Same symbols as for the radio-X-ray matches in Fig \ref{powersRX}.} \label{br_r}
\end{center}

\end{figure}

Figure \ref{bi_ik} displays  the $Bw-I$ colour  versus the $I-K$ colour. The optical point sources tend to occupy a different region from the optically extended ones in the colour-colour diagram. The clear separation between the two groups  reflects the effectiveness of the SExtractor parameter to separate extended objects (galaxies) and point-like objects (stars or QSOs). The blue colour of optical point-like objects is consistent with those of quasars. In addition to the point-like objects (13 sources)
spectroscopically classified as quasars (cyan filled circles), 20 of the identifications have stellar optical profiles and all of these lie in the AGN region of the X-ray optical plane (see Figure \ref{magR_flux}), suggesting that they are QSOs rather than stars. Figure \ref{br_r} shows the $Bw-R$ colour as a function of \textit{R} magnitude. A similar  trend is shown in this Figure. In this Figure a group of optically extended objects tend to have a bluer colour ($Bw -R <1$) at fainter magnitudes ($22<R<24$). \citet{akiyama2000} used this criterion to select candidates of BLAGNs in the LSS area.

\section{Final data table}

The final characteristics of the radio-X-ray matches are presented in Table \ref{infox},  the table contains the following information:

\begin{itemize}
\item (1) source ID number;
\item (2) X-ray object name (X-ray counterparts to FIRST  radio sources);
\item (3)-(4) radio coordinates in hours, minutes and seconds;
\item (5) radio flux in mJy;
\item (6)-(7)-(8) the X-ray flux in the full (0.5-7) keV, soft (0.5-2) keV and
 hard (2-7) keV bands in units of  erg s$^{-1}$ cm$^{-2}$;
\item (9) the hardness ratio HR, derived from the 0.5-2 and 2-8 keV band
  counts \citep{kenter05};
\item (10) X-ray-to-optical flux ratio;
\item (11)-(12) log of X-ray luminosity in 0.5-2 keV,  2-7 keV and 0.5-7
  keV  respectively in erg s$^{-1}$;\\    
\item (13) log of X-ray luminosity in  0.5-7 keV  in erg s$^{-1}$;
\item (14) log of 1.4 GHz luminosity in W Hz$^{-1}$;
\item (15)-(16) the apparent $R$ band magnitude associated with the
  corresponding errors and the $R-K$ colour respectively;
\item (17) the source extractor stellarity parameter used to classify sources
  into point-like objects and extended sources;
\item (18) the photometric redshift as derived from Hyperz  with the
  corresponding errors. Sources with spectroscopic redshift are printed in bold
  face;
\item (19) Classification obtained from the X-ray characteristics (based on
  the X-ray luminosity and the hardness ratio HR).

\end{itemize}

\section{Conclusions}

In this paper I have presented the broad band properties of a sample of 79 radio emitting
X-ray sources, obtained from a correlation of the FIRST 1.4 GHz survey and  the publicly available X-ray data of the  medium depth  XBo\"{o}tes field. Out of $\sim 900$ FIRST radio sources that lie in the Bo\"{o}tes field, 92  have an X-ray counterparts and 79 optical/infrareds sources are common to both  radio and X-ray sources.  All the 79 sources  are identified in the full X-ray band (0.5-7 keV), 76 in the hard X-ray band (2-7 keV) and 77 in the soft band (0.5-2 keV). The main conclusions can be summarised as follows:

Optical spectra were obtained from the SDSS for 22  radio-Xray matches. The majority of the spectroscopically identified sources are BL AGNs, $18\%$ are classified as LINERs, $14\%$ as star-forming galaxies and one source classified as BL Lac object. Among the sources with no spectroscopic classification 19 are point-like with blue colours and are most probably QSOs.

Photometric redshifts were calculated using the public code \textit{Hyperz}. The redshift distribution of the AGNs shows a peak at $z\sim 0.7$, supporting previous studies \citep{barger03} that show that the peak formation of super-massive black holes occurred at relatively recent times ($z<1$). This leads to the conclusion that medium depth X-ray surveys are well suited for studying and probing this epoch effectively. Photometric redshifts were also used to investigate the $K-z$ relation. It is found that the  $K-z$ relation obtained using photometric redshifts is similar to that obtained for brighter sources investigated by \citet{willott03} and \citet{brookes06}.  One should note that the $K$ band catalogue is not as deep compared to the study of \citet{willott03}.  Deep K band data and more use of spectroscopic redshifts is required before strong conclusions can be drawn.

 X-ray luminosities have been calculated  using photometric redshifts and also spectroscopic redshift 
wherever available. The majority ($88\%$) of the sources have a high  X-ray luminosity in the full band, $L_X>$ $10^{42}$ erg s$^{-1}$, and $11\%$ of  sources have low X-ray luminosity, $L_X<$ $10^{42}$ erg s$^{-1}$. The classification of the radio-X-ray matches  based on the hardness ratio and X-ray luminosity yielded $46\%$ AGN-1 (unobscured), $24\%$ AGN-2 (obscured), $15\%$ QSO-1, $3\%$ QSO-2 and $11\%$ normal galaxies. One should note that 13 sources were  spectroscopically classified as QSO (by SDSS) and 12/13 were classified as QSO-1 based on X-ray luminosity and hardness ratio. So it is interesting that the X-ray classification scheme is largely coincident with the classical AGN classification based on optical spectroscopic diagnostics.

The X-ray-to-optical flux ratios is a good discriminator between X-ray sources classes down to very faint optical magnitudes and X-ray fluxes, with AGN typically falling within the region defined by the loci $\log\, f_X/f_{opt}=0\pm1$, and star-forming galaxies and low luminosity AGNs have $\log\,f_{X} /f_{{\rm opt}}\leq -1$. The majority of the radio-X-ray matches  ($68\%$) are   AGN ($\log\, f_X/f_{opt}=0\pm1$). One notes again that all the sources classified as QSO (by SDSS) fall within this region. A significant population ($23\%$) exists  with high X-ray-to-optical flux ratio ($\log\, f_X/f_{opt}>1$) corresponding to high redshift or dust obscuration, and $8\%$  have low X-ray-to-optical flux ratio ($\log\, f_X/f_{opt}\le-1$) that comprise normal galaxies  and low X-ray luminosity sources. 

The $R-K$ colour of the radio-X-ray matches get redder towards fainter $R$ magnitudes, such trend is not present between $R-K$ and the $K$ magnitude. Five radio-X-ray matches fall in the class of Extremely Red Objects, $R-K>5$, with $17<K<19$. The majority (4/5) of the EROs tend to have soft X-ray hardness ratio, but the small sample size of the EROs and incompleteness of the infrared data do not allow for more conclusions.

Comparing the $R-K$ colours of the radio-X-ray matches with evolutionary tracks of various galaxy types and a QSO template as a function of redshift, I found that broad line AGNs, most of which exhibit soft X-ray spectra, have colours consistent with the QSO template prediction, while the extended objects follow the galaxy tracks indicating that these sources have colours dominated by the host galaxy.

No clear correlation found between the hardness ratio and optical/infrared colours. It is noted that the very hard sources tend to have low $R-K$ colour. This is a consequence of the correlation of the hardness ratio with redshift (see lower panel of figure \ref{hr_z-color}) and the fact that low $R-K$ galaxies are mostly at low redshift.  One should note, however, that, in general, sources optically classified as obscured AGN are redder than unobscured AGN and also have higher values of hardness ratio.

 Results form this study will be used to calculate the X-ray luminosity function and determine its 
evolution with redshift in a future study.

\section*{Acknowledgments}
I am very grateful to the referee Prof Simon Morris for the constructive and insightful suggestions. I would like to  thank Dr Matt Jarvis and Prof Chris Willott for kindly supplying the stellar evolution curves. I would like to thank Dr Matt Hilton for providing me with a code to calculate the stellar evolutionary tracks. Thanks are also due to  Prof Fethi Ahmed, Dr Kavilan Moodley and Dr Matt Hilton  for proof reading this manuscript. The funding of my Ph.D by  the South African Square Kilometer Array (SKA)  project is greatly appreciated.

This work makes  use of images data products provided by the NOAO Deep
Wide-Field Survey (Jannuzi and Dey $1999$), which is supported by the National
Optical Astronomy Observatory (NOAO). NOAO is operated by AURA, Inc., under a
cooperative agreement with the National Science  Foundation.

 This work also makes use of data products from
the FLAMEX survey. FLAMEX was designed and constructed by the
infrared instrumentation group (PI: R. Elston) at the University
of Florida, Department of Astronomy, with support from NSF grant
AST97-31180 and Kitt Peak National Observatory.

\begin{landscape}
\begin{table}
\scriptsize
 \caption{Characterisation of Faint Radio Sources}\label{infox}
\begin{center}
\begin{tabular}{lcccccccccccccccccc}
\hline
\hline
 IDs  & Object name  & Ra  &  Dec & $S_{1.4}$ &  flux & sflux &  hflux & HR & $\log\,(f_X/f_{opt})$ &$\log\, (L_X)$ &  $\log\,(L_X)$ & $\log\,\, L_X (0.5-7)$ &   $ \log \,L_{1.4\; GHz}$&R   &  R-K& Class & Photo-Z &  Class (X-ray)\\
 (1)  & (2) &  (3) &  (4) &  (5) &  (6) &  (7) &  (8) &  (9) &  (10) &  (11) &  (12) & (13) & (14) & (15) &  (16) & (17) & (18) & (19) \\
\hline\hline          
 1&J142449.3+340944 &14 : 24 :   49.3 &  34 : 09 :  44.7&  4.53 &  4.18 &  1.64 &  2.87& $-0.41_{-0.07 }  ^{+  0.07 } $&  -0.02  &         42.76  &         43.00  &         43.15  &         24.20  &   19.57 $\pm$     0.00  &    2.87 & 0.86&$  0.535_{-0.381} ^{+ 0.584}$&AGN-1\\
           2&J142456.3+351659 &14 : 24 :   56.5 &  35 : 16 :  59.9&  9.71 &  1.15 &  0.43 &  0.88& $-0.34_{-0.23 }  ^{+  0.21 } $&   1.10  &         42.57  &         42.88  &         43.00  &         24.92  &   23.78 $\pm$     0.09  &  ....... & 0.01&$  0.830_{-0.692} ^{+ 0.953}$&AGN-1\\
           3&J142507.3+323137 &14 : 25 :    7.3 &  32 : 31 :  37.5&  3.28 & 12.60 &  4.75 &  9.46& $-0.35_{-0.02 }  ^{+  0.02 } $&  -0.18  &         43.11  &         43.41  &         43.53  &         23.95  &   17.98 $\pm$     0.00  &  ....... & 1.00& $\bf{0.478\pm0.000}$&AGN-1\\
           4&J142516.5+345246 &14 : 25 :   16.5 &  34 : 52 :  46.6& 63.21 &  1.10 &  0.62 &  0.00& $-1.00_{-0.40 }  ^{+  0.00 } $&   0.00  &         41.48  & .......      &         41.72  &         24.48  &   21.06 $\pm$     0.01  &    3.01 & 0.97&$  0.205_{ +0.000} ^{+ 0.585}$&Gal\\
           5&J142524.2+340936 &14 : 25 :   24.2 &  34 : 09 :  35.7& 15.29 &  1.14 &  0.43 &  0.87& $-0.35_{-0.23 }  ^{+  0.22 } $&  -0.54  &         41.65  &         41.96  &         42.08  &         24.20  &   19.68 $\pm$     0.00  &    2.41 & 0.98&$  0.300_{-0.145} ^{+ 0.575}$&AGN-1\\
           6&J142532.8+330125 &14 : 25 :   32.8 &  33 : 01 :  25.4&  7.33 &  3.65 &  1.21 &  3.47& $-0.18_{-0.08 }  ^{+  0.08 } $&   0.23  &         43.36  &         43.82  &         43.84  &         25.15  &   20.34 $\pm$     0.00  &    2.65 & 0.95 &$\bf{1.201\pm0.001}$ &AGN-2\\
           7&J142533.5+345805 &14 : 25 :   33.5 &  34 : 58 :   5.7&  2.97 &  1.19 &  0.27 &  1.66& $ +0.20_{-0.26 }  ^{+  0.28 } $&   0.13  &         42.38  &         43.18  &         43.04  &         24.43  &   21.30 $\pm$     0.01  &    3.33 & 0.98&$  0.845_{-0.817} ^{+ 0.883}$&AGN-2\\
           8&J142543.9+335534 &14 : 25 :   43.9 &  33 : 55 :  33.2&298.44 &  3.54 &  0.33 &  6.86& $ +0.67_{-0.07 }  ^{+  0.08 } $&   1.12  &         42.65  &         43.97  &         43.68  &         26.61  &   22.61 $\pm$     0.05  &    3.02 & 0.08&$  1.030_{-0.987} ^{+ 1.047}$&AGN-2\\
           9&J142552.6+340239 &14 : 25 :   52.6 &  34 : 02 :  39.5&  1.25 &  3.37 &  1.57 &  1.31& $-0.67_{-0.08 }  ^{+  0.08 } $&  -0.13  &         42.20  &         42.15  &         42.54  &         23.11  &   19.54 $\pm$     0.01  &  ....... & 0.03&$  0.300_{+ 0.000} ^{+ 0.685}$&AGN-1\\
          10&J142601.8+343106 &14 : 26 :    1.9 &  34 : 31 :   7.1&  3.14 &  1.75 &  0.67 &  1.26& $-0.39_{-0.17 }  ^{+  0.16 } $&   0.17  &         43.26  &         43.53  &         43.67  &         24.92  &   20.98 $\pm$     0.01  &    2.75 & 0.98&$  1.415_{-1.295} ^{+ 1.625}$&AGN-1\\
          11&J142607.7+340425 &14 : 26 :    7.7 &  34 : 04 :  26.3& 35.70 & 49.30 & 23.90 & 15.30& $-0.73_{-0.01 }  ^{+  0.01 } $&   0.19  &         44.89  &         44.70  &         45.20  &         26.08  &   17.42 $\pm$     0.00  &    2.90 & 1.00&$  \bf{1.553\pm0.001}$&QSO-1\\
          12&J142611.1+354201 &14 : 26 :   11.0 &  35 : 42 :   1.7&  2.17 &  1.90 &  0.48 &  2.42& $ +0.09_{-0.16 }  ^{+  0.16 } $&   0.13  &         42.52  &         43.20  &         43.11  &         24.18  &   20.80 $\pm$     0.01  &    4.30 & 0.03&$  0.740_{-0.722} ^{+ 0.760}$&AGN-2\\
          13&J142611.1+333932 &14 : 26 :   11.1 &  33 : 39 :  32.7&  1.70 &  3.73 &  1.25 &  3.48& $-0.20_{-0.08 }  ^{+  0.08 } $&   1.44  &         43.08  &         43.53  &         43.57  &         24.23  &   23.35 $\pm$     0.11  &    5.30 & 0.03&$  0.880_{-0.848} ^{+ 0.964}$&AGN-1\\
          14&J142620.3+353707 &14 : 26 :   20.4 &  35 : 37 :   8.4&  6.02 & 23.80 & 10.20 & 12.70& $-0.54_{-0.01 }  ^{+  0.01 } $&  -0.10  &         42.41  &         42.51  &         42.78  &         23.18  &   17.49 $\pm$     0.00  &    3.18 & 0.03&$  0.150_{-0.086} ^{+ 0.392}$&AGN-1\\
          15& J142624.2+334632 &14 : 26 :   24.2 &  33 : 46 :  32.8&  3.44 &  0.97 &  0.33 &  0.88& $-0.22_{-0.28 }  ^{+  0.26 } $&   0.75  &         42.91  &         43.34  &         43.38  &         24.93  &   23.08 $\pm$     0.09  &    4.76 & 0.08&$  1.370_{-1.297} ^{+ 1.662}$&AGN-1\\
          16&J142632.2+350814 &14 : 26 :   32.2 &  35 : 08 :  14.7& 95.31 &  1.35 &  0.65 &  0.42& $-0.73_{-0.22 }  ^{+  0.16 } $&   0.86  &         42.97  &         42.78  &         43.28  &         26.15  &   23.00 $\pm$     0.07  &    4.86 & 0.27&$  1.050_{-0.987} ^{+ 1.156}$&AGN-1\\
          17&J142632.6+334621 &14 : 26 :   32.4 &  33 : 46 :  21.3&  1.67 &  0.76 &  0.10 &  1.33& $ +0.51_{-0.31 }  ^{+  0.39 } $&  -0.09  &         41.97  &         43.08  &         42.85  &         24.20  &   21.26 $\pm$     0.03  &    4.61 & 0.03&$  0.860_{-0.795} ^{+ 0.908}$&AGN-2\\
          18&J142633.6+333211 &14 : 26 :   33.6 &  33 : 32 :  11.8&  4.97 &  3.72 &  1.87 &  0.86& $-0.80_{-0.08 }  ^{+  0.07 } $&  -0.23  &         42.28  &         41.95  &         42.59  &         23.72  &   19.17 $\pm$     0.00  &    1.85 & 0.99&$  0.300_{-0.230} ^{+ 0.464}$&AGN-1\\
          19&J142651.5+351924 &14 : 26 :   51.5 &  35 : 19 :  24.7&  3.96 & 12.90 &  4.72 & 10.30& $-0.31_{-0.05 }  ^{+  0.05 } $&   0.19  &         44.30  &         44.64  &         44.74  &         25.23  &   18.88 $\pm$     0.00  &    1.95 & 0.83&$  \bf{1.758\pm0.002}$&QSO-1\\
          20&J142659.6+341158 &14 : 26 :   59.7 &  34 : 12 :   0.2&  9.71 &  3.38 &  1.91 &  0.00& $-1.00_{+0.00 }  ^{+0.00 } $&  -1.89  &         41.28  & .......      &         41.53  &         22.99  &   15.14 $\pm$     0.00  &    2.81 & 0.03&$  0.095_{-0.076} ^{+ 0.154}$&Gal\\
          21&J142708.6+331508 &14 : 27 :    8.7 &  33 : 15 :   8.9& 10.42 &  3.08 &  1.53 &  0.79& $-0.78_{-0.08 }  ^{+  0.07 } $&  31.62  &         40.85  &         40.57  &         41.15  &         22.68  &   ..............  &   ....... & 0.03&$  0.065_{ +0.000} ^{+ 7.000}$&Gal\\
          22& J142731.4+324435 &14 : 27 :   31.4 &  32 : 44 :  34.4&  2.25 &  1.06 &  0.10 &  2.07& $ +0.68_{-0.19 }  ^{+  0.25 } $&   0.77  &         41.99  &         43.30  &         43.00  &         24.34  &   23.04 $\pm$     0.05  &  ....... & 0.02&$  0.875_{-0.835} ^{+ 0.932}$&AGN-2\\
          23&J142735.5+341927 &14 : 27 :   35.6 &  34 : 19 :  28.1&  2.59 &  2.73 &  1.02 &  2.11& $-0.34_{-0.23 }  ^{+  0.21 } $&   1.06  &         42.75  &         43.04  &         43.18  &         24.15  &   22.73 $\pm$     0.13  &  ....... & 0.00&$  0.665_{-0.607} ^{+ 0.728}$&AGN-1\\
          24&J142741.4+325433 &14 : 27 :   41.6 &  32 : 54 :  32.1&  1.62 &  0.87 &  0.00 &  2.09& $ +1.10_{+0.00 }  ^{+  0.65 } $&   1.07  & .......      &         43.28  &         42.90  &         24.18  &   23.99 $\pm$     0.17  &  ....... & 0.04&$  0.855_{-0.770} ^{+ 1.051}$&AGN-2\\
          25&J142744.4+333828 &14 : 27 :   44.4 &  33 : 38 :  28.6& 16.24 & 11.50 &  4.96 &  6.12& $-0.54_{-0.03 }  ^{+  0.03 } $&   0.11  &         44.00  &         44.08  &         44.36  &         25.51  &   18.79 $\pm$     0.00  &    2.13 & 0.97& $ \bf{1.231\pm0.001}$&QSO-1\\
          26&J142750.4+344100 &14 : 27 :   50.3 &  34 : 41 :   0.8&  1.62 &  0.77 &  0.43 &  0.00& $-1.00_{ +0.00 }  ^{+  0.00 } $&  -0.34  &         42.43  & .......      &         42.68  &         24.00  &   20.60 $\pm$     0.03  &    3.31 & 0.83&$  0.710_{-0.693} ^{+ 0.743}$&AGN-1\\
          27&J142752.2+353031 &14 : 27 :   52.2 &  35 : 30 :  31.8&  2.28 &  1.33 &  0.32 &  1.77& $ +0.14_{-0.19 }  ^{+  0.19 } $&  -0.23  &         41.52  &         42.26  &         42.15  &         23.38  &   20.29 $\pm$     0.01  &    2.20 & 0.95&$  0.300_{-0.079} ^{+ 0.511}$&AGN-2\\
          28&J142810.3+353846 &14 : 28 :   10.3 &  35 : 38 :  47.0&  1.01 & 17.80 &  6.74 & 13.40& $-0.35_{-0.02 }  ^{+  0.02 } $&   0.19  &         43.74  &         44.04  &         44.15  &         23.91  &   18.52 $\pm$     0.00  &    2.10 & 0.97&$  \bf{0.804\pm0.001}$&QSO-1\\
          29&J142850.2+323248 &14 : 28 :   50.3 &  32 : 32 :  48.8&  1.46 &  1.27 &  0.20 &  2.12& $ +0.43_{-0.18 }  ^{+  0.20 } $&  -1.89  &         40.56  &         41.59  &         41.36  &         22.43  &   16.20 $\pm$     0.00  &  ....... & 0.03&$  \bf{0.128\pm0.000}$&AGN-2\\
          30&J142852.7+325026 &14 : 28 :   52.7 &  32 : 50 :  27.2&  3.07 &  3.46 &  0.68 &  5.21& $ +0.29_{-0.08 }  ^{+  0.08 } $&  -0.42  &         42.20  &         43.08  &         42.91  &         23.86  &   18.78 $\pm$     0.00  &  ....... & 0.03&$  \bf{0.446\pm0.000}$&AGN-2\\
          31&J142905.1+342640 &14 : 29 :    5.1 &  34 : 26 :  41.1&260.41 &  1.69 &  0.47 &  1.96& $ +0.00_{-0.17 }  ^{+  0.17 } $&   1.37  &         42.81  &         43.43  &         43.36  &         26.56  &   24.02 $\pm$     0.16  &  ....... & 0.51&$  1.030_{-0.997} ^{+ 1.046}$&AGN-2\\
          32&J142910.2+352946 &14 : 29 :   10.2 &  35 : 29 :  46.9& 17.81 &  1.89 &  0.59 &  1.94& $-0.12_{-0.15 }  ^{+  0.15 } $&  -0.69  &         43.61  &         44.11  &         44.11  &         26.08  &   18.76 $\pm$     0.00  &    2.26 & 0.98&$ \bf{2.224\pm0.001}$&QSO-2\\
          33&J142911.1+350320 &14 : 29 :   11.2 &  35 : 03 :  20.7&  2.37 &  8.22 &  3.94 &  2.71& $-0.72_{-0.04 }  ^{+  0.04 } $&   0.49  &         43.51  &         43.34  &         43.83  &         24.30  &   20.11 $\pm$     0.02  &    2.87 & 1.00&$  0.810_{-0.799} ^{+ 0.822}$&AGN-1\\
          34&J142916.1+335536 &14 : 29 :   16.1 &  33 : 55 :  38.8&  2.07 &  3.37 &  1.26 &  2.60& $-0.34_{-0.23 }  ^{+  0.21 } $&   1.05  &         43.23  &         43.56  &         43.67  &         24.46  &   22.48 $\pm$     0.04  &    5.04 & 0.04&$  1.035_{-1.024} ^{+ 1.102}$&AGN-1\\
          35&J142917.4+332626 &14 : 29 :   17.4 &  33 : 26 :  26.4&  6.79 &  1.45 &  0.12 &  2.88& $ +0.72_{-0.16 }  ^{+  0.21 } $&   1.37  &         42.15  &         43.53  &         43.23  &         24.90  &   24.20 $\pm$     0.12  &  ....... & 0.11&$  0.960_{-0.806} ^{+ 1.227}$&AGN-2\\
          36&J142922.8+351218 &14 : 29 :   22.8 &  35 : 12 :  19.7&  1.48 &  1.52 &  0.74 &  0.43& $-0.76_{-0.22 }  ^{+  0.17 } $&  31.31  &         42.90  &         42.66  &         43.20  &         24.20  &   ..............       &    0.00 & 0.50&$  0.920_{-0.743} ^{+ 1.102}$&AGN-1\\
          37&J142942.6+335654 &14 : 29 :   42.6 &  33 : 56 :  55.0&  3.64 & 11.30 &  4.55 &  7.43& $-0.44_{-0.03 }  ^{+  0.03 } $&   0.02  &         43.87  &         44.08  &         44.26  &         24.78  &   18.58 $\pm$     0.00  &    2.07 & 0.98&$  \bf{1.122\pm0.0001}$&QSO-1\\
          38&J142955.6+353707 &14 : 29 :   55.7 &  35 : 37 :   7.6& 11.78 &  2.38 &  1.12 &  0.88& $-0.69_{-0.14 }  ^{+  0.12 } $&   0.88  &         44.08  &         43.98  &         44.41  &         26.11  &   22.44 $\pm$     0.04  &  ....... & 0.35&$  2.760_{-2.424} ^{+ 3.001}$&QSO-1\\
          39&J143003.1+340649 &14 : 30 :    3.1 &  34 : 06 :  51.3&  4.43 &  1.27 &  0.12 &  2.47& $ +0.67_{-0.19 }  ^{+  0.25 } $&  -1.70  &         40.32  &         41.64  &         41.34  &         22.89  &   16.67 $\pm$     0.00  &    2.66 & 0.31&$  \bf{0.126\pm0.000}$&AGN-2\\
          40&J143008.8+344713 &14 : 30 :    8.9 &  34 : 47 :  13.8& 27.38 &  0.78 &  0.33 &  0.42& $-0.54_{-0.37 }  ^{+  0.29 } $&   0.27  &         43.76  &         43.87  &         44.15  &         26.68  &   22.13 $\pm$     0.04  &  ....... & 0.80&$  3.455_{-3.184} ^{+ 3.552}$&QSO-1\\
          41& J143011.5+350019 &14 : 30 :   11.5 &  35 : 00 :  19.7&  4.66 &  2.76 &  1.11 &  1.79& $-0.44_{-0.10 }  ^{+  0.10 } $&  -0.29  &         42.80  &         43.00  &         43.20  &         24.43  &   19.34 $\pm$     0.00  &    2.01 & 0.98&$  0.680_{-0.601} ^{+ 0.781}$&AGN-1\\
          42&J143103.6+334541 &14 : 31 :    3.4 &  33 : 45 :  41.6&  3.62 &  1.03 &  0.47 &  0.44& $-0.63_{-0.30 }  ^{+  0.23 } $&  -1.77  &         41.48  &         41.45  &         41.83  &         23.36  &   16.72 $\pm$     0.00  &    3.33 & 0.03&$  \bf{0.238\pm0.000}$&Gal\\
          43&J143112.4+353527 &14 : 31 :   12.5 &  35 : 35 :  27.3&  9.64 &  1.82 &  0.45 &  2.34& $ +0.11_{-0.15 }  ^{+  0.15 } $&   1.03  &         43.08  &         43.79  &         43.68  &         25.40  &   23.10 $\pm$     0.05  &    6.09 & 0.27&$  1.405_{-1.044} ^{+ 1.447}$&AGN-2\\
          44&J143114.4+323225 &14 : 31 :   14.3 &  32 : 32 :  25.6&170.73 &  3.60 &  1.52 &  2.07& $-0.51_{-0.07 }  ^{+  0.07 } $&  -1.10  &         42.43  &         42.57  &         42.81  &         25.49  &   17.03 $\pm$     0.00  &  ....... & 0.03&$  0.390_{-0.344} ^{+ 0.481}$&AGN-1\\
          45& J143121.8+344046 &14 : 31 :   21.9 &  34 : 40 :  46.6&  1.40 & 19.00 &  3.97 & 27.80& $ +0.25_{-0.03 }  ^{+  0.03 } $&   0.42  &         42.45  &         43.30  &         43.15  &         23.00  &   19.03 $\pm$     0.00  &    3.55 & 0.03&$  0.250_{-0.156} ^{+ 0.319}$&AGN-2\\
          46&J143125.0+331350 &14 : 31 :   25.3 &  33 : 13 :  48.3&  6.47 &  1.03 &  0.36 &  0.90& $-0.26_{-0.30 }  ^{+  0.28 } $&  31.14  &         39.32  &         39.72  &         39.77  &         21.57  &   ..............  &   ....... & 0.04&$  \bf{0.023\pm0.000}$&Gal\\
47&J143127.7+343740 &14 : 31 :   27.9 &  34 : 37 :  40.9&  2.83 &  1.39 &  0.56 &  0.91& $-0.43_{-0.20 }  ^{+  0.18 } $&   0.32  &         43.79  &         44.00  &         44.18  &         25.49  &   21.61 $\pm$     0.02  &  ....... & 0.98&$  2.755_{-2.353} ^{+ 2.902}$&QSO-1\\
          48&J143131.1+342721 &14 : 31 :   31.1 &  34 : 27 :  22.1&  1.19 &  2.69 &  0.25 &  5.23& $ +0.67_{-0.11 }  ^{+  0.12 } $&  -1.02  &         41.74  &         43.04  &         42.77  &         23.41  &   17.56 $\pm$     0.00  &    2.91 & 0.03&$  0.430_{-0.417} ^{+ 0.441}$&AGN-2\\
          
49&J143152.3+323213 &14 : 31 :   52.3 &  32 : 32 :  13.4&  4.83 &  7.96 &  2.94 &  6.29& $-0.32_{-0.03 }  ^{+  0.03 } $&  -0.07  &         42.49  &         42.82  &         42.92  &         23.70  &   18.74 $\pm$     0.00  &  ....... & 0.97&$  0.300_{-0.263} ^{+ 0.355}$&AGN-1\\
          50&J143156.4+325137 &14 : 31 :   56.4 &  32 : 51 :  37.8&  0.71 &  1.21 &  0.00 &  2.82& $ +1.00_{ +0.00 }  ^{+  0.35 } $&  -0.78  & .......      &         42.77  &         42.40  &         23.18  &   19.01 $\pm$     0.00  &  ....... & 0.40&$  \bf{0.420\pm0.001}$&AGN-2\\
          51&J143157.9+341649 &14 : 31 :   57.9 &  34 : 16 :  50.3&  0.65 & 42.90 & 18.00 & 25.10& $-0.50_{-0.01 }  ^{+  0.01 } $&  -0.07  &         44.04  &         44.20  &         44.43  &         23.61  &   16.91 $\pm$     0.00  &    1.80 & 0.99&$  \bf{0.715\pm0.001}$&QSO-1\\

\hline	
\hline
\end{tabular}
\end{center}
\end{table}
\end{landscape}

\begin{landscape}
\addtocounter{table}{-1}
\begin{table}
\scriptsize
\caption{\textit{Continued...}}
\begin{center}
\begin{tabular}{ccccccccccccccccccl}
\hline
\hline
     IDs  & Object name  & Ra  &  Dec & $S_{1.4}$ &  flux & sflux &  hflux & HR & $\log\,(f_X/f_{opt})$ &$\log\, L_X$ &  $\log\,L_X$ & $\log\,\, L_X $ &   $ \log \,L_{1.4\; GHz}$&\textit{R}   &  \textit{R-K}& Class & Photo-Z &  Class (X-ray)\\
\cline{6-8}
 &    &   &   &   (mJy) & \multicolumn{3}{c}{$\times 10^{-15}~{\rm (erg s^{-1} m^{-2})}$} &     &  &   (0.5-2) keV & (2-7) Kev & (0.5-7) Kev  & (W ${\rm Hz^{-1}}$) &  &  &  &  &  \\
 (1)  & (2) &  (3) &  (4) &  (5) &  (6) &  (7) &  (8) &  (9) &  (10) &  (11) &  (12) & (13) & (14) & (15) &  (16) & (17) & (18) & (19) \\
\hline\hline      
          52&J143214.3+342605 &14 : 32 :   14.3 &  34 : 26 :   5.7& 17.75 &  6.09 &  2.75 &  2.74& $-0.61_{-0.04 }  ^{+  0.04 } $&   0.33  &         42.72  &         42.72  &         43.08  &         24.53  &   20.05 $\pm$     0.01  &    3.46 & 0.03&$  0.405_{-0.345} ^{+ 0.505}$&AGN-1\\
          53&J143308.1+353151 &14 : 33 :    8.1 &  35 : 31 :  51.4&  2.33 &  1.55 &  0.67 &  0.79& $-0.56_{-0.16 }  ^{+  0.14 } $&   0.06  &         42.56  &         42.62  &         42.92  &         24.08  &   20.85 $\pm$     0.01  &    3.83 & 0.03&$  0.660_{-0.582} ^{+ 0.675}$&AGN-1\\
          54&J143311.0+335828 &14 : 33 :   11.1 &  33 : 58 :  28.8& 22.72 &  2.74 &  1.32 &  0.87& $-0.73_{-0.10 }  ^{+  0.09 } $&  -0.43  &         42.87  &         42.69  &         43.18  &         25.11  &   19.00 $\pm$     0.00  &    3.63 & 0.03&$  0.675_{-0.647} ^{+ 0.685}$&AGN-1\\
          55& J143315.7+332858 &14 : 33 :   15.7 &  33 : 28 :  58.4&  5.81 &  3.02 &  1.27 &  1.75& $-0.50_{-0.09 }  ^{+  0.08 } $&   1.06  &         43.49  &         43.63  &         43.87  &         25.15  &   22.62 $\pm$     0.03  &    4.53 & 0.52&$  1.360_{-1.297} ^{+ 1.548}$&AGN-1\\
          56&J143318.5+344404 &14 : 33 :   18.5 &  34 : 44 :   4.3&  2.69 & 16.10 &  6.51 & 10.40& $-0.44_{-0.02 }  ^{+  0.02 } $&  32.34  &         40.91  &         41.11  &         41.30  &         21.53  &   ..............   &   ....... & 0.03&$  \bf{0.034\pm0.001}$&Gal\\
          57&J143355.1+340933 &14 : 33 :   55.2 &  34 : 09 :  34.3&  3.06 &  1.43 &  0.69 &  0.46& $-0.73_{-0.22 }  ^{+  0.16 } $&   0.06  &         41.15  &         40.97  &         41.46  &         22.79  &   20.94 $\pm$     0.01  &  ....... & 0.98&$  0.135_{+0.000} ^{+ 0.236}$&AGN-1\\
          58&J143413.9+345506 &14 : 34 :   13.8 &  34 : 55 :   6.2&  2.60 &  0.98 &  0.44 &  0.44& $-0.61_{-0.30 }  ^{+  0.23 } $&  -0.31  &         42.32  &         42.32  &         42.66  &         24.08  &   20.42 $\pm$     0.01  &    3.54 & 0.03&$  0.620_{-0.592} ^{+ 0.640}$&AGN-1\\
          59&J143423.3+342610 &14 : 34 :   23.4 &  34 : 26 :   9.9&  3.78 &  1.62 &  0.46 &  1.87& $ +0.00_{-0.17 }  ^{+  0.17 } $&   0.99  &         42.76  &         43.38  &         43.32  &         24.68  &   23.13 $\pm$     0.06  &  ....... & 0.28&$  0.995_{-0.870} ^{+ 1.039}$&AGN-2\\
          60&J143428.0+331102 &14 : 34 :   28.0 &  33 : 11 :   2.0& 23.23 &  1.26 &  0.48 &  0.92& $-0.37_{-0.24 }  ^{+  0.22 } $&   1.30  &         43.32  &         43.59  &         43.73  &         26.00  &   24.18 $\pm$     0.11  &    5.30 & 0.22&$  1.770_{-1.753} ^{+ 1.802}$&AGN-1\\
          61&J143434.2+351009 &14 : 34 :   34.2 &  35 : 10 :   9.5& 67.51 &  7.06 &  2.71 &  5.15& $-0.37_{-0.04 }  ^{+  0.04 } $&   0.34  &         42.94  &         43.23  &         43.36  &         25.34  &   19.90 $\pm$     0.00  &    3.92 & 0.98&$  0.520_{-0.449} ^{+ 0.543}$&AGN-1\\
          62& J143445.3+332820 &14 : 34 :   45.3 &  33 : 28 :  20.6& 30.03 & 52.00 & 18.90 & 42.60& $-0.29_{ +0.00 }  ^{+  0.00 } $&   0.01  &         42.92  &         43.28  &         43.36  &         24.11  &   16.90 $\pm$     0.00  &    3.31 & 0.01&$  \bf{0.198\pm0.001}$&AGN-1\\
          63& J143513.2+333117 &14 : 35 :   13.3 &  33 : 31 :  18.3&  2.15 &  2.02 &  0.92 &  0.89& $-0.62_{-0.15 }  ^{+  0.13 } $&   0.58  &         42.82  &         42.81  &         43.18  &         24.20  &   21.86 $\pm$     0.04  &    4.07 & 0.04&$  0.760_{-0.747} ^{+ 0.782}$&AGN-1\\
          64& J143527.3+331240 &14 : 35 :   27.3 &  33 : 12 :  40.1&  1.30 &  2.09 &  0.96 &  0.88& $-0.64_{-0.13 }  ^{+  0.11 } $&  -0.33  &         42.71  &         42.66  &         43.04  &         23.84  &   19.55 $\pm$     0.00  &    2.14 & 0.98&$  0.655_{-0.598} ^{+ 0.701}$&AGN-1\\
          65&J143527.9+350928 &14 : 35 :   27.9 &  35 : 09 :  28.3&  1.66 &  3.92 &  1.51 &  2.83& $-0.38_{-0.07 }  ^{+  0.07 } $&  -0.35  &         44.04  &         44.30  &         44.45  &         25.08  &   18.82 $\pm$     0.00  &    2.08 & 0.99&$  \bf{2.247\pm0.002}$&QSO-1\\
          66&J143528.0+331145 &14 : 35 :   28.0 &  33 : 11 :  45.5& 39.60 &  1.04 &  0.47 &  0.48& $-0.61_{-0.29 }  ^{+  0.23 } $&   1.19  &         43.26  &         43.28  &         43.61  &         26.20  &   24.10 $\pm$     0.15  &    6.65 & 0.09&$  1.700_{-1.595} ^{+ 1.802}$&AGN-1\\
          67&J143528.4+331931 &14 : 35 :   28.4 &  33 : 19 :  31.5& 23.27 &  6.68 &  3.21 &  2.21& $-0.72_{-0.04 }  ^{+  0.04 } $&  -0.11  &         42.90  &         42.74  &         43.20  &         24.76  &   18.83 $\pm$     0.00  &    1.85 & 0.99&$  0.455_{-0.234} ^{+ 0.529}$&AGN-1\\
          68&J143541.6+345055 &14 : 35 :   41.6 &  34 : 50 :  55.3&  1.46 &  1.50 &  0.42 &  1.74& $ +0.00_{-0.17 }  ^{+  0.17 } $&   0.48  &         42.52  &         43.15  &         43.08  &         24.08  &   21.93 $\pm$     0.03  &  ....... & 0.08&$  0.795_{-0.781} ^{+ 0.816}$&AGN-2\\
          69& J143547.6+335310 &14 : 35 :   47.6 &  33 : 53 :   9.9&  2.22 &  1.07 &  0.49 &  0.44& $-0.65_{-0.31 }  ^{+  0.24 } $&  -0.44  &         43.48  &         43.43  &         43.83  &         25.15  &   20.01 $\pm$     0.00  &    2.30 & 0.98&$  \bf{2.112\pm0.001}$&AGN-1\\
          70&J143608.7+350613 &14 : 36 :    8.7 &  35 : 06 :  13.8&  0.68 & 46.20 & 19.20 & 27.40& $-0.49_{-0.01 }  ^{+  0.01 } $&   1.56  &         44.18  &         44.34  &         44.57  &         23.73  &   20.91 $\pm$     0.01  &    3.62 & 0.79&$  0.795_{-0.779} ^{+ 0.819}$&QSO-1\\
          71& J143623.4+352710 &14 : 36 :   23.4 &  35 : 27 :  10.8& 17.48 &  1.51 &  0.63 &  0.88& $-0.50_{-0.18 }  ^{+  0.16 } $&  -1.58  &         41.08  &         41.20  &         41.45  &         23.51  &   16.77 $\pm$     0.00  &  ....... & 0.03&$  \bf{0.129\pm0.000}$&Gal\\
          72&J143644.2+350626 &14 : 36 :   44.2 &  35 : 06 :  27.1&  4.81 &  1.32 &  0.21 &  2.19& $ +0.43_{-0.18 }  ^{+  0.20 } $&   1.13  &         43.40  &         44.41  &         44.18  &         25.75  &   23.69 $\pm$     0.09  &  ....... & 0.00&$  2.850_{-1.916} ^{+ 3.284}$&QSO-2\\
          73&J143651.3+333909 &14 : 36 :   51.3 &  33 : 39 :  10.2&  1.88 &  1.67 &  0.59 &  1.44& $-0.26_{-0.17 }  ^{+  0.17 } $&  -0.21  &         40.94  &         41.32  &         41.40  &         22.45  &   20.10 $\pm$     0.00  &    1.75 & 0.99&$  0.115_{ +0.000} ^{+ 0.143}$&Gal\\
          74&J143713.5+350553 &14 : 37 :   13.6 &  35 : 05 :  54.8& 59.62 &  6.18 &  2.45 &  4.19& $-0.42_{-0.05 }  ^{+  0.05 } $&   0.05  &         42.87  &         43.11  &         43.28  &         25.26  &   19.33 $\pm$     0.00  &    2.08 & 0.99&$  0.500_{-0.206} ^{+ 0.557}$&AGN-1\\
          75&J143715.3+340652 &14 : 37 :   15.3 &  34 : 06 :  52.5&  4.48 &  5.55 &  2.73 &  1.56& $-0.76_{-0.04 }  ^{+  0.04 } $&  -0.05  &         43.15  &         42.91  &         43.46  &         24.36  &   19.19 $\pm$     0.00  &    3.38 & 0.06&$  0.650_{-0.634} ^{+ 0.660}$&AGN-1\\
          76&J143752.0+351939 &14 : 37 :   52.0 &  35 : 19 :  40.1& 11.48 &  1.82 &  0.69 &  1.38& $-0.35_{-0.16 }  ^{+  0.15 } $&  -0.51  &         42.23  &         42.53  &         42.65  &         24.45  &   19.25 $\pm$     0.00  &  ....... & 0.03&$  0.455_{-0.443} ^{+ 0.469}$&AGN-1\\
          77&J143756.4+351936 &14 : 37 :   56.5 &  35 : 19 :  36.5& 53.48 & 32.80 & 13.60 & 19.70& $-0.48_{-0.01 }  ^{+  0.01 } $&   0.27  &         43.67  &         43.84  &         44.04  &         25.28  &   18.07 $\pm$     0.00  &  ....... & 0.74&$  \bf{0.537\pm0.01}$ &QSO-1\\
          78&J143821.8+344000 &14 : 38 :   21.8 &  34 : 40 :   1.1& 12.28 &  0.99 &  0.44 &  0.46& $-0.60_{-0.29 }  ^{+  0.23 } $&  -0.29  &         42.30  &         42.32  &         42.64  &         24.74  &   20.46 $\pm$     0.00  &  ....... & 0.98&$  0.605_{-0.499} ^{+ 0.727}$&AGN-1\\
          79&J143830.2+353915 &14 : 38 :   30.2 &  35 : 39 :  15.9&  3.59 & 55.50 & 22.00 & 37.60& $-0.42_{-0.01 }  ^{+  0.01 } $&   0.37  &         43.23  &         43.48  &         43.64  &         23.45  &   17.73 $\pm$     0.00  &  ....... & 0.03&$  \bf{0.262\pm0.001}$&AGN-1\\
\hline
\hline

\end{tabular}
\end{center}

\end{table}
\end{landscape}

\begin{figure*}
  \begin{center}
    \begin{tabular}{c} 
   \includegraphics[height= 75mm, width=130mm]{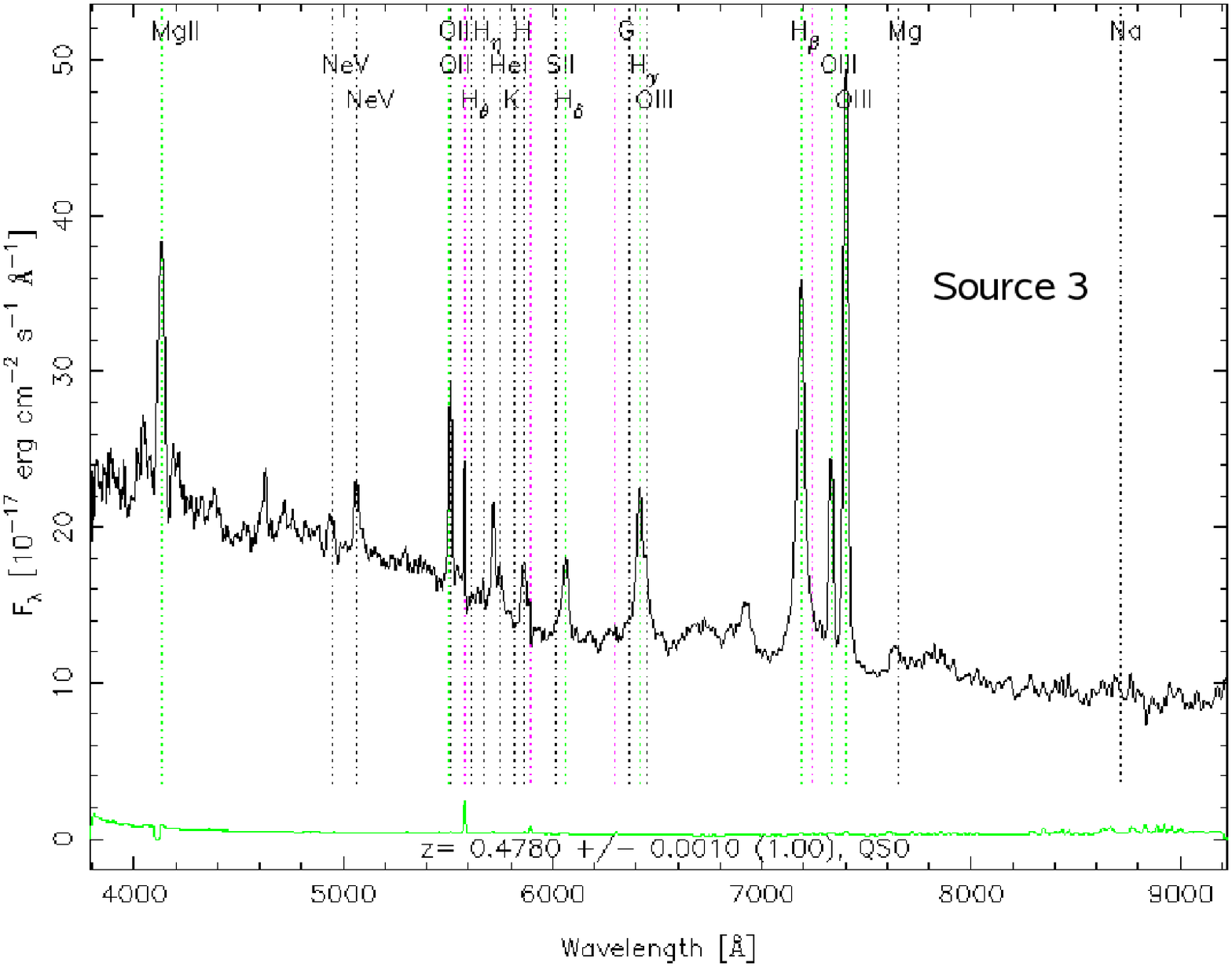} \\
  \includegraphics[height= 75mm, width=130mm]{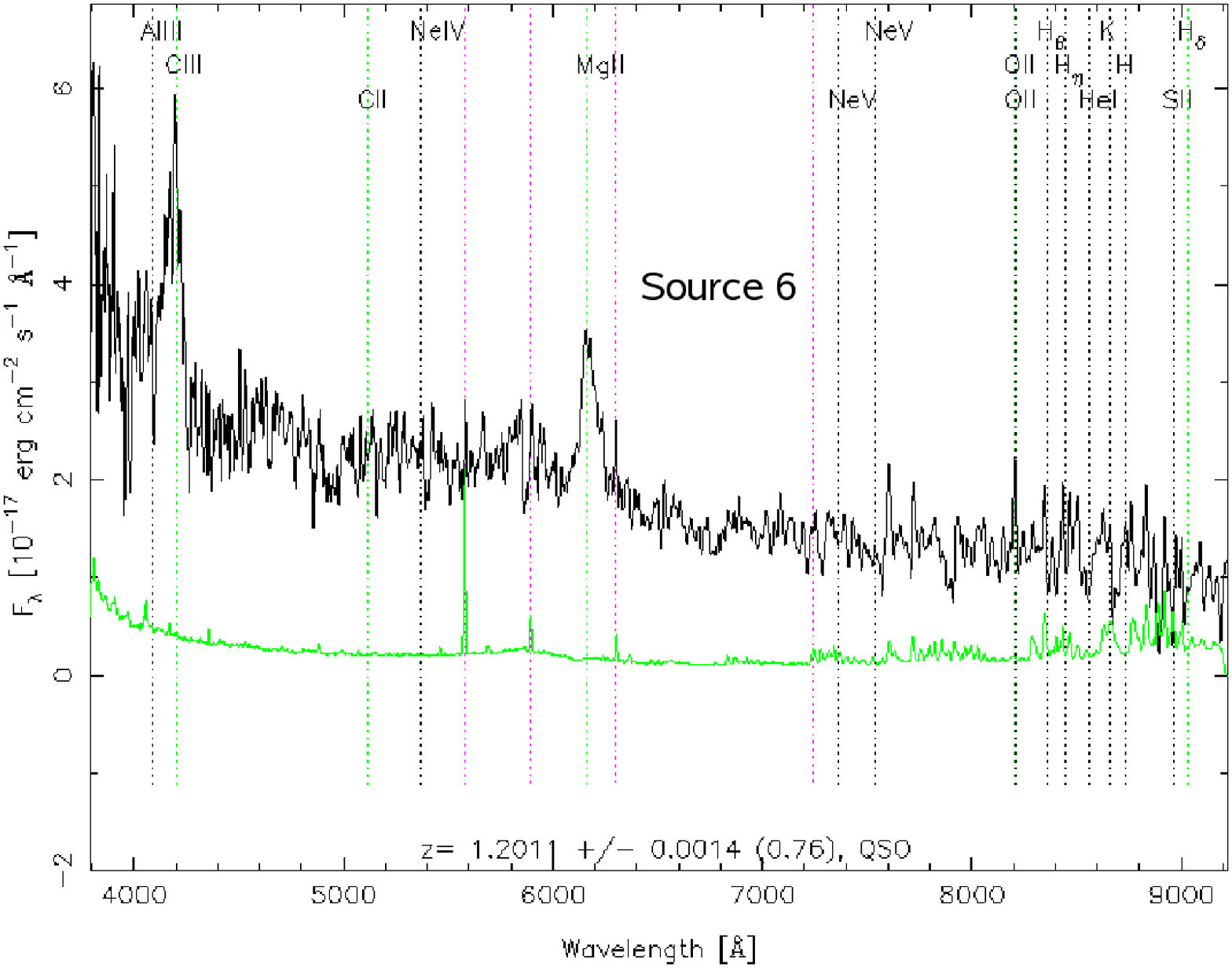}\\
  \includegraphics[height= 75mm, width=130mm]{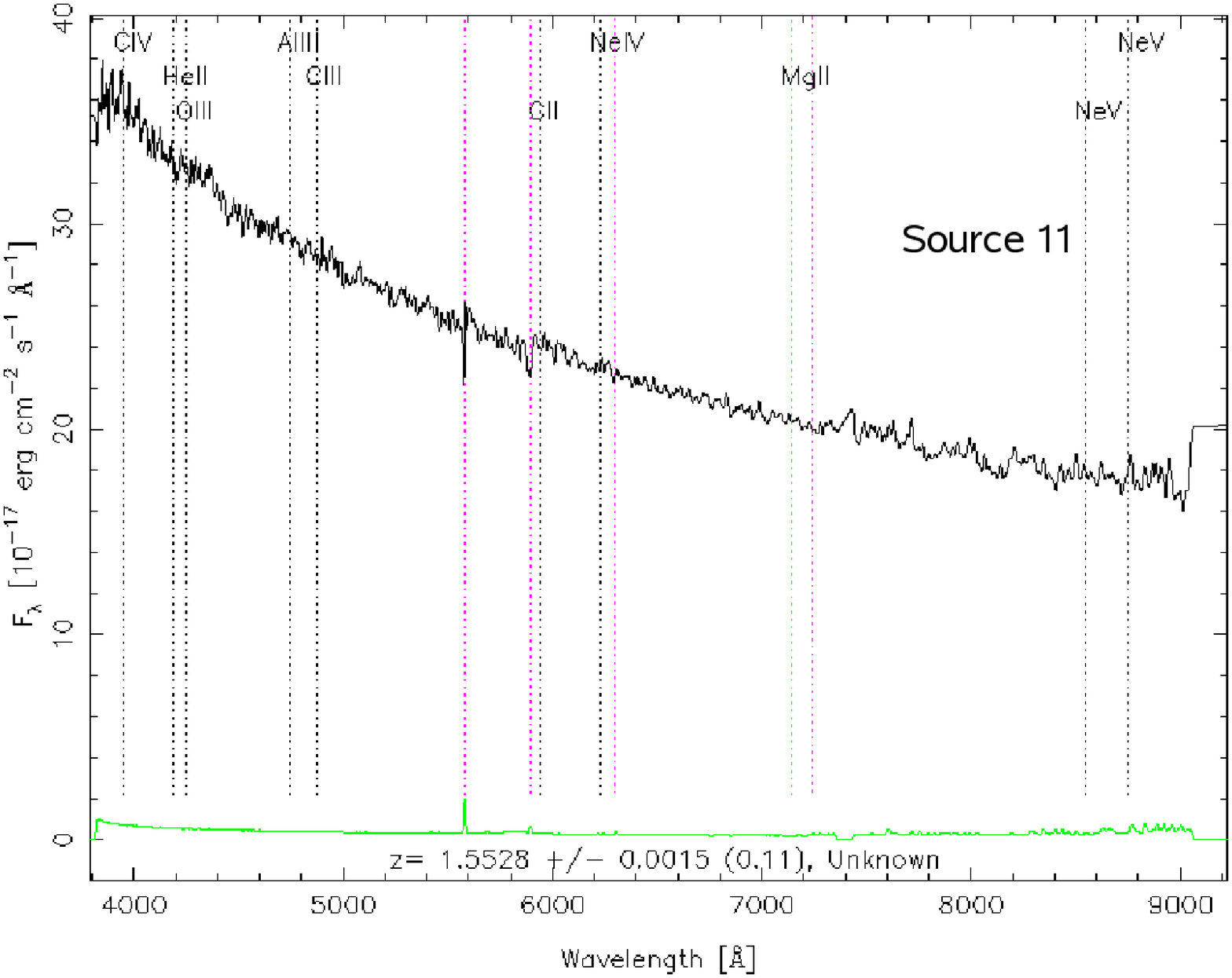} \\
\end{tabular}
  \end{center}
\caption{Optical spectra of 22 radio-X-ray matches obtained from the SDSS
  survey. The spectrum of each object is shown with the identification label,
  spectroscopic redshift and classification.}
\label{sdss-bootes}
\end{figure*}%

\begin{figure*}
  \begin{center}
    \begin{tabular}{c} 
 \includegraphics[height= 75mm, width=130mm]{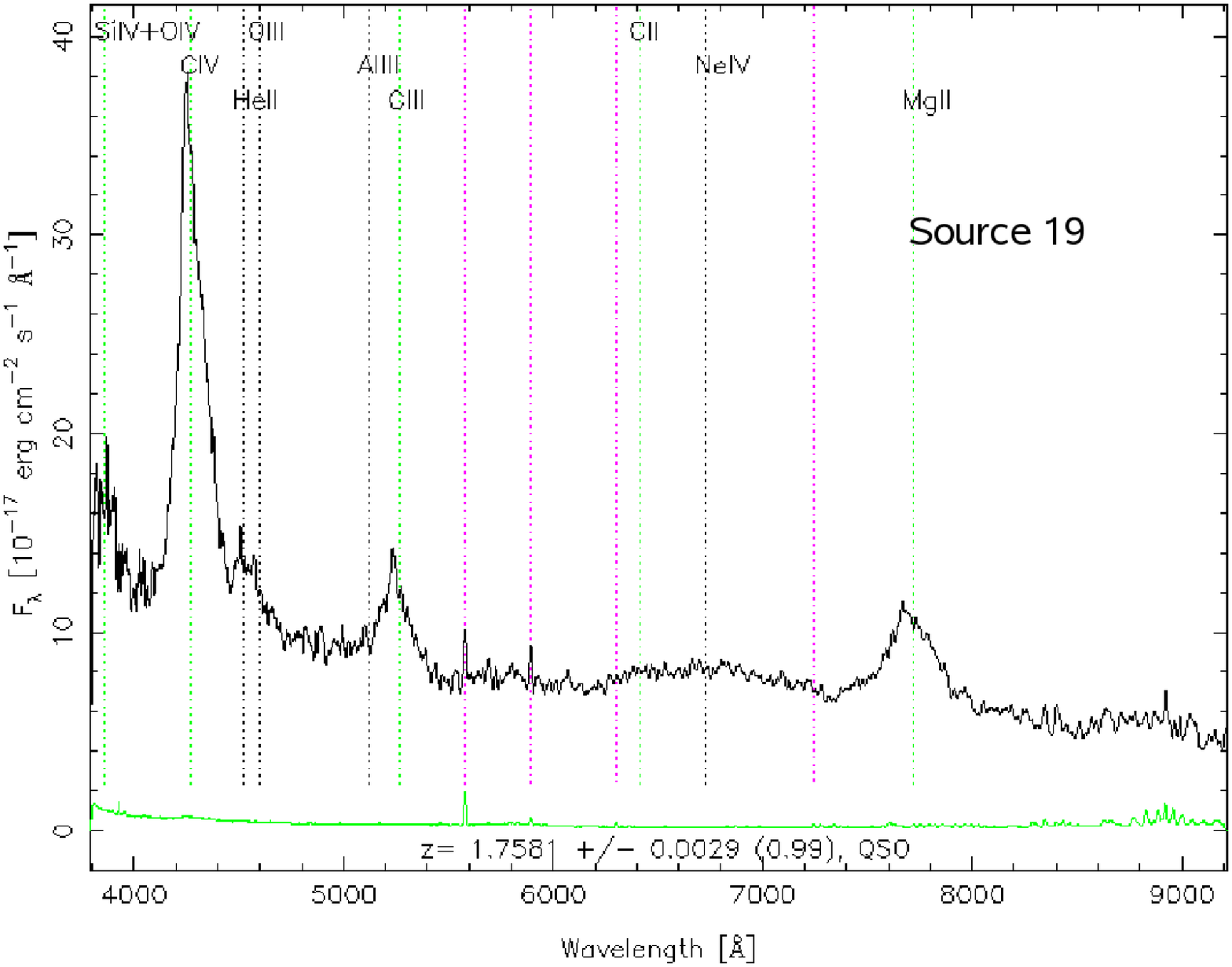} \\
    \includegraphics[height= 75mm, width=130mm]{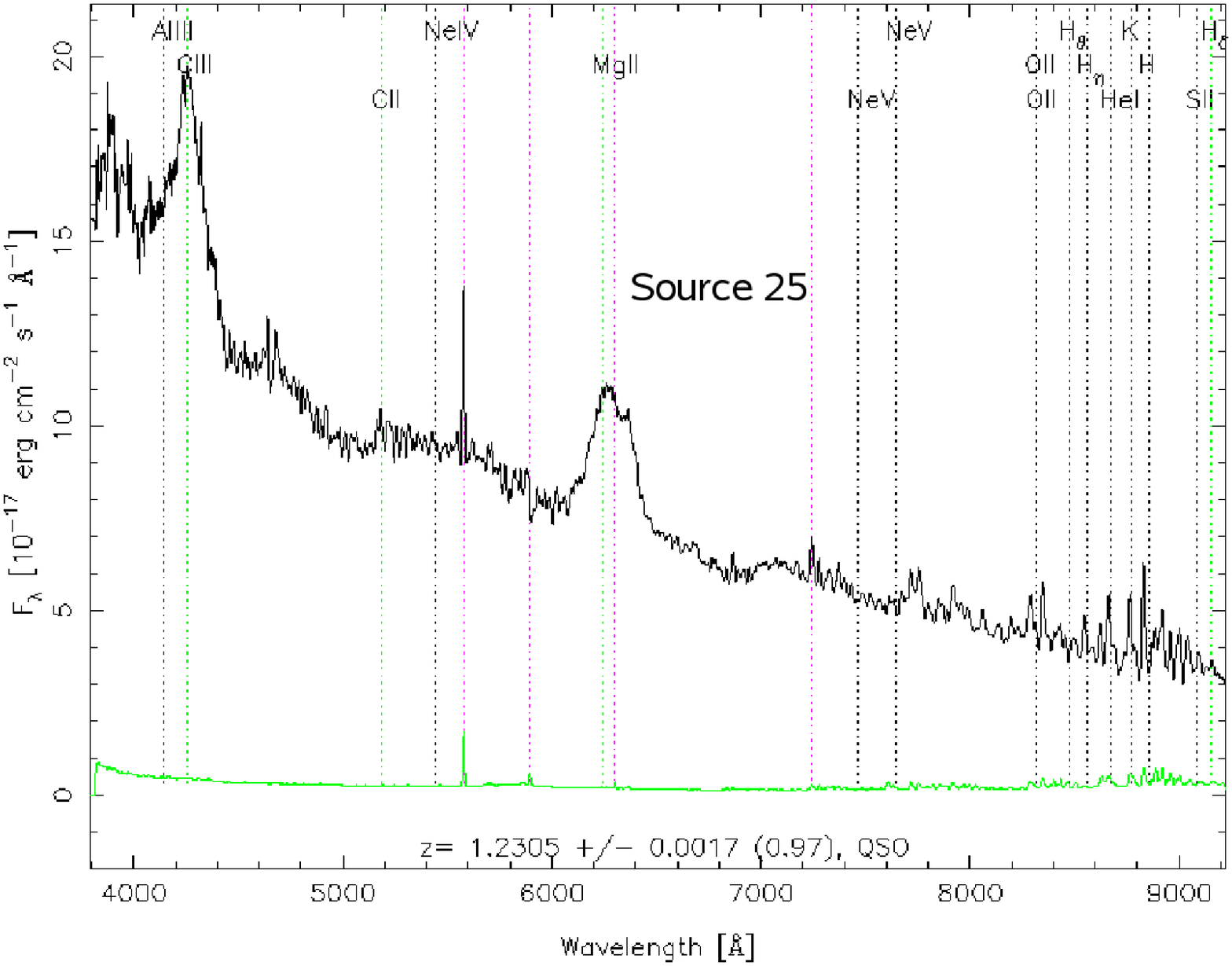} \\
  \includegraphics[height= 75mm, width=130mm]{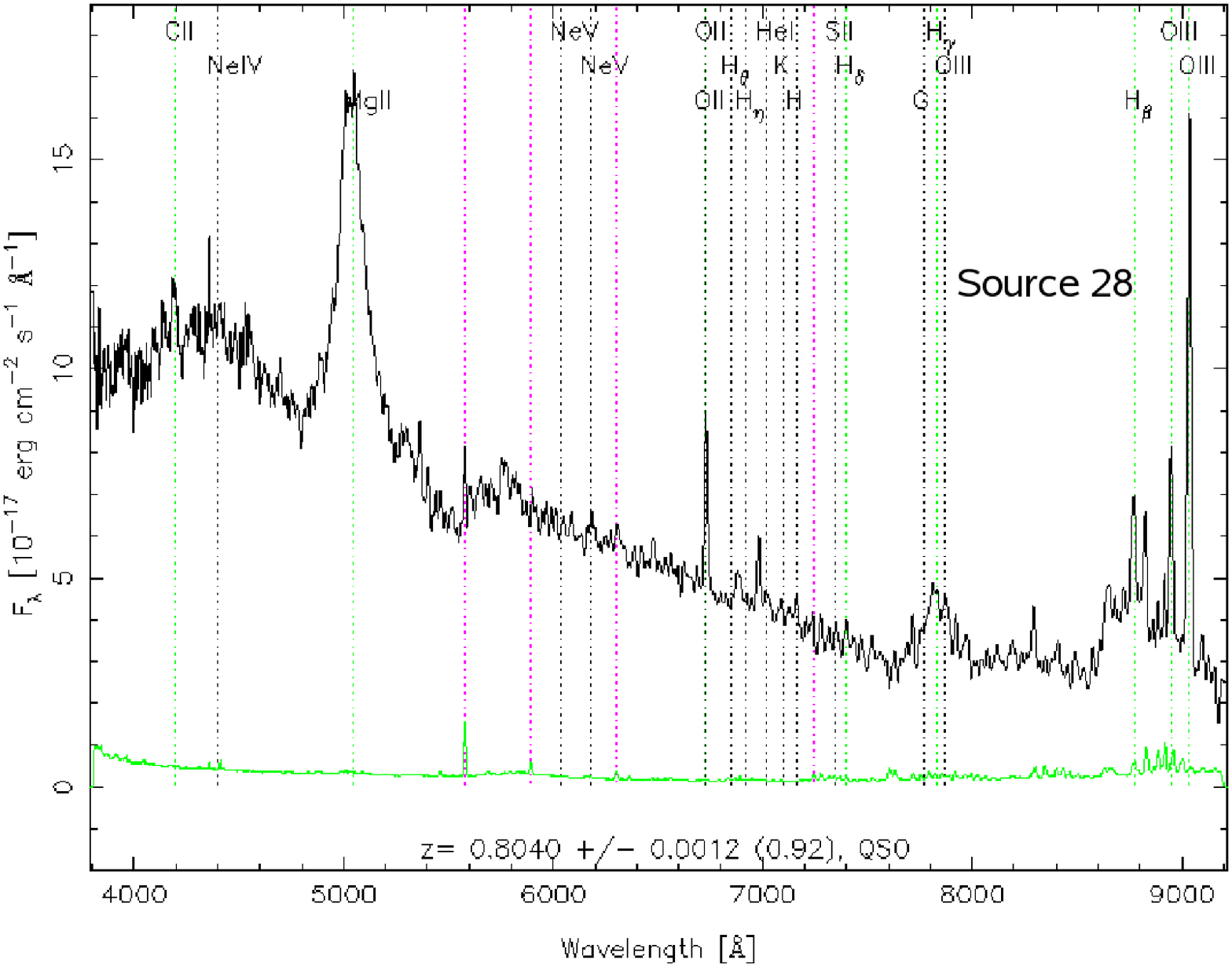} \\
\end{tabular}
  \end{center}
\contcaption{ }
\end{figure*}%

\begin{figure*}
  \begin{center}
    \begin{tabular}{c} 
 \includegraphics[height= 75mm, width=130mm]{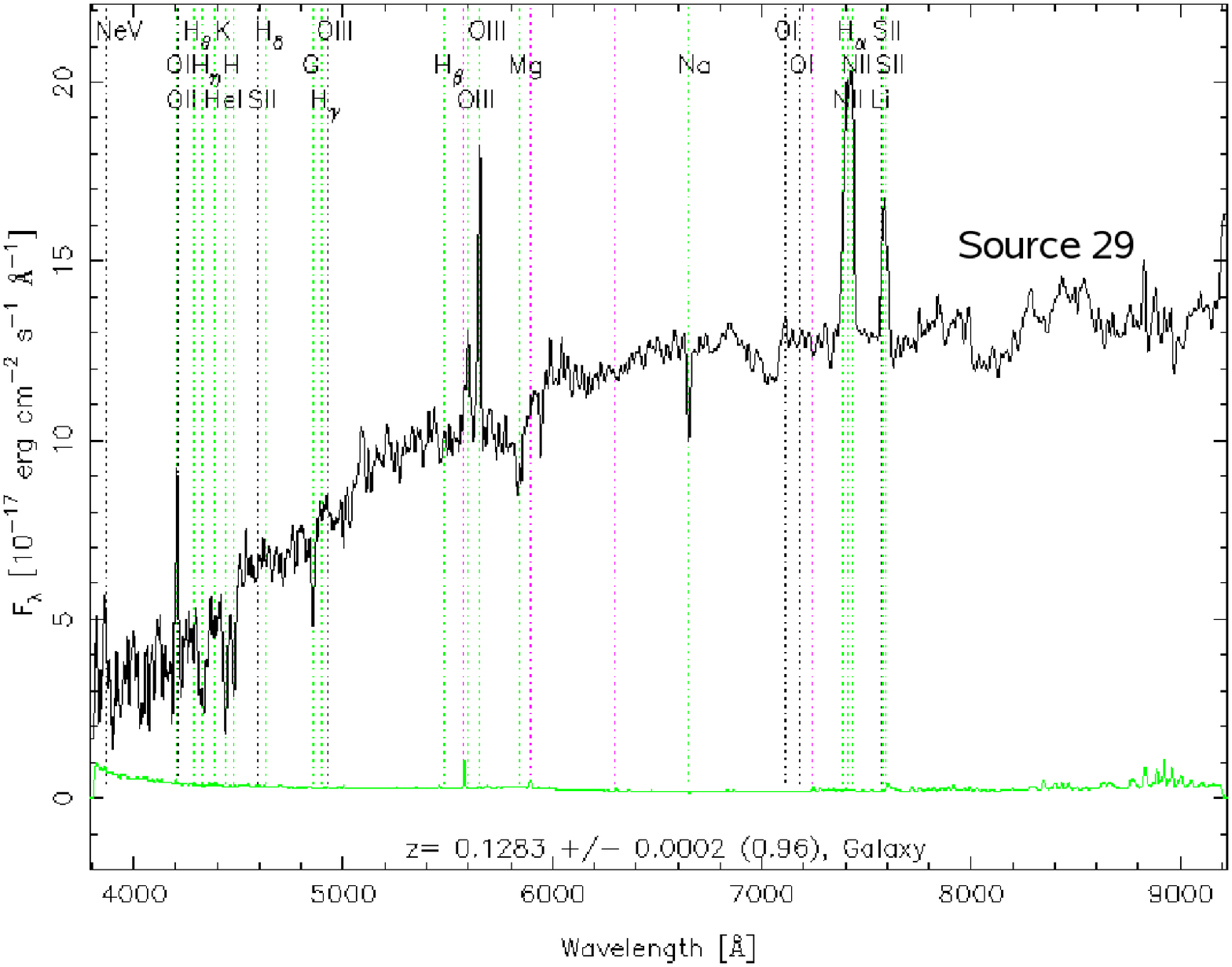}\\
 \includegraphics[height= 75mm, width=130mm]{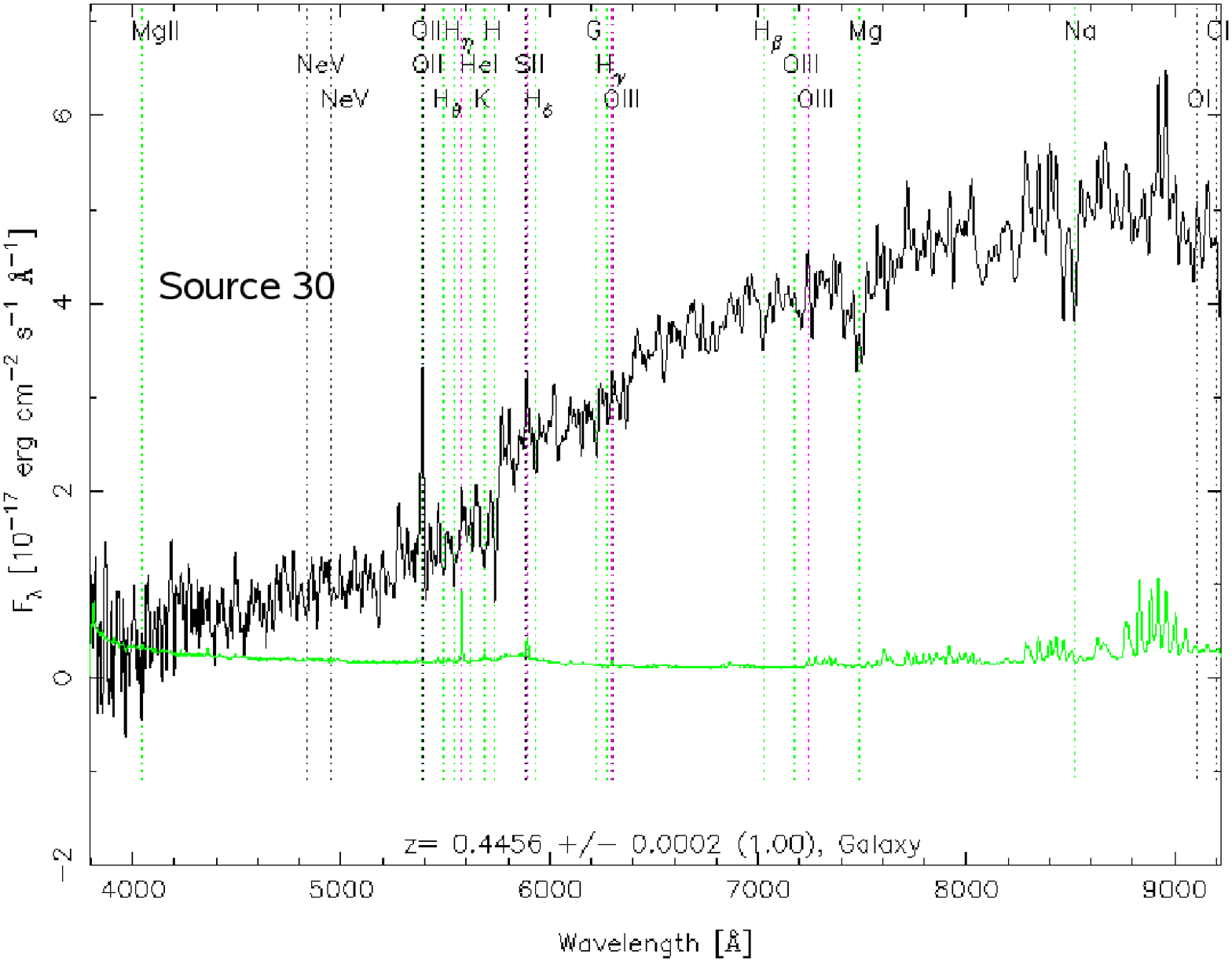} \\
\includegraphics[height= 70mm, width=130mm]{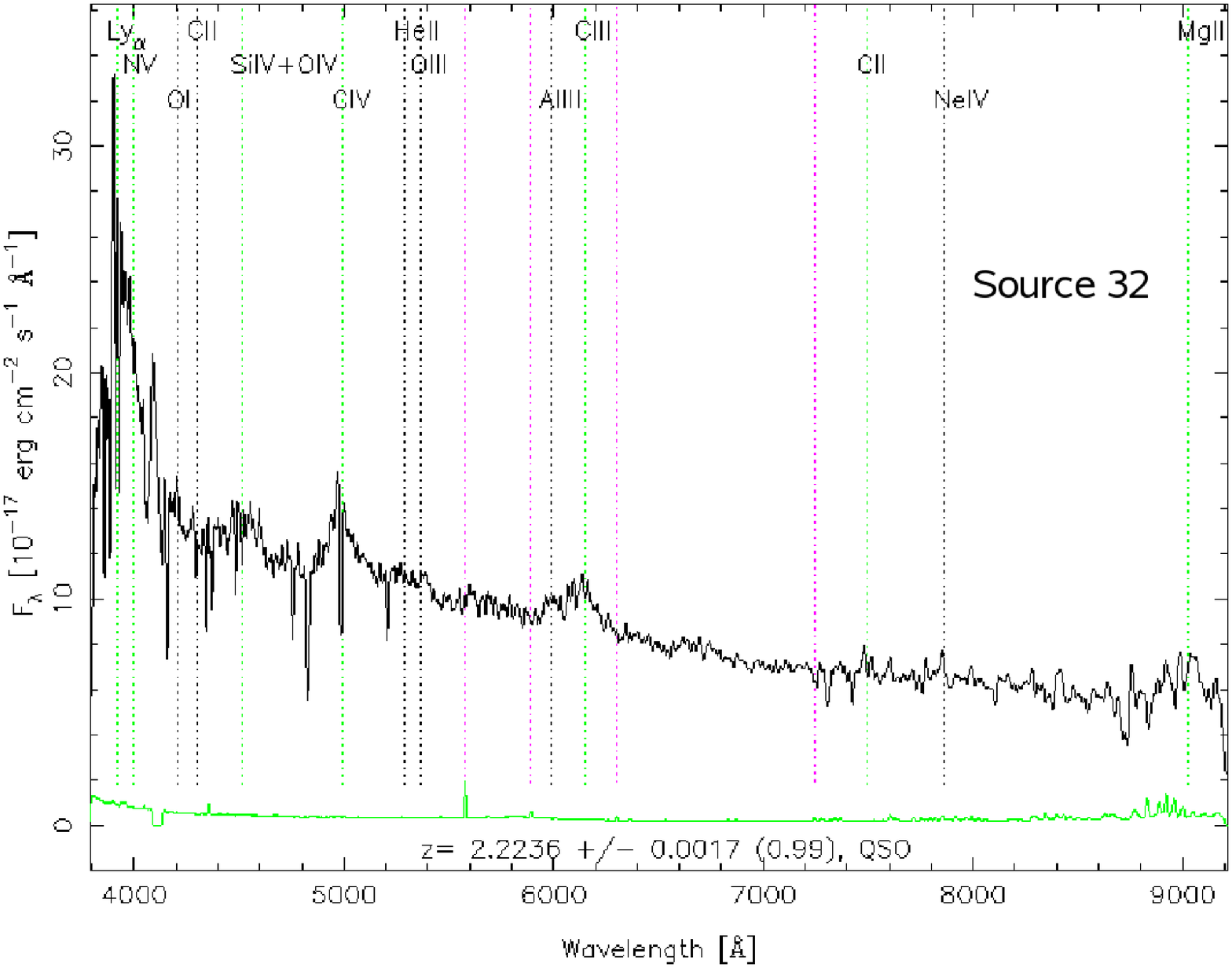} 
\end{tabular}
  \end{center}
\contcaption{ }
\end{figure*}%

\begin{figure*}
  \begin{center}
    \begin{tabular}{c} 
  \includegraphics[height= 75mm, width=130mm]{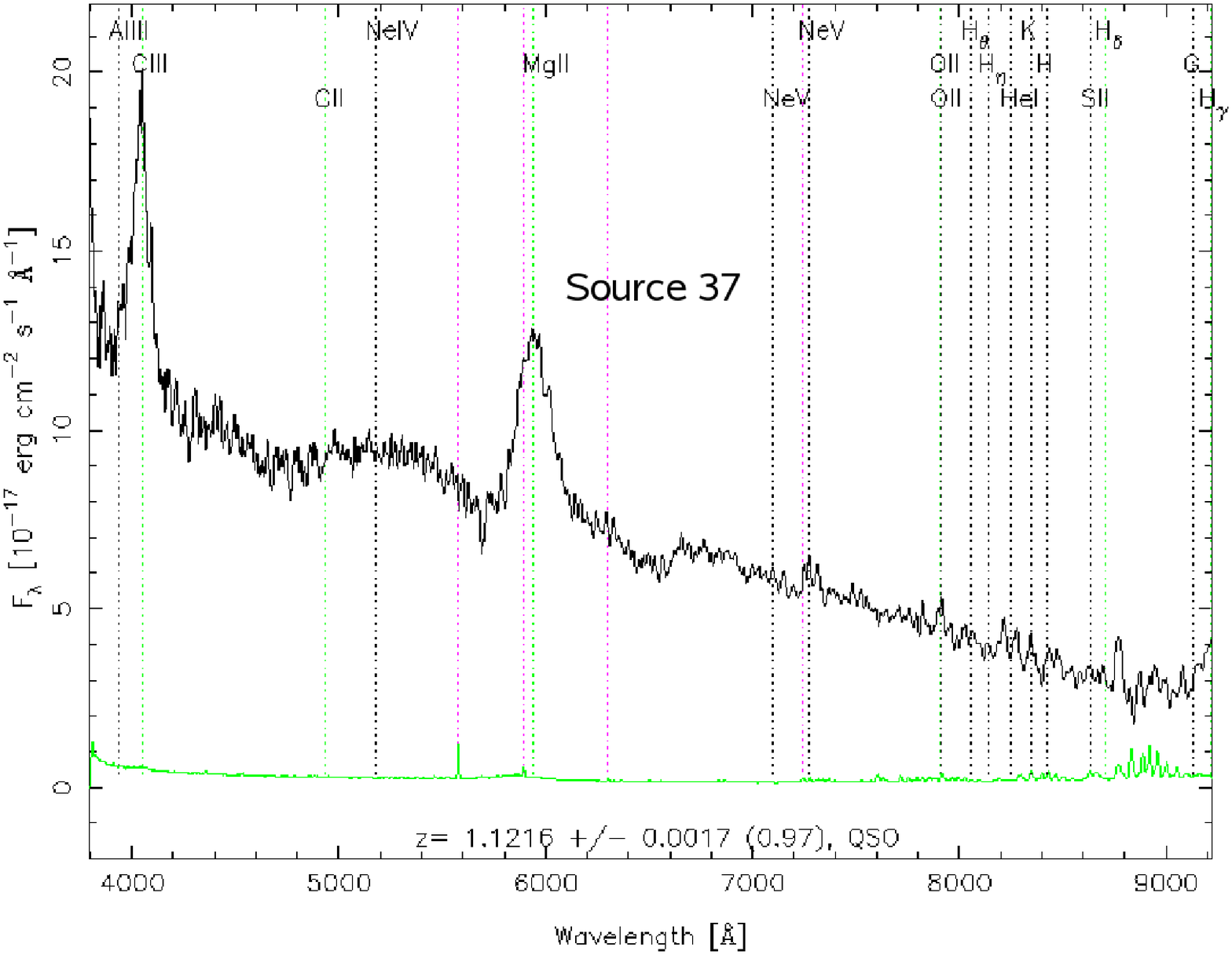} \\
   \includegraphics[height= 75mm, width=130mm]{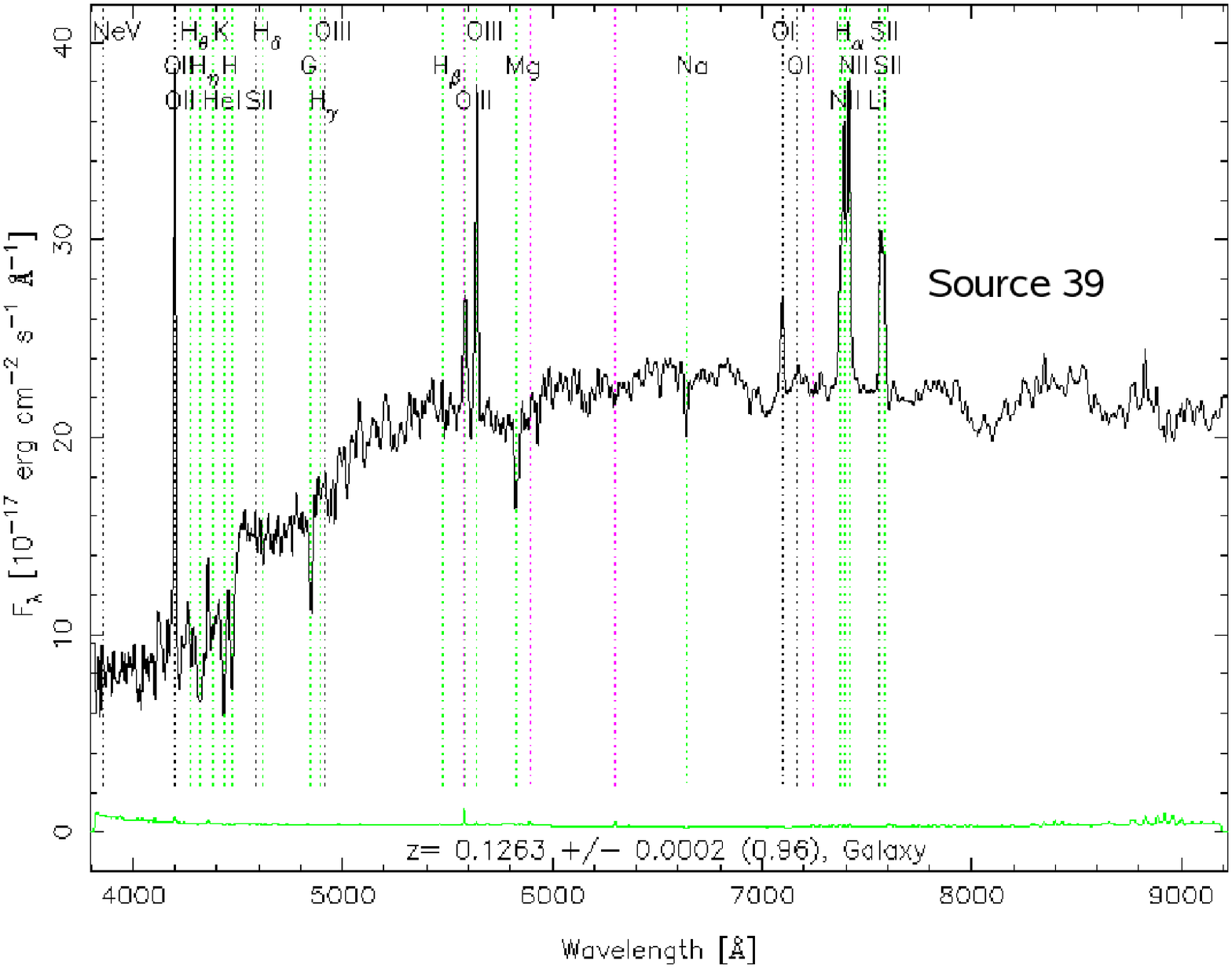}  \\
 \includegraphics[height= 90mm, width=130mm]{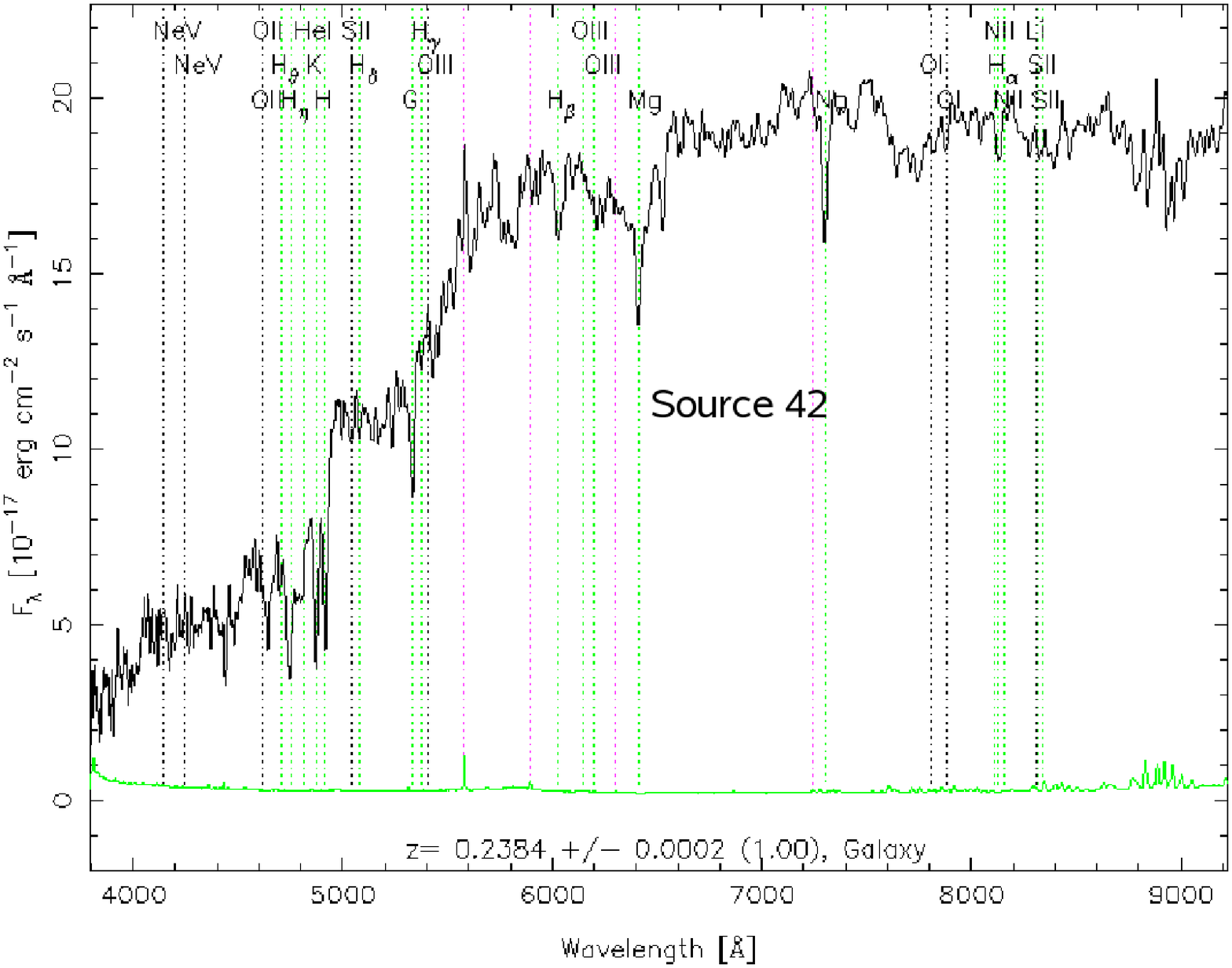}  \\
\end{tabular}
  \end{center}
\contcaption{ }
\end{figure*}%

\begin{figure*}
  \begin{center}
    \begin{tabular}{c} 
\includegraphics[height= 75mm, width=130mm]{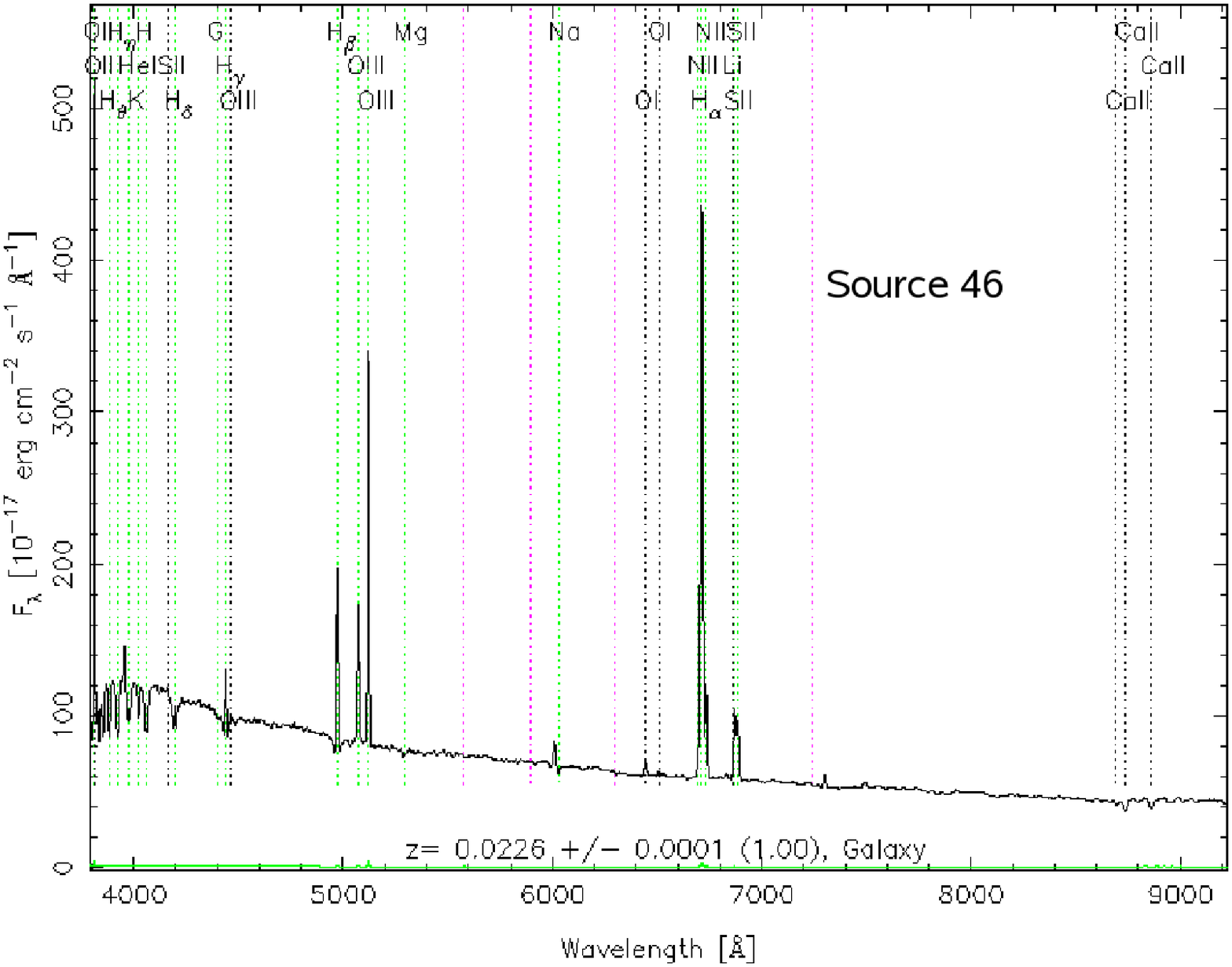} \\
   \includegraphics[height= 75mm, width=130mm]{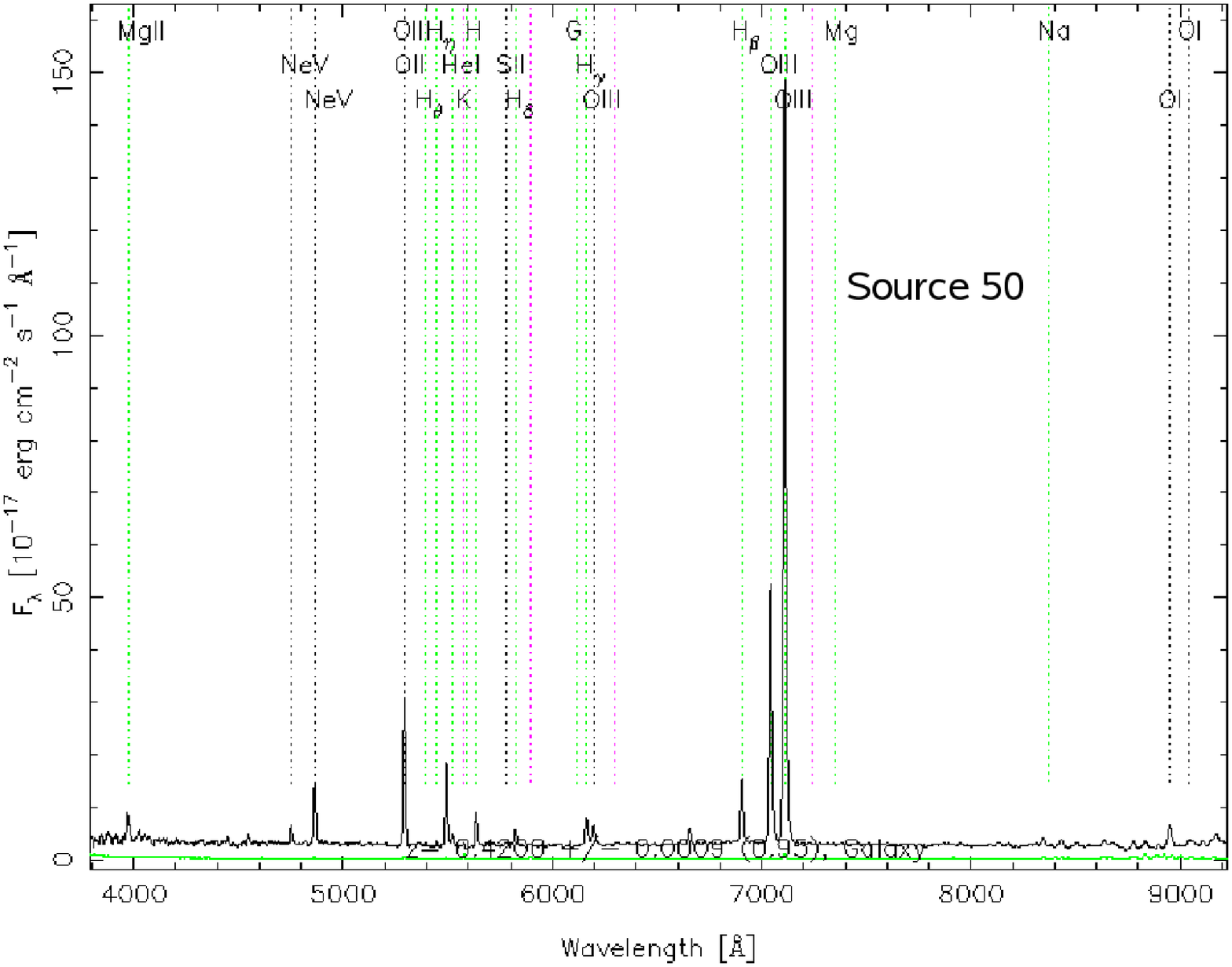} \\
\includegraphics[height= 75mm, width=130mm]{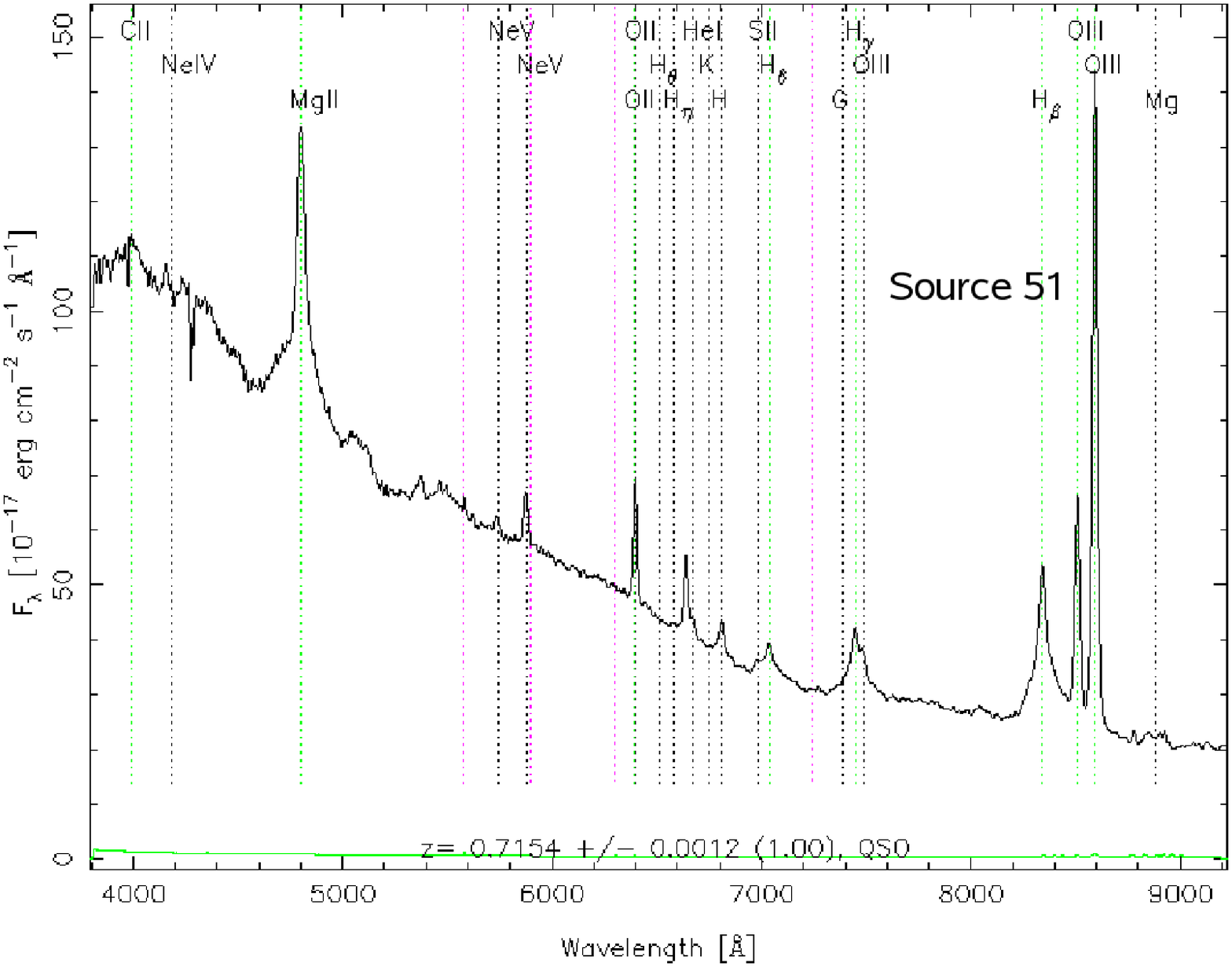} \\
    \end{tabular}
  \end{center}
\contcaption{ }
\end{figure*}
\newpage

\begin{figure*}
  \begin{center}
    \begin{tabular}{c} 
 \includegraphics[height= 75mm, width=130mm]{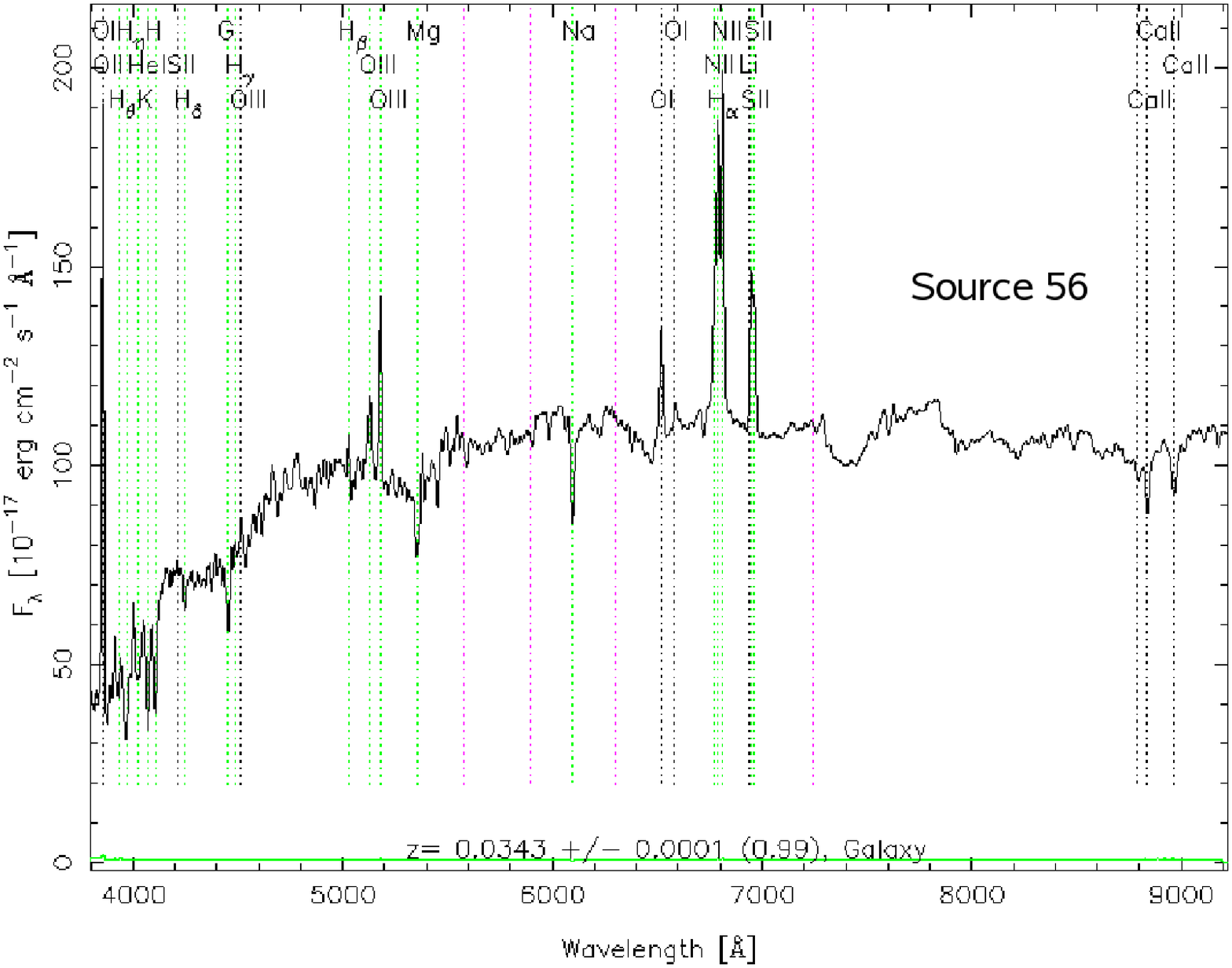} \\
 \includegraphics[height= 75mm, width=130mm]{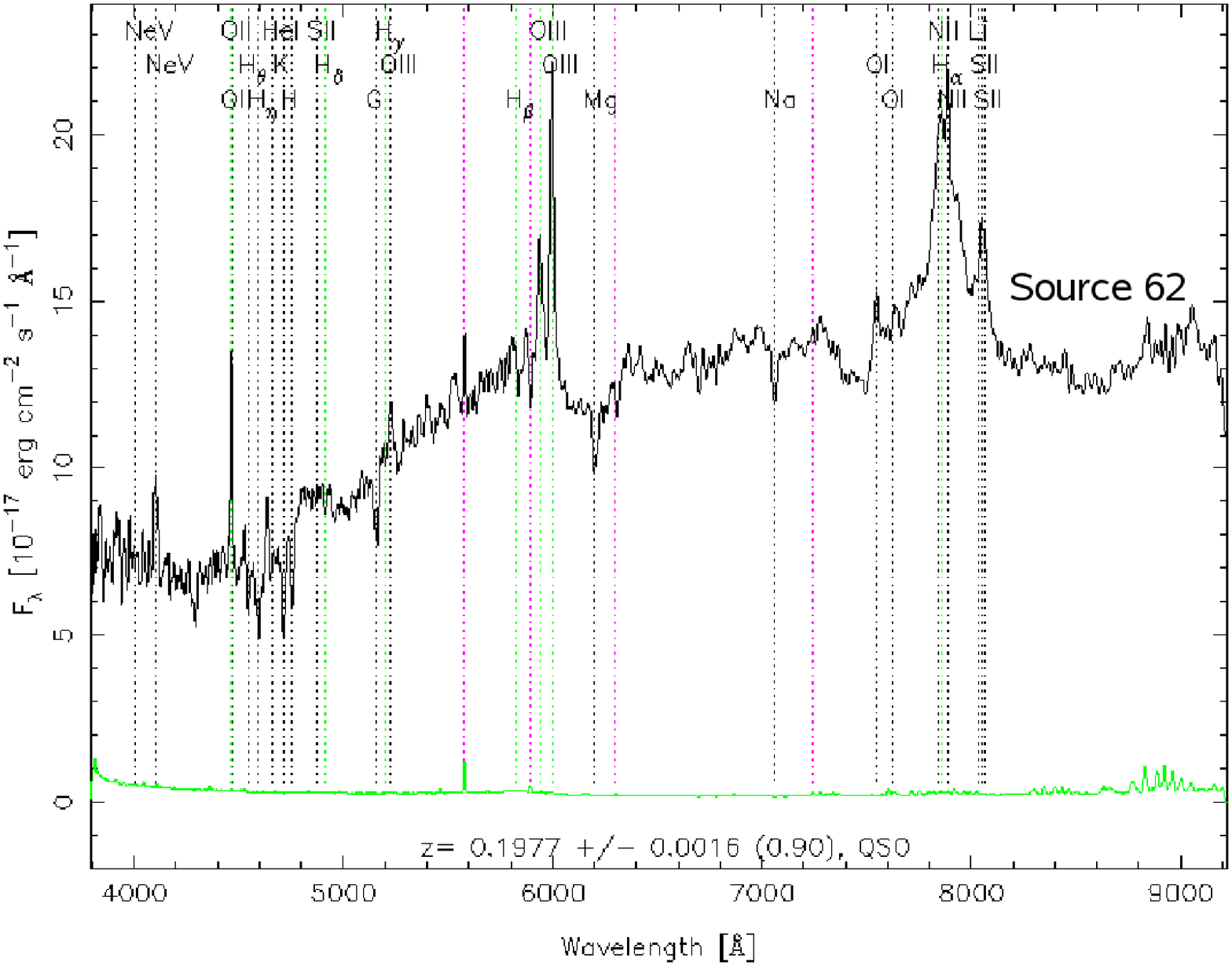} \\
 \includegraphics[height= 75mm, width=130mm]{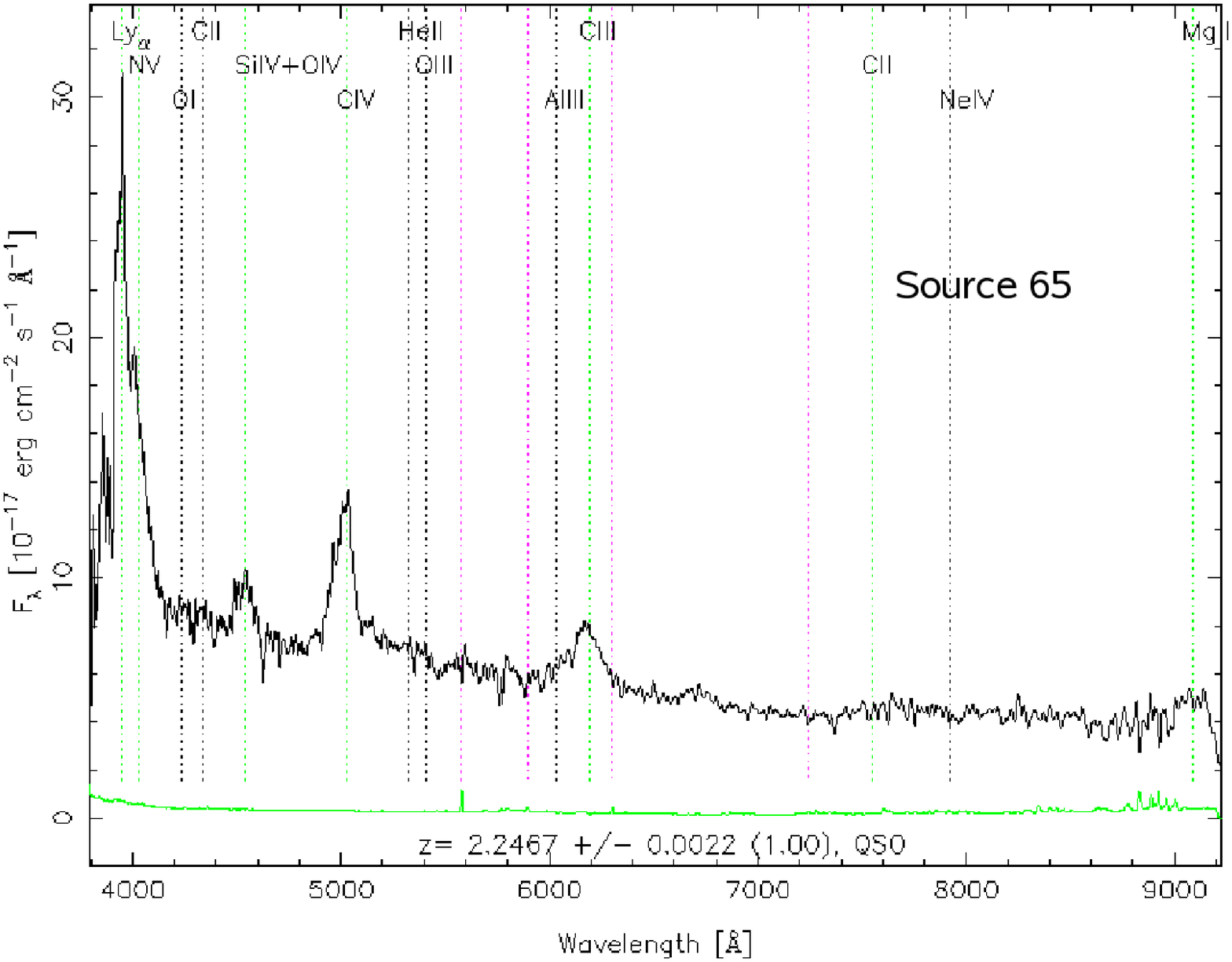}\\
  \end{tabular}
  \end{center}
\contcaption{ }
\end{figure*}

\begin{figure*}
  \begin{center}
    \begin{tabular}{c}
  \includegraphics[height= 75mm, width=130mm]{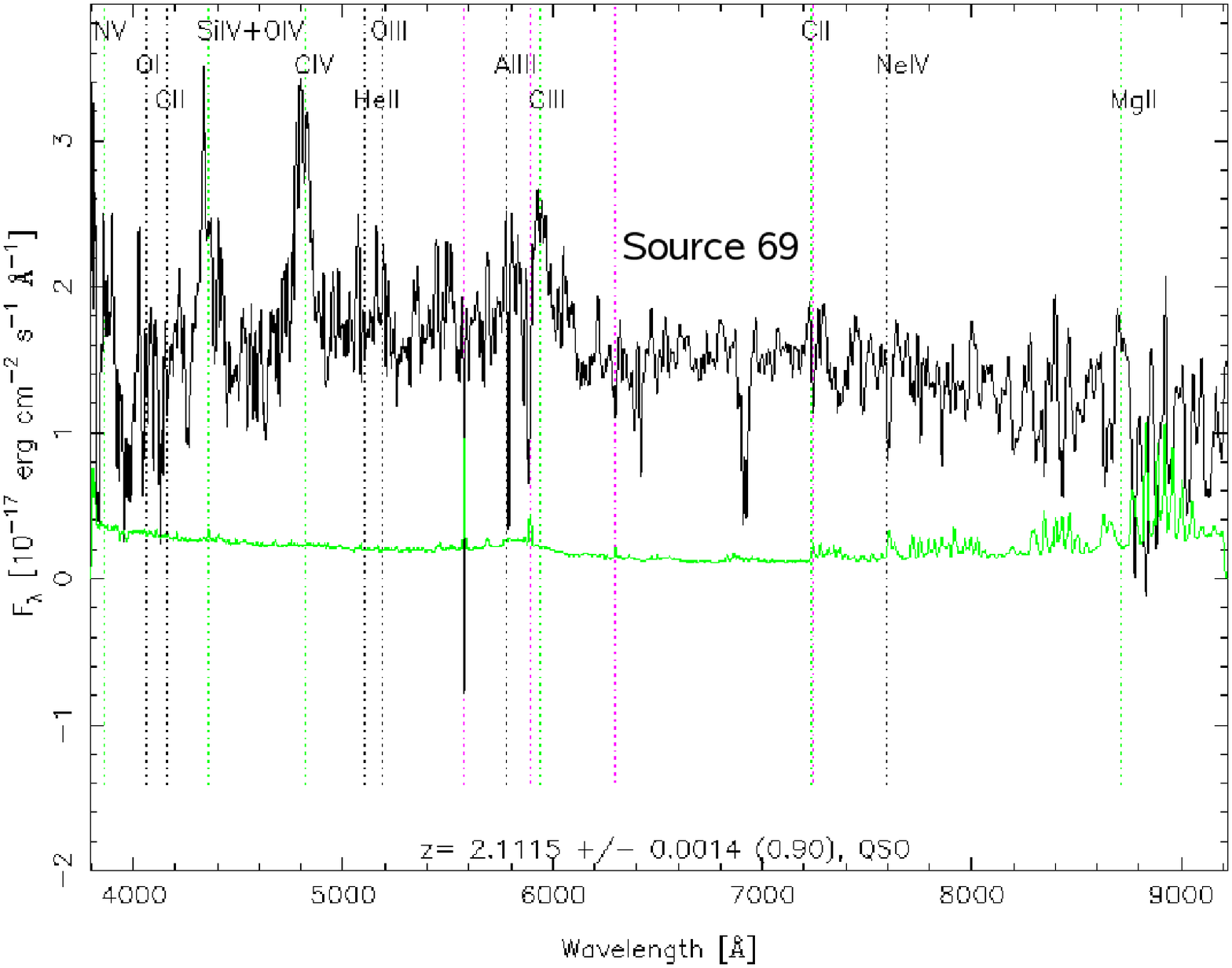} \\
\includegraphics[height= 75mm, width=130mm]{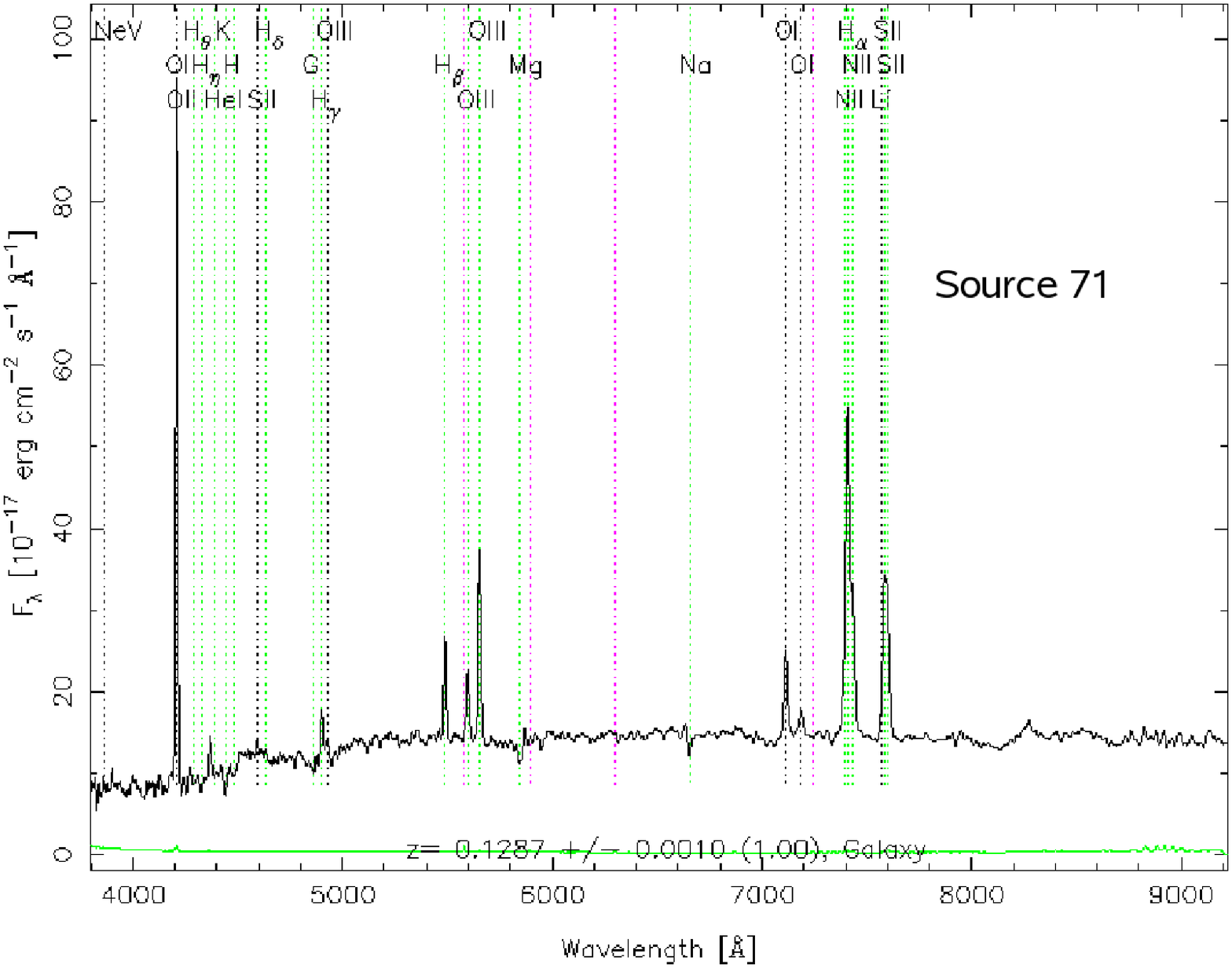} \\
    \includegraphics[height= 75mm, width=130mm]{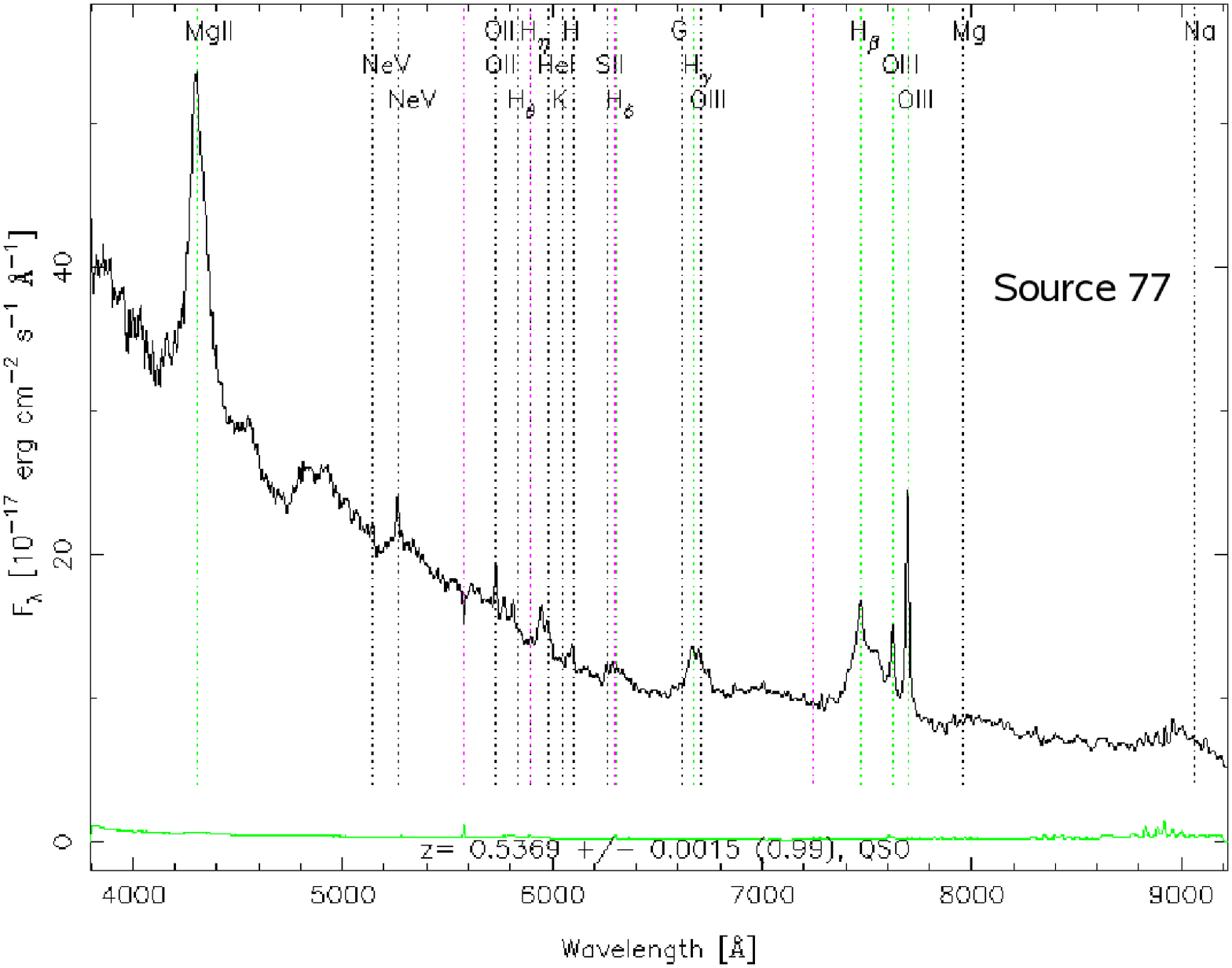}\\
 \end{tabular}
  \end{center}
\contcaption{ }
\end{figure*}
\begin{figure*}
  \begin{center}
    \begin{tabular}{c}
 \includegraphics[height= 75mm, width=130mm]{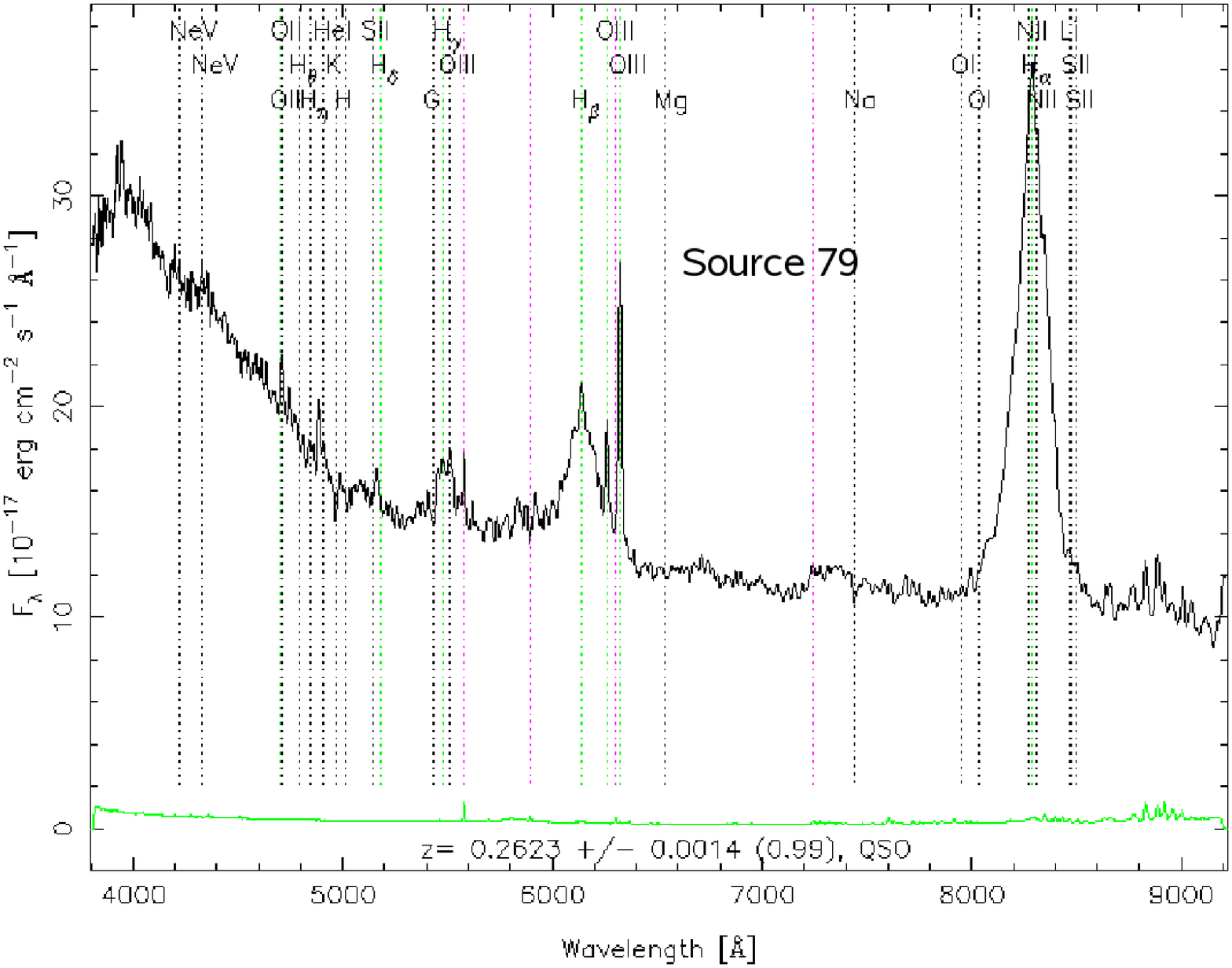}   \\
    \end{tabular}
  \end{center}
 \contcaption{ }
\end{figure*}

\end{document}